\documentclass[usenatbib,letter]{mnras}

\usepackage{graphicx,rotating}
\usepackage{amssymb,amsmath,pdflscape}
\usepackage{natbib,cases}

\newcommand{\re}{R_{\rm e}}

\newcommand{\rhoatm}{\rho_3}

\newcommand{\rpol}{R_{\rm p}}

\newcommand{\irot}{{i_{\rm rot}}}

\newcommand{\aenv}{a_2}
\newcommand{\ellenv}{\epsilon_2}
\newcommand{\ellcor}{\epsilon_1}

\newcommand{\compell}{\bar{\epsilon}}
\newcommand{\compellenv}{\bar{\epsilon}_2}
\newcommand{\compellcor}{\bar{\epsilon}_1}

\newcommand{\omobs}{\Omega_{\rm obs}}

\newcommand{\rhoenv}{\rho_2}

\begin{document}

\title[]{Nested spheroidal figures of equilibrium\\ III. Connection with the gravitational moments $J_{2n}$}

\author[B. Basillais and J.-M. Hur\'e]
       {B. Basillais$^{3}$\thanks{E-mail:baptiste.boutin-basillais@uzh.ch} and J.-M. Hur\'e$^{1,2}$\thanks{E-mail:jean-marc.hure@u-bordeaux.fr}\\
$^{1}$Univ. Bordeaux, LAB, UMR 5804, F-33615, Pessac, France\\
$^{2}$CNRS, LAB, UMR 5804, F-33615, Pessac, France\\
$^{3}$Institute for Computational Science, University of Zurich, Winterthurerstr. 190, 8057, Zurich, Switzerland}
       
\date{Received ??? / Accepted ???}
 
\pagerange{\pageref{firstpage}--\pageref{lastpage}} \pubyear{???}

\maketitle

\label{firstpage}

\begin{abstract}
We establish, in the framework of the theory of nested figures, the expressions for the gravitational moments $J_{2n}$ of a systems made of ${\cal L}$ homogeneous layers separated by spheroidal surfaces and in relative rotational motion. We then discuss how to solve the inverse problem, which consists in finding the equilibrium configurations (i.e. internal structures) that reproduce ``exactly''  a set of observables, namely the equatorial radius, the total mass, the shape and the first gravitational moments. Two coefficients $J_{2n}$ being constrained per surface, ${\cal L}=1+\frac{n}{2}$ layers ($n$ even) are required to fix $J_2$ to $J_{2n}$.
  As shown, this problem already suffers from a severe degeneracy, inherent in the fact that two spheroidal surfaces in the system confocal with each other leave unchanged all the moments. The complexity, which increases with the number of layers involved, can be reduced by considering the rotation rate of each layer. Jupiter is used as a test-bed to illustrate the method, concretely for ${\cal L}=2,3$ and $4$. For this planet, the number of possible internal structures is infinite for ${\cal L} > 2$. Intermediate layers can have smaller or larger oblateness, and can rotate slower or faster than the surroundings. Configurations with large and massive cores are always present. Low-mass cores (of the order a few Earth masses) are predicted for ${\cal L} \ge 4$.
The results are in good agreement with the numerical solutions obtained from the Self-Consistent-Field method.
\end{abstract}


\section{Introduction}

The exterior gravitational potential of a body is usually expandable in series over the spherical distance $r$ from the center. Under axial and equatorial symmetries, this is
\citep[e.g.][]{kellogg29,heiskanen1976physical}
\begin{flalign}
 \Psi_{\rm ext.}(\boldsymbol{r}) = -\frac{GM}{r}\left[1 -\sum_{n=1}^\infty{ J_{2n}\left(\frac{a}{r}\right)^{2n} P_{2n}(\mu)}\right],
 \label{eq:psiext}
\end{flalign}
where $P_{2n}$ is the Legendre polynomial of order $n$, $\mu = \cos \theta$ ($\theta$ being the colatitude measured from the $z$-axis), $a \le r$ is the reference equatorial radius of the system, $M$ is the total mass, and the coefficents $J_{2n}$ are defined by
\begin{equation}
M a^{2n} J_{2n} = 2 \upi \iint_{\rm body}{\rho(r,\mu) r^{2n} P_{2n}(\mu) r^2d\mu dr}.
\label{eq:j2ndef}
\end{equation}
As this potential fully governs the motion of any test-particle orbiting around the body, there is a tremendeous opportunity of setting contraints on the mass-density distribution $\rho(r,\mu)$ inside the body if certain data are known, namely, $a$, $M$ and especially the $J_{2n}$'s. On this basis, our understanding of planetary interiors has been largely improved since the $1970$'s through space probes travelling the Solar System. The recent measurements of the first gravitational moments of Jupiter \citep{fo17,durante20} combined with complex models based on sophisticated equation-of-states, transport mechanisms, surface winds etc. offer yet new opportunities to better unveil the properties of material rotating at depth \citep{wahl17,helled18}. The capabilitiy of models to correctly outputing the $J_{2n}$'s | despite these coefficients are not the most appropriate quantities to probe the very central regions | remains a major challenge \citep{gui99,mgf16,nettelmann2017,cmm19,Neuenschwander21,net21}. The case of Jupiter is singular in the sense that the even and odd moments up to $J_{12}$ have been measured, with a relatively high level of precision compared to Saturn for instance \citep{iess19}, ranging from $10^{-8}$ for $J_2$ to $1 \%$ for $J_8$ (error bars are larger beyond).

\subsection{Rotation and surface bounding layers}

A major difficulty in the modeling of the internal structure of planets, already known from classical theories \cite[e.g.,][]{1903volterra,veronet12,alma990001099760306161}, comes from rotation, which is the result of both formation and evolution of the body towards some quasi stable-state. As in most studies dealing with self-gravitating systems, the rotation profile is generally prescribed, with a preference for simple laws like rigid rotation \citep[e.g.][]{hachisu86}. A famous solution is due to Maclaurin (based on Newton's theorem about homoeoids): a homogeneous body bounded by a spheroidal surface in rigid rotation is a perfect equilibrium. Unfortunately, it is not possible to construct a composite body in global rotation just by using pieces of Maclaurin spheroids. Actually, piling layers up does not automatically ensure a null force-budget everywhere, and pressure balance at each interface. As soon asserted by \cite{poincare88}, {\it only layers bounded by confocal spheroidal surfaces are strictly compatible with a rigidly rotating body}. \cite{hamy90} has completed this theorem by showing that the mass-density profiles required for confocal equilibria, regardless of the number ${\cal L}$ of layers involved, are not physically pertinent. This result has been established in another context by \cite{ak74} and by \cite{mmc83}. As quoted in \cite{veronet12}, and more recently shown in \cite{chambat1994non} and by recursion in \cite{pohanka11} and in \cite{,h2022b,h2022a}, this holds in the continuous limit where ${\cal L} \rightarrow \infty$. These classical results are robusts, and they have been fully confirmed by the numerical solutions obtained by solving the ${\cal L}$-problem from the Self-Consistent-Field method, in the incompressible case and for non-zero polytropic indices as well \citep{kiu10,ka16,bh21}. Poincar\'e's theorems and others clearly state that globally rotating, inhomogeneous bodies having spheroidal bounding-layers that are not confocal to each others are necessarily approximate. Exact (or, at least, self-consistent) solutions can be recovered by relaxing one of the underlying hypothesis. In fact, for global rotation, not only isobaric/isopycnic surfaces are generally not similar surfaces, but these slightly deviate from ellipses, rigorously  \citep{1903volterra,veronet12}; see for instance \cite{cmm17} who consider depressed spheroids. On this basis, \cite{zt70} have shown, from a bivariate expansion of the surface levels for the pressure (or effective potential) combined with a minimization procedure, that the first moments of Jupiter are fully accessible. A $7$th-order treatment enables to reach $J_{12}$ with only four layers \citep{net21}. Deviations with respect to perfect ellipses are also enabled in the model of Concentric Maclaurin Spheroid (CMS) proposed by \cite{hub13}. These approaches are widely applied to model Jupiter and Saturn \citep{net17,mwh19,ni20} in global rotation.
  
\subsection{Comments on estimating the moments $J_{2n}$'s}

As (\ref{eq:j2ndef}) suggests, the accurate determination of the gravitational moments can be problematic. This is especially the case when $\rho(\mathbf{r})$ is not analytical, but results from some discretization. A first difficulty comes from the fact that the coefficients are expected to rapidly decrease with the order $n$. This is, for instance, $J_2/J_4 \approx 2 \times 10^{3}$ for the Sun \citep{rk2021}, and $J_2/J_{10} \approx 10^5$ for Jupiter \citep{durante20}. High resolutions are undisputably required to get both large and tiny values \citep[e.g.][]{dc18}. In order to get $J_2$ with a relative accuracy of a few $10^{-8}$ (and subsequently a few $10^{-6}$ for $J_4$ which is smaller than $J_2$ by a factor about $25$) from second-order discretization schemes, a numerical resolution better than about $10^{-4}$ is in principle sufficient. This requires a computational grid carrying at least $10^3 \times 10^3$ nodes in total (under axial symmetry), which may become prohibitive in terms of computing time for a recurrent use. Second, the presence of sharp density-gradients and even jumps between layers (due to changes in the equations-of-state) is another severe obstacle, in particular for spectral methods, which are in principle more powerful than local methods provided the mass-density is a smooth, derivable function of the coordinates. Another important source of technical dificulty comes from the specific form of the kernels (namely, the presence of Legendre polynomials $P_{2n}$) \citep[e.g.][]{mm21}. For any functions taking positive and negative values (the case of the $P_{2n}$'s), the efficiency of numerical quadratures is generally reduced, due to substractive cancellations, and this becomes critical for relatively tiny quantities. Finally, the interface between layers, deformed by rotation, must be localized with an extreme precision, especially in the purpose of this kind of numerical integrations. It follows that any mass-density profile, even defined as a piece-wise function of spatial coordinates, capable of removing or reducing the uncertainties in estimating the $J_{2n}$'s are of great importance. This is the case of homogeneous layers considered here.

\subsection{Note on the theory of nested figures. Motivation}

In this article, we use the theory of nested spheroidal figures reported in  \cite{h2022b,h2022a} (hereafter, Paper I and Paper II respectively) to establish the link between the gravitational moments $J_{2n}$ and the internal structure of a body made of ${\cal L}$ heteroeoidal layers  (i.e. layers bounded by {\it non-similar spheroidal surfaces}) in asynchronous motion (i.e. layers can rotate at {\it different rates}). By structure, we mean the set of ellipses $\{E_i(a_i,b_i)\}_{\cal L}$ bounding the layers, the set of mass-densities  $\{\rho_i\}_{\cal L}$ from the center to the surface, and the individual rotation rates $\{\Omega_i\}_{\cal L}$. Actually, the equilibrium of the system requires a specific dynamical setup, and it is not possible for all $\{E_i,\rho_i,\Omega_i\}_{\cal L}$. While the formalism of nested spheroids is an approximate theory, it is designed for spheroids sharing {\it small confocal parameters $c_{i,j}$}, i.e
  \begin{equation}
   |c_{i,j}|=\frac{1}{a_j^2}\left|a_i^2-b_i^2-(a_j^2-b_j^2)\right| \ll 1,
  \end{equation}
  which, in the case of slow rotations, corresponds to small ellipticities $\epsilon_i$, with $\epsilon_i^2 =1-b_i^2/a_i^2$.

\subsection{Preliminary note about the relevance of the method to Jupiter (and to other planets)}

Some assumptions underlying the theory of nested figures are quite restrictive, and may not be nominal for all applications. The hypothesis of incompressible layers (null polytropic index) is probably well suited for telluric or ocean-planets. This is more critical for a gaseous planets or stars. In addition, meridional circulations (i.e. ascending/descending fluid currents) may perturb significantly the pure hydrostatic balance, close to the body's surface for instance. Some layers can exhibit a latitude-dependent rotation state (possibly, as a consequence of the Coriolis force and geostrophic streams), which is out of range of the present study. In order to make the problem tangible and illustrate the method, a concrete case must be considered. We use Jupiter\footnote{Other planets in the Solar System can be analyzed. In terms of avalaible data, the situation is very similar for Saturn, although the uncertainty level in the $J_{2n}$'s is globally a little bit higher \citep{iess19}. For Neptune and Uranus, there are much less constraints.} as a ``test-bed'' (the even moments are known with accuracy up to $n=6$). Actually, as the mass-density is prescribed (and constant in each layer), we can address the usual problem of the $J_{2n}$'s in the reverse sense: these coefficients, together with the radius, the mass, the flattening and the rotation rate of the planet core, are treated as {\it input parameters}. This is a considerable simplification. In most studies \citep[e.g.][]{zt70,hub13}, the best structures are found through a minimization procedure over the moments, which are output quantities.

  It is well admitted that Jupiter is mainly composed of hydrogen and helium, but there are still uncertainties regarding the interior of Jupiter, and in particular at great depth, which is not accessible through photon emission. We do not know the chemical composition in radius in detail, including the proportion and distribution of heavier elements. Another matter of debate is the mass density profile: are there well separated layers, or is the transition from the center to the surface rather smooth \cite[e.g.][]{miguel2022arXiv220301866M} ? Recent models depicts Jupiter as being made of $3$ main layers with a dense core of heavy elements, an inner region of metallic hydrogen and helium, and an outer envelope composed of molecular hydrogen and depleted in helium \citep{militzer2016understanding,helled18}. Observations from the \textit{Juno} probe seems to indicate the presence of a large and dilute core which extends up to about half of the planet, thereby adding a possible new layer to the structure \citep{wahl17,ni2019understanding}. This brings new questions regarding the formation of such a dilute core \citep{liu2019formation,muller2020challenge}.

At the surface, Jupiter exhibits a complex dynamical structure, with fast leading and trailing zonal winds spread over $\sim 3000$ km in depth \citep{guillot2018Nat,kaspi2018Nat,kaspi20}, superimposed to a dominant rigid motion. Even if the dynamics of this layer is slightly altered by Coriolis forces, it is very thin at the scale of the planet, e.g. $\lesssim 4 \%$ in size, and it contains a tiny fraction of the total mass. The depth where zonal winds are predicted is probably not large enough to change drastically the low-order gravitational moments $J_{2n}$ ($n <5$ typically). The contribution to the first gravitational moments of the winds relative to the deep interior is $1$ to $3$ orders of magnitude lower, and there is no clear indication of the rotation state, beleved to be solid-like \citep{kaspi13,kaspi17}. At great depth, the density seems to varies very slowly with the radius \cite[e.g.][]{wahl17,Neuenschwander21}.

We easily understand that, given the complexity of the planet, the present approach (in its current state at least), is not supposed to compete with or even to replace other models nor to bring new decisive insights. However, we show that unlocking the actual constraints on the ellipticity of spheroidal surfaces \citep[see, e.g.,][]{cmm19}, and by desynchronizing the motion of layers \citep{1903volterra} open new horizons in terms of variety of solutions. As we will see, ${\cal L}=\frac{n}{2}+1$ layers are sufficient to reproduce the even zonal harmonics up to $J_{2n}$ (i.e. four layers are sufficient to match even values $J_{2}$ to $J_{12}$). There is a single two-layer solution compatible with Jupiter's data, but an inifinity of equilibria with $3$ layers and more, even if we exclude the solutions that imply a surface layer with relative extent of a few purcents.

\subsection{Content of the article}

The article is organized as follows. We give in Sec. \ref{sec:theory} the equation set for the mass $M$ and for the $J_{2n}$'s of a inhomogeneous ${\cal L}$-layer body made of homogeneous heteroeoids. Some constraints about the relative geometry of the spheroidal surfaces in the sample and the mass-density stratification are given. Next, we recall the conditions of equilibrium of the entire structure, which are defined by the expressions for the rotation rate of all layer in the sample. In particular, we get a fundamental equation by matching the rotation rate of one of the layers in the composite body to an known value. All the problem is presented in a scale-free version. Sections \ref{sec:2layers} to \ref{sec:4layers} are devoted to the $2$-layer, $3$-layer and $4$-layer problems, respectively, by using Jupiter's data. The impact of errors bars in the $J_{2n}$'s are considered.  The {\tt DROP}-code, that numerically solves the ${\cal L}$-layer problem from the Self-Consistent-Field method, is used in support \citep{bh21}, and some examples are given. The last section is devoted to a discussion. A few perspectives are also given.

\section{Theoretical background}
\label{sec:theory}

\subsection{Mass and gravitational moments of a homogeneous spheroid}

For certain density distributions, the coefficients $J_{2n}$ in (\ref{eq:j2ndef}) are fully analytical. In particular, when the body is a homogeneous ellipsoid of revolution (i.e. a spheroid), an appropriate form for the integral in (\ref{eq:j2ndef}) is
\begin{flalign}
  \iint_{E(a,b)}{ r^{2n+2} P_{2n}(\mu) d\mu dr} =  \frac{2}{3} a^{2n+3} \compell \times j_{2n} \epsilon^{2n},
\label{eq:j2n}
\end{flalign}
where $\epsilon$ is the ellipticity of the bounding surface $E(a,b)$, $a$ and $b$ are the semi-major and semi-minor axis respectively, $\compell=\sqrt{1-\epsilon^2}=b/a$ is the axis ratio, the mass is $M = \rho V$ (the volume is $V=\frac{4}{3}\upi a^3 \epsilon$), and the coefficient $j_{2n}$ is given by \citep[see also][]{heiskanen1976physical,cis19}
\begin{flalign}
  j_{2n} = -\frac{3(-1)^n}{(2n+1)(2n+3)}, \qquad n > 0.
\label{eq:smallj2n}
\end{flalign}
As (\ref{eq:j2ndef}) is linear in $\rho$ (this property is intrinsic to the gravitational potential), the superposition principle applies: the contribution to $M$ and $J_{2n}$ to any extra mass density distribution $\delta \rho$ is simply obtained by adding the corresponding extra mass $\delta M$ and the extra moment $\delta J_{n}$, respectively.

\begin{figure}
    \centering
    \includegraphics[width=8.2cm,bb=0 0 545 531]{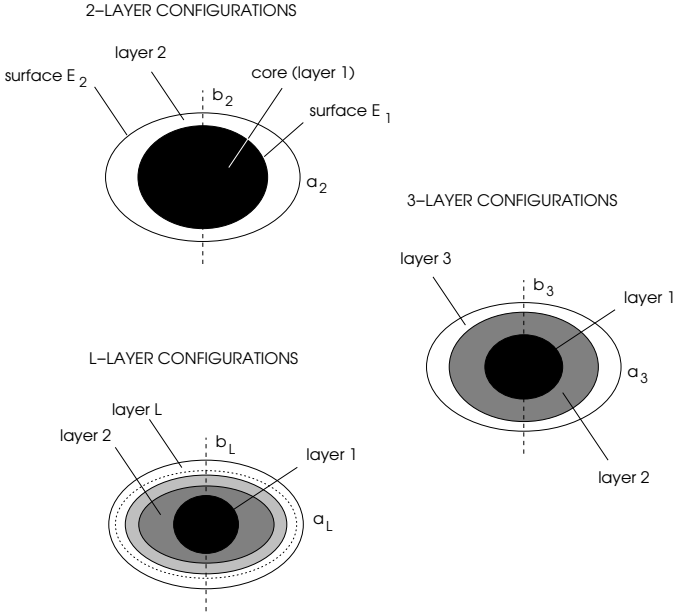}
    \caption{The composite body made of homogeneous layers bounded by spheroidal surfaces and in relative rotation.}
    \label{fig:configs.eps}
\end{figure}

\subsection{Case of a body made of ${\cal L}$ layers bounded by spheroidal surfaces}

We consider, on top a homogeneous spheroid with mass density $\rho_1$ and bounded by the surface $E_1(a_1,b_1)$ (also called a ``layer'' for convenience), the piling up of ${\cal L}-1$ homogeneous layers sharing the same axis of revolution and the same plane of symmetry, as depicted in Fig. \ref{fig:configs.eps}. Each layer (index $i \ge 1$) mechanically supports a larger layer (index $i+1$) with mass density $\rho_{i+1}$, externally bounded by a spheroidal surface $E_{i+1}(a_{i+1},b_{i+1})$. The outermost layer (index ${\cal L}$) has mass density $\rho_{\cal L}$, and is externally bounded by $E_{\cal L}(a_{\cal L},b_{\cal L})$. In these conditions, the total mass $M$ and the total $n$-order gravitational moment $J_{2n}$ of this composite system made of ${\cal L}$ heteroeoidal layers ($E_i$ and $E_{i+1}$ are not homothetical or similar) are respectively given by
\begin{flalign}
  M = \frac{4}{3}\upi \rho_{\cal L} a_{\cal L}^3 \bar{\epsilon}_{\cal L}  \left(1 + \sum_{i=1,{\cal L}-1} C_i \right),\label{eq:mj2nmassa}
  \end{flalign}

and
\begin{flalign}
M J_{2n} = \frac{4}{3} \upi j_{2n} \rho_{\cal L} a_{\cal L}^3 \bar{\epsilon}_{\cal L}  \epsilon_{\cal L}^{2n} \left(1 + \sum_{i=1,{\cal L}-1} C_i y_i^n \right),\label{eq:mj2nmassb}
  \end{flalign}

where
\begin{flalign}
  y_i=q_i^2 \frac{\epsilon_i^2}{\epsilon_{\cal L}^2} \ge 0,
  \label{eq:yell}
\end{flalign}
\begin{flalign}
  q_i=\frac{a_i}{a_{\cal L}} \le 1
\end{flalign}
is the equatorial radius of layer $i$ relative to the equatorial radius $a_{\cal L}$ of the body,
\begin{flalign}
  \alpha_i=\frac{\rho_i}{\rho_{i+1}},
\end{flalign}
is the mass-density jump between layer $i$ and layer $i+1$, and the ``coefficients'' $C_i$ are
\begin{flalign}
  C_i = (\alpha_i-1)q^3_i \frac{\bar{\epsilon}_i}{\bar{\epsilon}_{\cal L}} \times
  \begin{cases}
    \prod_{\ell=i+1}^{{\cal L}-1} \alpha_\ell,\\\\
    1, \quad \text{if } i = {\cal L}-1,
  \end{cases}
  \label{eq:cell}
\end{flalign}
where $i \in [1,{\cal L}-1]$ from (\ref{eq:yell}) to (\ref{eq:cell}). Note that $C_i$ is proportional to the excess of mass density of layer $i+1$ with respect to layer $i$, and we see that $C_i=0$ removes the $y_i$ variable in the problem, as expected (two adjacent layers merge).

Equations (\ref{eq:mj2nmassa}) and (\ref{eq:mj2nmassb}) can be written in dimensionless form, in terms of $\rho_{\cal L}/\bar{\rho}$, where is $\bar{\rho}=M/V$ is the mean mass-density, and by using the $\eta_{2n}$-parameter, defined as
\begin{flalign}
\eta_{2n} j_{2n} \epsilon^{2n} = J_{2n}, \quad n \ge 0.
\label{eq:eta2n}
\end{flalign}

Note that (\ref{eq:mj2nmassb}) is to be replicated as much as necessary to express $J_2$, $J_4$ to $J_{2n}$, which means $n$ equations.

\subsection{Conditions of immersion. Stability principle}

We assume that the spheroidal surfaces are perfectly nested and do not intersect (by more than one point at the equator $\theta=\frac{\upi}{2}$ or at the pole $\theta=0$). This enables to avoid density inversions (see below), and especially zones with negative mass densities which would be inconceivable. These ``immersion conditions'' require $q_{i+1} - q_i \ge 0$ and $b_{i+1}-b_i \ge 0$, simultaneously. The latter inequality, equivalent to $q_{i+1}\bar{\epsilon}_{i+1} - q_i\bar{\epsilon}_i >0$, can also be written in terms of the $y_i$'s. By using (\ref{eq:yell}), it follows that the immersion conditions write
\begin{subnumcases}{}
  q_{i+1} - q_i > 0, \label{eq:immersionconditionsa}\\
 q_{i+1}^2 - q_i^2 - (y_{i+1}-y_i) \epsilon_{\cal L}^2 >0, \label{eq:immersionconditionsb}
\end{subnumcases}
which mainly sets a lower limit for all the $q_i$'s. Actually, if $y_{i+1} \ge y_i$, then $q_{i+1}$ must be strictly larger than $q_i$. This is the case for instance if the oblateness decreases with depth (at worst, $E_i$ and $E_{i+1}$ are in contact at the pole). At the opposite, if the $y_{i+1} \le y_i$, then $q_{i+1}$ can be equal to $q_i$  (at worst, $E_i$ and $E_{i+1}$ are in contact at the equator). As $q_i$ can not exceed unity, we see that a large value for one of the $y_i$'s is potentially incomfortable (see Sec. \ref{sec:4layers}). This depends on the planet properties. Note that (\ref{eq:immersionconditionsa}) and (\ref{eq:immersionconditionsb}) can advantageously be combined into a single inequality, namely
\begin{flalign}
  q_{i+1} \ge \sqrt{\max\{q_i^2,q_i^2 - (y_{i+1}-y_i) \epsilon_{\cal L}^2 \}} \equiv  q_{i+1, \rm min},
  \label{eq:immersionconditionsc}
\end{flalign}
which depends strongly on $y_i$'s (see below). If all ellipticities are deliberately set to the surface value $\epsilon_{\cal L}$, then $y_i= q_i$ for all layers. This is the hypothesis of {\it coellipticity}, which, however is not compatible with a state of global rotation; see Paper II and \cite{cmm17}.

Further, for stability reasons, we impose a negative gradient of the mass density from the centre to the surface (i.e., no density inversion), which means
\begin{flalign}
  \alpha_i >1, \quad i \in [1,{\cal L}-1].
  \label{eq:alphastable}
\end{flalign}

\subsection{The $y_i$-problem}
\label{subsec:yellproblem}

Assuming that $a_{\cal L}$, $\epsilon_{\cal L}$, $M$ and the even moments up to $J_{2n}$ are known, then (\ref{eq:mj2nmassa}) and (\ref{eq:mj2nmassb}) represent $n+1$ equations in total with $2{\cal L}-1$ unknowns, namely $\rho_{\cal L}$, the $C_i$'s and the $y_i$'s. Unless redundant equations, {\it it is in principle possible to solve this equation set} if
\begin{equation}
  n+1 = 2 {\cal L}-1,
  \label{eq:nl}
\end{equation}
regardless of any consideration about the equilibrium of the structure. This is what we call the ``$y_i$-problem'' in the following. In other words, the set of data $(a_{\cal L}, \epsilon_{\cal L}, M, J_2, \dots  J_{2n})$ can eventually be reproduced by a ${\cal L}$-layer body, provided $\frac{n}{2}={\cal L}-1$ is an integer ($n$ must be even). This condition is necessary but not sufficient.

As we shall see, the solution of the $y_i$-problem is obtained by finding the roots of a ${{\cal L}-1}$ degree polynomial. Clearly, these roots depends directly on the $J_{2n}$'s through (\ref{eq:mj2nmassb}), and must be real and positive, which is not guaranteed. Any large root can disqualify a given solution, simply based on (\ref{eq:immersionconditionsc}). This is also the case if all roots are large and very different from each other. Not only $y_i$ is allowed to vary from one layer to the other, but it can take small and large values, depending mainly on the ellipticities. It is close to zero for small or/and spheroidal surfaces, and close to unity if the fractional radius of layer $i$ is close to unity (the layers above are therefore very thin) and if $E_i$ and $E_{\cal L}$ have similar ellipticities. But $y_i$ can eventually be much larger than one. This occurs if $E_{\cal L}$ is close to spherical while  $E_i$, in contrast, is very oblate and $q_i$ close to unity. However, as $q_i \epsilon_i$ can not exceed unity, we must have
\begin{flalign}
 0 \le y_i \le \frac{1}{\epsilon_{\cal L}^2}, \qquad i \in [1,{\cal L}-1].
  \label{eq:maxyl}
\end{flalign}
The upper limit for the $y_i$'s, which is larger than one, is therefore imposed by the oblateness of the outermost layer.

\subsection{Canonical set $\{Y_k\}$ and permutations}
\label{subsec:canonicalset}

The system of equation (\ref{eq:mj2nmassa}) and (\ref{eq:mj2nmassb}) possesses a certain symmetry. As a consequence, if a solution $(y_1,y_2,\dots, y_{{\cal L}-1})$ to the $y_i$-problem is found (this includes $\bar{\rho}/\rho_{\cal L}$ and the $C_i$'s), then any permutation of values inside this set is admissible. The reason is that the layers are not interchangeable. For a three-layer problem, there are at most two possible permutations, namely $(y_1,y_2)$ and $(y_2,y_1)$, and $6$ for in the four layer-case. The net number depends on the planet's data. It is especially convenient to define, among the $({\cal L}-1) !$ possible permutations, a {\it canonical set} $(Y_1, Y_2, \dots, Y_{{\cal L}-1})$ which serves as a basis to generate all the $y_i$-solutions by permutations, namely
\begin{flalign}
  (y_1,y_2,\dots, y_{{\cal L}-1})=
  \begin{cases}
 (Y_1, Y_2, \dots, Y_{{\cal L}-1}) \equiv S_{1,2,\dots,{{\cal L}-1}},\\
 (Y_2, Y_1,\dots, Y_{{\cal L}-1}) \equiv S_{2,1,\dots,{{\cal L}-1}},\\
  \nonumber
  \dots\\
 (Y_{{\cal L}-1},\dots, Y_2,Y_1) \equiv S_{{{\cal L}-1},\dots,2,1},\\
  \end{cases}
\end{flalign}
In practice, we form this special set by sorting the $Y_k$'s in ascending order, i.e. $Y_1<Y_2< \dots <  Y_{{\cal L}-2} < Y_{{\cal L}-1}$. This is not fully arbitrary: this distribution roughly corresponds to a ``standard'' configuration fulfilling (\ref{eq:immersionconditionsa}) and hosting spheroidal surfaces having a increasing ellipticity from the center to the surface.

\subsection{Degeneracy vs. overdetermination}

As $C_i$ and $y_i$ hide three parameters in total (the fractional radius $q_i$, the ellipticity $\epsilon_i$ and the jump $\alpha_i$), we see that any solution of the $y_i$-problem can give rise to a {\it continuum of configurations} matching the same set of observables $(a_{\cal L},\epsilon_{\cal L}, M, J_2,..., J_{2n})$, but these configurations are not necessarily in equilibrium; see below. Without any additional constraint, the problem therefore appears highly degenerate. First, there is an infinite number of pairs $(q_i,\epsilon_i)$ corresponding to a given value for $y_i \epsilon_{\cal L}^2$, according to (\ref{eq:yell}). As a consequence, each layer, through the two bounding surfaces, is a source of degeneracy. Note that $\sqrt{y_i} \epsilon_{\cal L}$ basically represents the {\it location of the focus of the ellipse} $E_i$. Second, as seen above, the permutations obtained from the canonical set $\{Y_k\}$ is another source of multiple solutions. Degeneracy is important at this level as it leaves degrees of freedom to fix other observables, like the rotation rate in one or more layers (see below). In contrast, if some parameters are imposed, the problem may be overdetermined, which is therefore much less adequate. This is the case for instance if we set $y_i=1$ for $i \in [1,{\cal L}-1]$. Then, we have $q_i^2 \epsilon_i^2-{q_j}\epsilon_{j}^2$ from (\ref{eq:yell}). All the spheroidal surfaces are {\it confocal} to each other \citep{poincare88,ak74,mcm90}. The system of equations  (\ref{eq:mj2nmassa}) and (\ref{eq:mj2nmassb}) is therefore overdetermined as soon as ${\cal L} > 1$, unless either i) $C \equiv \sum_{i=1,{\cal L}-1} C_i=-1$, which implies density inversions inside the structure, or ii) $C \ne -1$ but $\eta_{2n}=1$ for all $n$, which is possible only with a single layer, i.e. for ${\cal L}=1$. If all $y_i$ (still except $y_{\cal L}$) are equal, then the system of equations is, again, overdetermined as soon as i) ${\cal L} > 2$ if $y_i \ne 1$, or ii) ${\cal L} > 1$ if  $y_i = 1$. Another case of overdetermination which is worth to mention is {\it coellipticity}. Actually, for the scale-free problem, the series of heteroeoids is described by $3{\cal L}-2$ parameters. But if we impose $\epsilon_i=\epsilon_{\cal L}$ for $i \in [1,{\cal L}-1]$, this number is reduced to $2{\cal L}-1$, precisely as in (\ref{eq:nl}). It means that there is, at most, a unique solution to the $y_i$ problem for a given set of data $(a_{\cal L}, \epsilon_{\cal L}, M, J_2, \dots  J_{2n})$, regardless of the rotation state or equilibrium of the system (see next Sect.).

\subsection{Conditions for the dynamical equilibrium}
\label{subsec:dyn}

In order to decide which values for the $q_i$'s, the $\epsilon_i$'s and the $\alpha_i$'s and fulfilling (\ref{eq:immersionconditionsa}),(\ref{eq:immersionconditionsb}) and (\ref{eq:alphastable}) represent an equilibrium, we must consider the dynamical equilibrium of the structure as a whole. There is no argument to decide if all layers can rotate at the same rate, or not. Each rotating layer must satisfy the Bernoulli-like equation (resulting from the integration of the Euler equation) and the pressure along all interfaces $E_i$ must be in balance. As shown in \cite{hamy90}, the fact that all layers (internally and externally bounded by spheroidal surfaces) can be in rigid rotation is acceptable in a first approximation, if all the ellipticities are close to zero.  As shown in Papers I and II, this approximation holds in a more general case where the confocal parameters $c_{i,j}$ defined by
\begin{flalign}
  c_{i, j}=q_i^2 \epsilon_i^2-\epsilon_{j}^2 , \qquad i, j \in [1,{\cal L}]^2
  \label{eq:cij}
\end{flalign}
are close to zero. This prolongation enables to consider highly flattened systems. In such a case, the rotation rate $\Omega_i$ of each layer is perfectly determined and is of the form 
  \begin{flalign}
    \Omega_i \equiv \Omega_i(\rho_{\cal L},q_1,\dots,q_{{\cal L}-1},\epsilon_1,\dots,\epsilon_{{\cal L}},\alpha_1, \dots, \alpha_{{\cal L}-1}),
    \label{eq:omegal}
  \end{flalign}
for $i \in [1,{\cal L}]$. In practice, we use (32) and (33) of Paper II (not reported here) which yield the series of rotation rates $\Omega_i^2$ in the form of (\ref{eq:omegal}), by recursion. The number of parameters is apparently large, but it can advantageously be reduced in the present context. First, $\epsilon_{{\cal L}}$ is known (this is the ellipticity of the outermost layer). Second, the $y_i$-problem is supposed to be solved; see Sect. \ref{subsec:yellproblem}. It means that the $\alpha_i$ depends the $\epsilon_i$ and on $q_i$ from (\ref{eq:cell}), and $\epsilon_i$ depends on the $q_i$'s from (\ref{eq:yell}). It turns out that the rotation rate $\Omega_i$ of layer $i$ depends only on ${\cal L}-1$ variables (this is true for all layers). If we select the $q_i$'s as variables, then (\ref{eq:omegal}) formally becomes
  \begin{flalign}
    \Omega_i \equiv \Omega_i(q_1,\dots,q_{{\cal L}-1}),
    \label{eq:omegal_bis}
  \end{flalign}
If the observed rotation rate $\omobs$ can be attributed to a single layer $\irot$ in the sample (usually, it is attributed to the core), then we have to consider a supplementary equation
\begin{equation}
\chi_{\cal L}(q_1,\dots,q_{{\cal L}-1})=0,
\label{eq:chiLlayers}
\end{equation}
where
\begin{flalign}
\chi_{\cal L}(q_1,\dots,q_{{\cal L}-1}) = \frac{\Omega_\irot^2(q_1,\dots,q_{{\cal L}-1})}{\omobs^2}-1,
\label{eq:omobs}
\end{flalign}
and $\irot=1$ to assign the core (this assumption is retained in the paper throughout). It is clear that the roots of (\ref{eq:chiLlayers}) in this ${\cal L}-1$ dimension space are not analytical, at least in the general case, but these can be found by numerical means form standard techniques.

\subsection{Equation set for the full problem. Strategy}

The physical configurations compatible with the set of observational data $(a_{\cal L},\epsilon_{\cal L}, M, J_2,..., J_{2n},\omobs)$ can therefore be deduced from (\ref{eq:mj2nmassa}), (\ref{eq:mj2nmassb}) and (\ref{eq:chiLlayers}), satisfying (\ref{eq:maxyl}), (\ref{eq:immersionconditionsc}) and (\ref{eq:alphastable}). The full problem is solved in two steps:
\begin{itemize}
\item first, we determine the solution of the $y_i$-problem, from (\ref{eq:mj2nmassa}) and (\ref{eq:mj2nmassb}), and define the canonical set $\{Y_k\}$; see Sect. \ref{subsec:canonicalset}. Then, we consider all permutations in the canonical set compatible with (\ref{eq:immersionconditionsc}) and (\ref{eq:maxyl}); some sets $\{y_i\}$ can be rejected.
\item second, we form the $\chi_{\cal L}$-function which compares the theoretical rotation rate of the core to the observed value, from (\ref{eq:chiLlayers}). The relationships between the solutions of the $y_i$-problem and the $q_i$'s, the $\epsilon_i$'s and the $\alpha_i$'s are considered in this second step to reduce the number of variables; see (\ref{eq:omobs}). By finding the zeros of (\ref{eq:omobs}), and provided (\ref{eq:immersionconditionsc}) and (\ref{eq:alphastable}) hold, we get the collection of equilibria that exactly reproduces the data, not only $(a_{\cal L}, \epsilon_{\cal L}, M, J_2, \dots  J_{2n})$, but also $\omobs$. This procedure is to be repeated for all relevant permutations.
\end{itemize}

It is clear that the absence of any solution for the $y_i$-problem or the rejection of all permutations (see Sec. \ref{subsec:yellproblem}) is a dead end and invalidate the execution of the second step. In this case, no equilibrium configuration is possible with the actual set of observational data. The reverse is not true: any solution to the $y_i$-problem does not necessarily lead to a physically relevant configuration, in the sense that the $q_i$'s, the $\epsilon_i$'s and the $\alpha_i$'s can be out of their domain of interest/definition.

\subsection{Fractionnal masses, moment of inertia and higher-order gravitational moments}

For a given solution of the full problem, we can calculate the fractional mass of each layer. We have
\begin{flalign}
    \nu_1 = q_1^3  \frac{\bar{\epsilon}_1}{\bar{\epsilon}_{\cal L}}\frac{\prod_{\ell=1}^{{\cal L}-1} \alpha_\ell}{1+\sum_{\ell=1,{\cal L}-1} C_\ell},
    \label{eq:nu1}
  \end{flalign}
for the deepest layer, and
  \begin{flalign}
    \nu_i = \frac{q_i^3\bar{\epsilon}_i - q_{i-1}^3\bar{\epsilon}_{i-1}}{\bar{\epsilon}_{\cal L}}\frac{\prod_{\ell=i}^{{\cal L}-1} \alpha_\ell}{1+\sum_{\ell=1,{\cal L}-1} C_\ell}
    \label{eq:nuell}
  \end{flalign}
  for $i \in [2,\cal L]$. It can be shown that the immersion conditions ensure $\nu_i \ge 0$. Another quantity of importance is the normalized moment of inertia, namely
\begin{equation}
   \frac{I_\Delta}{M\re^2}= \frac{2}{5} \frac{{1+\sum_{i=1,{\cal L}-1} C_i q_i^2}}{1+\sum_{i=1,{\cal L}-1} C_i}.
    \label{eq:nmoi}
\end{equation}
Interestingly enough, due to the assumption of incompressibility and to the specific form of the term in parenthesis in (\ref{eq:mj2nmassb}), the gravitational moments beyond $J_{2n}$ can be derived, by recursion; see below.

\section{The two-layer case}
\label{sec:2layers}

\subsection{The equation set and the key-function $\chi_2(q_1)$}

According to Sect. \ref{sec:theory} for ${\cal L}=2$, the total mass and the first two gravitational moments $J_2$ and $J_4$ are given by
\begin{subnumcases}{}
M = \frac{4}{3}\upi \rhoenv \aenv^3 \compellenv (1 + C_1), \label{eq:j2j4mass:a}\\
M a_2^2 J_2 = \frac{4}{3} \upi j_2 \rhoenv \aenv^5  \compellenv (1 + C_1 y_1),\label{eq:j2j4mass:b}\\
M a_2^4 J_4 = \frac{4}{3} \upi j_4 \rhoenv \aenv^7 \compellenv (1 + C_1 y_1^2),\label{eq:j2j4mass:c}
\end{subnumcases}
where $q_1=\frac{a_1}{a_2}$, $\alpha_1=\frac{\rho_1}{\rho_2}$,
\begin{flalign}
y_1 \epsilon_2^2=q_1^2 \epsilon_1^2 \in [0,1].
\label{eq:y2layers}
\end{flalign}
and
\begin{flalign}
C_1 = (\alpha_1-1)q_1^3\frac{\compellcor}{\compellenv}.
\label{eq:alpha2layers}
\end{flalign}
Note that $q_1 \in [q_{1,\rm min},q_{1,\rm max}]$ with
  \begin{flalign}
\begin{cases}
    q_{1,\rm min}^2=y_1 \ellenv^2,\\
    q_{1,\rm max}^2=1+(y_1-1)\ellenv^2,
\end{cases}
  \end{flalign}
where the lower bound is obtained for $\ellcor=1$ (the maximum allowed value), and  the upper bound corresponds to $b_1=b_2$ (the envelope surrounding the core has null thickness at the pole). By using (\ref{eq:eta2n}), the above equation set can be written in the following compact form
\begin{subnumcases}{}
\bar{\rho} = \rhoenv (1 + C_1),\\
\eta_2 (1 + C_1) = 1 + C_1 y_1,\\
\eta_4 (1+C_1)= 1 + C_1 y_1^2.
\end{subnumcases}
and solve for $y_1$, $C_1$ and $\rho_2$ provided $\bar{\rho}$, $\eta_2$ and $\eta_4$ are known. The unique relevant solution of tye $y_i$-problem is
\begin{subnumcases}{}
  y_1=\frac{\eta_2-\eta_4}{1-\eta_2}, \label{eq:y2minusa}\\
  C_1=-\frac{1-\eta_2}{y_1-\eta_2}, \label{eq:y2minusb}\\
  \frac{\rho_2}{\bar{\rho}}=\frac{1}{1+C_1}, \label{eq:y2minusc}
\end{subnumcases}
where $y_1$ must satisfy (\ref{eq:maxyl}). The mass-density jump is then deduced from (\ref{eq:alpha2layers}), namely
  \begin{flalign}
    \alpha_1=1+\frac{C_1 \compellenv}{q_1^2\sqrt{q_1^2- \ellenv^2 y_1}} = \alpha_1(q_1).
    \label{eq:alphaqom}
  \end{flalign}

 We see from (\ref{eq:y2layers}) that $q_1$ and $\epsilon_1$ are not known yet. Without additional constraint, there is a priori an infinity of solutions standing in between two extreme configurations : i) the core is a flat equatorial disk with fractional radius $q_{1,\rm min}$, and ii) the core is naked at the poles and occupies the largest volume with $q_{1, \rm max}$ as the relative equatorial extension. As discussed in Sec. \ref{subsec:dyn}, the conditions of dynamical equilibrium of the composite structure must be incorporated. For ${\cal L}=2$, the rotation rates for the core and for the surrounding envelope depend basically on five parameters, namely
  \begin{equation}
    \Omega_i \equiv \Omega_i(\rho_2,q_1,\epsilon_1,\epsilon_2,\alpha_1), \quad i \in \{1,2\},
  \end{equation}
where $\epsilon_2$ is known. Besides, we have $\epsilon_2^2/\epsilon_1^2=q_1^2/y_1$ from (\ref{eq:y2layers}) and $\alpha_1$ is given by (\ref{eq:alphaqom}). As $\rho_2$, $C_1$ and $y_1$ are known from (\ref{eq:y2minusa})-(\ref{eq:y2minusc}), we see that the $\Omega_i$'s depend only on one variable, for instance $q_1$. If the observed rotation rate $\omobs$ is due to the core, then (\ref{eq:chiLlayers}) writes 
\begin{equation}
\chi_2(q_1)=0,
\label{eq:chi2layers}
\end{equation}
where
\begin{equation}
  \chi_2(q_1) = \frac{\Omega_1^2(q_1)}{\omobs^2}-1.
\end{equation}
The configuration that reproduces $(a_2,\epsilon_2,M,J_2,J_4,\omobs)$ is basically obtained by finding the roots of (\ref{eq:chi2layers}). In the case of small ellipticities, $\Omega_1$ takes a simple form (see Paper I; the associated function $\chi_2$ is given Appendix \ref{app:chi2}; see below). Note that, if  the reference rotation rate $\Omega_{obs}$ is assigned to the envelope, a different $\chi_2$-function would be obtained.

\subsection{Beyond $J_4$}

The moments beyond $J_4$ are obtained from (\ref{eq:mj2nmassb}) by recursion. Actually, at order $n$, we have
 \begin{flalign}
   M a_2^{2n} J_{2n} = \frac{4}{3} \upi j_{2n} \rhoenv \aenv^{2n+3} \epsilon_2^{2n}\compellenv (1 + C_1 y_1^{2n}),
   \label{eq:j2n2layers}
 \end{flalign}
or $\eta_{2n} (1+C_1)= 1 + C_1 y_1^{2n}$ in a dimensionless form, by using (\ref{eq:eta2n}). By comparing $\eta_{2n+2}$, $\eta_{2n}$ and $\eta_{2n-2}$, we have
 \begin{flalign}
 \eta_{2n+2}-\eta_{2n} = y_1(\eta_{2n}-\eta_{2n-2}),
 \label{eq:eta6estimated}
 \end{flalign}
 which formula yields $\eta_6$ for $n=2$, $\eta_8$ for $n=3$ etc., and susequently the series $J_6$, $J_8$, $\dots$; see Tab. \ref{tab:jup2layers}. These values do not depend on $q_1$.

\subsection{Results with Jupiter's data}

\begin{table}
  \centering
  \begin{tabular}{rrrr}
    &reference data\\\hline\hline
    $\bar{\epsilon}_{\cal L}=\frac{\rpol}{\re}$   & $\frac{66854}{71492} \approx 0.93512560$  \\
    & $\rightarrow \epsilon_{\cal L} \approx 0.35431637$\\
    $M$ & $1.898124 \times 10^{30}$ g\\
    $\omobs^\dagger$ & $\frac{2 \upi}{35729.704}$ s$^{-1}$\\\\\\ 
    $2n$ &  $J_{2n}^\star (\times 10^6)$ & $j_{2n}$ & $\eta_{2n}$ \\ \hline
    $2$ & $+14696.5735 \pm 0.0017$ & $+\frac{1}{5}$ & $+0.58533384$\\
    $4$ & $-586.6085 \pm 0.0024$ & $-\frac{3}{35}$ & $+0.43424039$ \\
    $6$ & $+34.2007 \pm 0.0067$ & $+\frac{1}{21}$ & $+0.36300024$\\ 
    $8$ & $-2.422 \pm 0.021$  & $-\frac{1}{33}$ & $+0.32177940$\\
    $10$ & $+0.181 \pm 0.065$ & $+\frac{3}{143}$ & $+0.27668210$\\
    $12$ & $+0.062 \pm 0.190$ & $-\frac{1}{65}$ & $-1.02946195$ \\
    $14$ &                    & $+\frac{3}{255}$ & \\\hline \hline
  \end{tabular}\\
  \raggedright
  \qquad $^\dagger$from \cite{hig97,yr09}\\
  \qquad $^\star$from \cite{durante20}
  \caption{Reference data for Jupiter. Values of $j_{2n}$ and $\eta_{2n}$ are found from (\ref{eq:smallj2n}) and (\ref{eq:eta2n}) respectively.}
  \label{tab:jupdata}
\end{table}

\begin{table}
   \begin{tabular}{rr}
    \multicolumn{2}{r}{the $y_i$-problem$^\dagger$}\\ \hline
    $y_1$ & $0.36437$\\
    $C_1$   & $1.87665$\\
    $\rho_2/\bar{\rho}$ & $0.34762$\\
    $J_6 \; (\times 10^6)$ & $+ 35.7256$\\ 
    $J_8 \; (\times 10^6)$ & $-2.703$  \\
    $J_{10} \; (\times 10^6)$ & $+0.230$  \\
    $J_{12} \; (\times 10^6)$ & $-0.021$\\
    $J_{14} \; (\times 10^9)$ & $+2.01$ \\ \hline
  \end{tabular}\\
   $^\dagger$input data: $a_2,\epsilon_2,M,J_2,J_4$
  \caption{The solution of the $y_i$-problem for the $2$-layer model for Jupiter and the gravitational moments beyond $J_4$ estimated from (\ref{eq:eta6estimated}); see Tab. \ref{tab:jupdata} for the reference data.}
  \label{tab:jup2layers}
\end{table}

We apply the above method to Jupiter. For this planet, the reference data, namely $a_2$,$\epsilon_2$, $M$, $J_2$, $J_4$ and $\omobs$, are gathered in Tab. \ref{tab:jupdata}. We alse give the coefficients $j_{2n}$ and $\eta_{2n}$ computed from (\ref{eq:smallj2n}) and (\ref{eq:eta2n}) for $n \in [1,7]$, respectively. As quoted above, it is assumed that the rotation rate deduced from decametric observations is due to the core \citep{hig97,yr09}. From this data, we can calculate the solution $(y_1,C_1,\rho_2)$ of the $y_i$-problem from (\ref{eq:y2minusa}) to (\ref{eq:y2minusc}). The results are gathered in the Tab. \ref{tab:jup2layers}. Despite the simplicity of the model, we notice that the predictions for $J_6$ to $J_{12}$, deduced from (\ref{eq:eta6estimated}), are remarkably close to the observed values.

\begin{figure}
    \centering
    \includegraphics[height=8.7cm,bb=50 50 554 770, clip==,angle=-90]{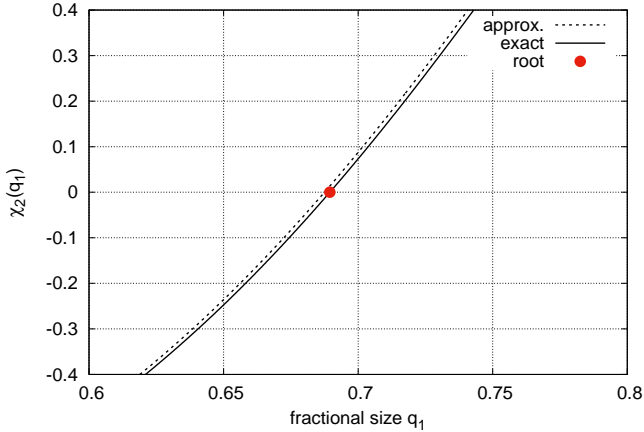}
    \caption{The function $\chi_2(q_1)$ for Jupiter ({\it plain line}) has a single root in the interval of interest at $q_1 \approx 0.689$. Also shown ({\it dashed line}) is the approximation for $\chi_2$ resulting from the rotation rates expanded in the small ellipticity limit (according to (\ref{eq:chi2order0_typev}), the approximate root is then $0.687$).}
    \label{fig:chi2.ps}
\end{figure}

The $\chi_2(q_1)$-function is perfectly defined from the solution of the $y_i$-problem and from the reference rotation rate $\omobs$. It is plotted versus $q_1$ in Fig. \ref{fig:chi2.ps}, together with its approximation from (\ref{eq:chi2order0_typev}). It goes to zero for $q_1 \approx 0.689$. This is the unique root on the interval. This value is therefore the relative equatorial extension of the core. The envelope has a relative extension of the order of $0.31$, and therefore, it fully includes the observed zonal winds. The core ellipticity $\epsilon_1$ and the mass-density jump $\alpha_1$ follow from (\ref{eq:y2layers}) and (\ref{eq:alphaqom}) respectively, and we can calculate the mass fractions and the moment of interia. The results are listed in Tab. \ref{tab:drop2layers} (column 2). We see that the core is slightly more spherical than the surrounding envelope. The fractional mass of the core from (\ref{eq:nu1}) is about $77 \%$, and the normalized moment of inertia is $0.263$ from (\ref{eq:nmoi}). The core rotates slighlty faster than the envelope. We are therefore very close to global rotation. The uncertainties in $J_2$ and $J_4$ produce errors in $q$, $\alpha$ and $\epsilon$, which are typically of the order of $10^{-5}$ with respect to central values. These uncertainties are not sufficient to significantly change $q_1$ and to reverse the relative motion of the two layers. With a rotation period slightly increased, however, the synchroneous motion can be reached; see Sect. \ref{sec:discussion}.
 
  \begin{table}
   \begin{tabular}{rrr}\\
     & this work$^\dagger$  & {\tt DROP}-code$^\star$ \\ \hline
   $q_{1, \rm min}$ & $0.21387$\\
    $q_{1, \rm max}$ & $0.92020$\\
    $q_1$ & $0.68948$&  $0.68924$\\
    $\epsilon_1$ & $0.31019$ & $0.30820$\\
    $b_1/a_2=q_1 \bar{\epsilon}_1$ & $0.65547$ & $0.65569$ \\
    $c_{1,2}$ & $-0.07979$ & $-0.08041$ \\
    $\alpha_1$ &  \multicolumn{2}{c}{$6.63183$}\\
    $\Omega_1^2/\Omega_2^2$ & \multicolumn{2}{c}{$1.00017$} \\
    $\nu_1$ & $0.76821$ & $0.76837$\\ 
    $I_\Delta/MR_e^2$ & $0.26310$  & $0.26300$ \\ \hline
   \end{tabular}\\
   \raggedright
   $^\dagger$input data: $a_2,\epsilon_2,M,J_2,J_4,\omobs$\\
   $^\star$input data: $a_2,\epsilon_2,M,b_1/a_2,\alpha_1,\Omega_1/\Omega_2$\\
   \caption{Solution of the 2-layer problem (column 2) for Jupiter, and values obtained from the numerical SCF-method (column 3; see note \ref{note:drop}); see Tab. \ref{tab:jupdata} for the reference data and Tab. \ref{tab:jup2layers} for the solution of the $y_i$-problem.}
  \label{tab:drop2layers}
\end{table}

\begin{figure*}
  \centering
  \includegraphics[height=17.9cm,bb=60 60 230 730,clip==, angle=-90]{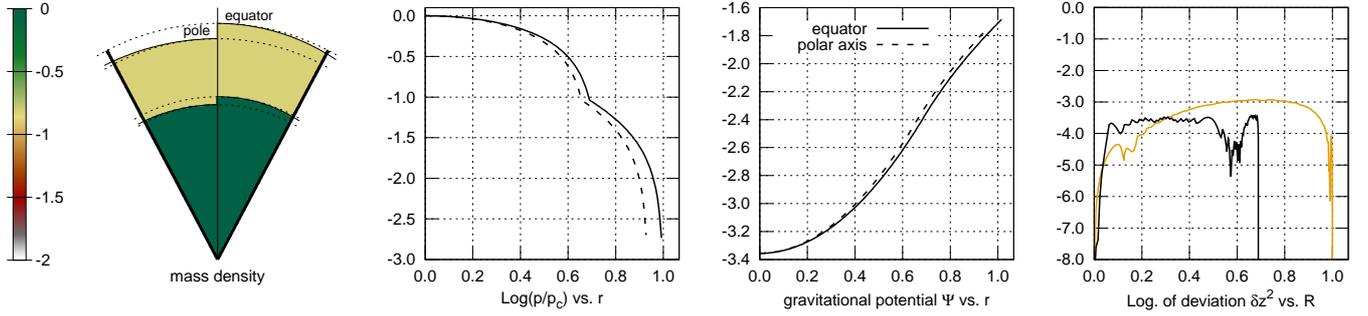}
  \caption{The internal structure as computed with the {\tt DROP}-code for the ${\cal L}=2$ ({\it from left to right}):  the normalized mass-density in color code, the normalized pressure along the polar axis ({\it dotted lines}) and at the equator ({\it plain lines}), the gravitational potential, and the deviations between the ellipses $E_\ell$ and the ``true'' interfaces. The data are listed in Tab. \ref{tab:jup2layers}.}
  \label{fig:drop2layers.ps}
  \end{figure*}

\subsection{Comparison with the Self-Consistent-Field method (the {\tt DROP}-code)}

It is interesting to compare this analytical method with the solution obtained by numerically solving the Bernoulli-like equation coupled with the Poisson equation. In this purpose, we use the {\tt DROP}-code\footnote{In its current version, the set of input parameters of the {\tt DROP}-code includes i) the axis ratios $b_i/a_{\cal L}$ of each layer $i$, ii) the mass-density jumps $\alpha_i$ at the pole, and iii) the ratios $\Omega^2_i/\Omega^2_{i+1}$ for $i \in [1, {\cal L}-1]$. The surfaces bounding layers are properly determined at second order at each step of the SCF-cycle and can therefore be compared to the ellipses $E_\ell$. The fractional radii $q_i$ are not prescribed in advance but output by the code. In the paper througout, the code is deliberately run at a low resolution corresponding to $128^2$ nodes in a cylindrical $(R,Z)$-computational box, which enables fast runs. Accordingly, the error level in output quantities is expected to be of the order of $10^{-3}$ in relative. This holds for the estimate of the $J_{2n}$'s. \label{note:drop}} that is capable of finding the ${\cal L}$-layer configuration from the Self-Consistent-Field (SCF) method at second-order in the grid spacing from a set of $2{\cal L}-1$ input data \citep{bh21}. The configuration retained for ${\cal L}=2$ is shown in Fig. \ref{fig:drop2layers.ps}. The two approches share the same mass-density jump and the same ratio $\Omega_1/\Omega_2$. The graphs show the mass-density in color code, the pressure (normalized to the central value) at $R=0$ and at $Z=0$, the gravitational potential along these two directions and the deviations between the ``true'' surface levels and the ellipses $E_i$. The number of SCF-cycle required for convergence is $17$, and the relative virial parameter is $2.8 \times 10^{-5}$. The data output by the code are gathered in Tab. \ref{tab:drop2layers} (column 3). We see that the fractional radii $q_1$ and fractional mass $\nu_1$ are in very good agreement, within a few $10^{-4}$ typically. The confocal parameter $c_{1,2}$ is much smaller than unity in absolute, which validates the use of the formalism (see Paper I).

\section{The three-layer case}
\label{sec:3layers}

\subsection{The solution of the $y_i$-problem}

As discussed in Sect. \ref{sec:theory}, matching the next two gravitational moments is possible with an additional layer. For ${\cal L}=3$, the mass $M$ and the four even moments $J_2$ to $J_8$, in dimensionless form, are given by
\begin{subnumcases}{}
\bar{\rho} = \rhoatm (1+C),\label{eq:j2toj8massa}\\
\eta_2 (1+C) = 1+C_1y_1 + C_2 y_2,\label{eq:j2toj8massb}\\
\eta_4 (1+C)= 1+C_1y_1^2 + C_2 y_2^2,\label{eq:j2toj8massc}\\
\eta_6 (1+C)= 1 +C_1y_1^3+ C_2 y_2^3\label{eq:j2toj8massd}, \\
\eta_8 (1+C)= 1 +C_1y_1^4 + C_2 y_2^4\label{eq:j2toj8masse},
\end{subnumcases}
where we have set $C=C_1+ C_2$ for convenience, $y_1 \epsilon_3^2=q_1 \epsilon_1^2$, $y_2 \epsilon_3^2=q_2 \epsilon_2^2$, $q_1=\frac{a_1}{a_3}$, $q_2=\frac{a_2}{a_3}$, $\alpha_1=\frac{\rho_1}{\rho_2}$,  $\alpha_2=\frac{\rho_2}{\rho_3}$ and

\begin{subnumcases}{}
  C_1 = (\alpha_1-1) q_1^3 \frac{\bar{\epsilon}_1}{\bar{\epsilon}_3} \alpha_2,\\
  C_2 = (\alpha_2-1) q_2^3 \frac{\bar{\epsilon}_2}{\bar{\epsilon}_3 }.
\end{subnumcases}

If the set of data $(a_3,\epsilon_3, M, J_2,J_4,J_6,J_8)$ is known, then (\ref{eq:j2toj8massa})-(\ref{eq:j2toj8massd}) can be solved for $(y_1,y_2,C_1,C_2,\rho_3)$ as there are $5$ unknowns and the same amount of equations. The solutions are obtained, for instance, by substracting the quantity $1+ C$ to both sides of (\ref{eq:j2toj8massb}) to (\ref{eq:j2toj8masse}). Then, from  (\ref{eq:j2toj8massb}) to (\ref{eq:j2toj8massd}), we get after some algebra a first relationship between $y_1$ and $y_2$, namely
\begin{flalign}
y_1y_2+A_1(y_1+y_2)+A_2=0,
\label{eq:solution3}
\end{flalign}
where
\begin{flalign}
\begin{cases}
  A_1=\frac{\eta_4-\eta_2}{1-\eta_2}\\
  A_2=\frac{\eta_4-\eta_6}{1-\eta_2}
\end{cases}
\label{eq:a1a2}
\end{flalign}
 The second link between $y_1$ and $y_2$ is obtained by considering for instance (\ref{eq:j2toj8masse}) instead of (\ref{eq:j2toj8massd}). We find
\begin{flalign}
y_1y_2+A_3(y_1+y_2)+A_4=0,
\label{eq:solution3bis}
\end{flalign}
where
\begin{flalign}
\begin{cases}
  A_3=\frac{\eta_6-\eta_4}{\eta_2-\eta_4}\\
  A_4=\frac{\eta_6-\eta_8}{\eta_2-\eta_4}.
\end{cases}
\label{eq:a3a4}
\end{flalign}



 These relationships can be put in matrix form
\begin{flalign}
	\begin{pmatrix} 
		1 & A_1\\ 
		1 & A_3 \\
	\end{pmatrix}
        \begin{pmatrix} 
		y_1y_2\\
                y_1+y_2\\
	\end{pmatrix}
        =     -   \begin{pmatrix} 
		A_2\\
                A_4\\
	\end{pmatrix}.
\end{flalign}
and solved for the product of the roots $p=y_1 y_2$ and for the sum $s=y_1+y_2$ with
\begin{flalign}
\begin{cases}
 s= \frac{A_4-A_2}{A_1-A_3}\\
 p = \frac{A_2A_3-A_1A_4}{A_1-A_3}.
\end{cases}
\label{eq:asp}
\end{flalign}

By eliminating $y_2$ (or $y_1$), we can see that both quantities are the roots of the second degree polynomial
\begin{flalign}
  Y^2-s Y+p=0,
 \label{eq:p2y}
\end{flalign}
where $Y$ stands for $y_1$ or $y_2$. If $\Delta=s^2-4p \ge 0$, then the roots $Y_1$ and $Y_2$ are positive and we have
\begin{flalign}
\label{eq:solution3y1y2}
  Y_k = &\frac{s - \sqrt{\Delta}}{2}+(k-1)\sqrt{\Delta}, \quad k=\{1,2\}.
\end{flalign}
There are $2$ possible combinations for $(y_1,y_2)$ (see \ref{subsec:yellproblem}): the canonical set $(Y_1,Y_2 \ge Y_1) \equiv S_{1,2}$ and $(Y_2,Y_1) \equiv S_{2,1}$; See Sect. \ref{subsec:canonicalset}. For each pair, the three contants $C_1$, $C_2$ and $\rho_3$ are easily deduced from (\ref{eq:j2toj8massa})-(\ref{eq:j2toj8massc}), namely
\begin{flalign}
  \begin{cases}
 C_1=\frac{(1-\eta_2)(\eta_4-y_2^2)-(1-\eta_4)(\eta_2-y_2)}{(\eta_2-y_2)(y_1^2-\eta_4)-(y_1-\eta_2)(\eta_4-y_2^2)},\\
 C_2=\frac{(1-\eta_2)(\eta_4-y_1^2)-(1-\eta_4)(\eta_2-y_1)}{(\eta_2-y_1)(y_2^2-\eta_4)-(y_2-\eta_2)(\eta_4-y_1^2)},\\
 \rho_3= \frac{\bar{\rho}}{1+C_1+C_2},
\label{eq:c1c2rho3}
\end{cases}
\end{flalign}

\begin{table}
   \begin{tabular}{rrrr}
      \multicolumn{2}{r}{the $y_i$-problem$^\dagger$} \\ \hline
      $A_1$ & $-0.36437$ & $A_2$ & $+0.17180$\\
      $A_3$ & $-0.47149$ & $A_4$ & $+0.27281$\\
    $s$   & $+0.94298$\\
    $p$ & $+0.17179$\\
    $Y_1$ & $+0.24675$\\
    $Y_2$ & $+0.69622$\\
    $C_1$ & $+1.71976$\\
    $C_2$ & $+1.51145$\\
    $\rho_3/\bar{\rho}$ & $+0.23633$\\
    $J_{10} \; (\times 10^6)$ & $+0.193$ \\
     $J_{12} \; (\times 10^6)$ & $-0.016$ \\
     $J_{14} \; (\times 10^9)$ & $+1.46$ \\ \hline
   \end{tabular}\\
   \raggedright
   $^\dagger$input data: $a_3,\epsilon_3,M,J_2,J_4,J_6,J_8$\\
  \caption{The canonical solution of the $y_i$-problem for the 3-layer model for Jupiter and the gravitational moments beyond $J_8$ computed from (\ref{eq:eta8estimated}); see Tab. \ref{tab:jupdata} for the reference data; see Fig. \ref{fig:y1y2.ps}.}
  \label{tab:jup3layers}
\end{table}

\begin{figure}
\includegraphics[height=8.8cm,bb=40 100 544 740, clip==,angle=-90]{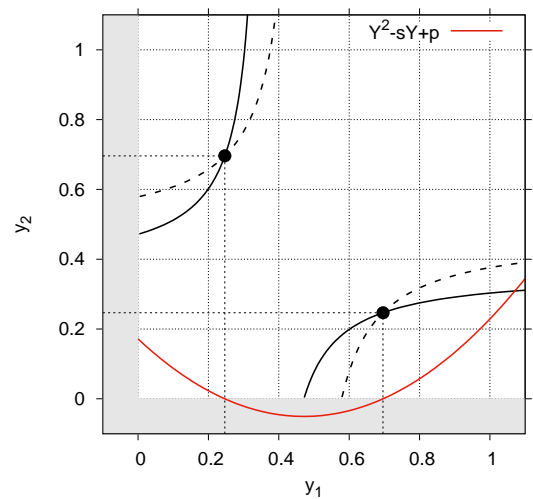}
\caption{The two relationships between $y_1$ and $y_2$ from (\ref{eq:solution3}) and (\ref{eq:solution3bis}) involved in the $3$-layer model ({\it black lines}), and the equivalent, second degree polynomial ({\it red line}); see (\ref{eq:p2y}).}
\label{fig:y1y2.ps}
\end{figure}

\begin{figure}
  \includegraphics[height=8.8cm,bb=40 100 544 740, clip==,angle=-90]{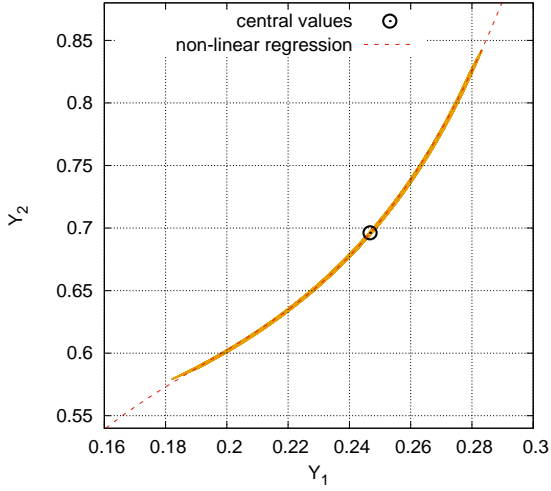}
  \caption{Image of the error box $(J_6 \pm \Delta J_6) \times (J_8 \pm \Delta J_8)$ in the $(Y_1,Y_2)$-plane (canonical set) for the $3$-layer problem applied to Jupiter ({\it orange}). See Tab. \ref{tab:jup3layers} for the central values  ({\it black circle}). See (\ref{eq:y2y1fit}) for the non-linear regression ({\it red dashed line}).}
\label{fig:3layersy1y2.ps}
\end{figure}

\subsection{Results with Jupiter's data}

We give in Tab. \ref{tab:jup3layers} the results obtained for Jupiter's data. Figure \ref{fig:y1y2.ps} shows $y_2$ versus $y_1$, as given by (\ref{eq:solution3}) and (\ref{eq:solution3bis}), and the second degree polynomial in $Y$; see (\ref{eq:p2y}). As $Y_1$ and $Y_2$ have both small magnitude and fulfill (\ref{eq:maxyl}), they autorize the determination of equilibria by solving $\chi_3(q_1,q_2)=0$ (see below). The moments beyond $J_8$ are obtained from (\ref{eq:mj2nmassb}) with ${\cal L}=3$ and $n > 4$. As we have 
\begin{flalign}
  M a_2^{2n} J_{2n} = \frac{4}{3} \upi j_{2n} \rho_3 a_3^{2n+3} \epsilon_3^{2n}\bar{\epsilon_3} (1 + C_1 y_1^{2n}+C_2 y_2^{2n}),
  \label{eq:j2n3layers}
\end{flalign}
or
\begin{flalign}
\eta_{2n} (1+C)= 1 +C_1y_1^{2n} + C_2 y_2^{2n}\label{eq:j2j4j6masse},
\end{flalign}
in dimensionless form, a recurrence relation can obtained by comparing the expressions for $(\eta_{2n}-\eta_{2n-2})(y_1+y_2)$ and for $(\eta_{2n-2}-\eta_{2n-4})y_1y_2$. We find
\begin{flalign}
  \nonumber
  &\left(\eta_{2n-2}-\eta_{2n-4}\right)y_1 y_2\\
  &\qquad - \left(\eta_{2n}-\eta_{2n-2}\right)(y_1+y_2)-(\eta_{2n+2}- \eta_{2n})=0,
\label{eq:eta8estimated}
\end{flalign}
which logically yields (\ref{eq:solution3}) and (\ref{eq:solution3bis}) for $n=2$ and $n=3$. We deduce $J_{10}$ to $J_{14}$ by setting $n=4$, $5$ and $6$, respectively; see (\ref{eq:eta2n}). The values obtained are given in Tab. \ref{tab:jup3layers}. Note that both pairs of solution lead to the same moments (as $s$ and $p$ are unchanged by permutation of $y_1$ and $y_2$). We notice that these two moments are close to measured values and well inside the error bars. This is not the case for ${\cal L}=2$. A value of $J_{14} \approx +1.46 \times 10^{-9}$ is predicted, which is close to the value obtained in the two-layer case. The uncertainties in $J_6$ and especially in $J_8$ (of the order of one purcent in relative) leads to significant shifts in the $y_i$'s. Fig. \ref{fig:3layersy1y2.ps} displays all the canonical sets $(Y_1,Y_2)$ obtained by scanning the entire error box $(J_6 \pm \Delta J_6) \times (J_8 \pm \Delta J_8)$. We can expand (\ref{eq:solution3y1y2}) in $\Delta J_6/J_6$ and in $\Delta J_8/J_8$ to get the trend analytically, but the formula is quite complicated. More directly, we can correctly fit of the image of the error box by a third degree polyomial, namely
\begin{flalign}
  \label{eq:y2y1fit}
  Y_2&=-0.818716+19.0979 Y_1\\
  \nonumber
  & \qquad -91.7828 Y_1^2+159.111 Y_1^3,
\end{flalign}
for $Y_1 \in [0.19,0.28]$; see again Fig. \ref{fig:3layersy1y2.ps}.

\begin{figure}
    \centering
    \includegraphics[height=8.7cm,bb=50 50 554 770, clip==,angle=-90]{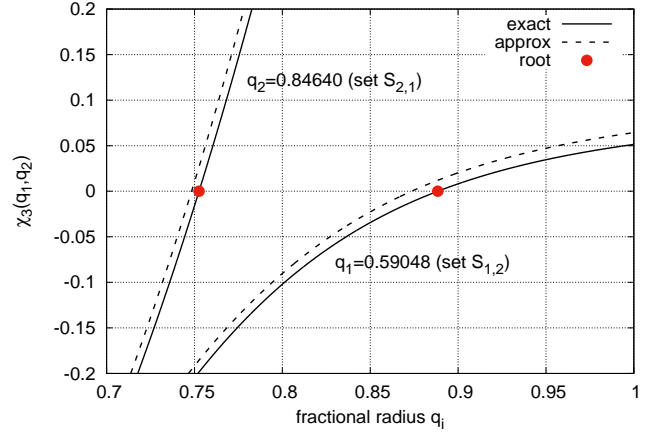}
    \caption{The function $\chi_3(q_1,q_2)$ and its root for $q_1=0.59048$ obtained from $S_{1,2}$ and for $q_2=0.84640$ obtained from $S_{2,1}$ in the $3$-layer model for Jupiter ({\it plain lines}). Also shown ({\it dashed lines}) is the approximation of the function at small ellipticities (see Paper II); see also Figs. \ref{fig:q1q2_perm12.ps} and \ref{fig:q1q2_perm21.ps}.}
    \label{fig:chi3.ps}
\end{figure}

\subsection{The key-function $\chi_3(q_1,q_2)$. Results}

In order to get the physical configurations, we must now consider the equations for the dynamical equilibrium. At thsi level, $y_1$ and $y_2$ are known (along with $C_1$, $C_2$ and $\rho_3$). The fractional radii, ellipticities and mass-density jumps are yet to be determined. An infinite number of physical equilibria potentially follows. As in the two-layer problem, the rotation rate of each layer cannot be prescribed arbitrarily. In this purpose, we still use (32) and (33) of Paper II. Each rate depends on $3{\cal L}-1=8$ variables, i.e.
\begin{flalign}
\Omega_i \equiv \Omega_i(\rho_3,q_1,q_2,\epsilon_1,\epsilon_2,\epsilon_3,\alpha_1,\alpha_2), \quad i \in \{1,2,3\},
\end{flalign}
but the $\epsilon_3$ is an input, and the $y_i$-problem is solved; see above. Thus, $2{\cal L}=6$ variables can be eliminated. The $\Omega_i$'s depend ultimately on two variables, for instance $q_1$ and $q_2$. If we assume that the observed rotation rate is due to the core, then (\ref{eq:omobs}) still holds. The $\chi$-function for the three-layer problem is
\begin{equation}
  \chi_3(q_1,q_2) \equiv \frac{\Omega_1^2(q_1,q_2)}{\omobs^2}-1,
  \label{eq:chi3layers}
\end{equation}
and it is again a root-finding problem in a two dimensional space. We see that, in contrast with the two-layer case, we can not isolate a single value for the $q_1$ and $q_2$. For the canonical set $S_{1,2}$ and its unique permutation $S_{2,1}$, we have therefore determined the zeros of (\ref{eq:chi3layers}) by varying $q_1$ in the full range $[0,1]$, and $q_2 \in [q_{2,\rm min},1]$ where
\begin{flalign}
  q_{2,\rm min} = \sqrt{\max\{q_1,q_1^2 + (y_2-y_1) \epsilon_3^2\}},
\end{flalign}
according to (\ref{eq:immersionconditionsc}). We show in Fig. \ref{fig:chi3.ps} $\chi_3(q_1,q_2)$ versus $q_2$ for $q_1=0.59048$ obtained from the canonical solution $S_{1,2}$, and versus $q_1$ for $q_2=0.84640$ obtained from the other set $S_{2,1}$. We have also plotted the results obtained when the rotation rates are expanded in the limit of small ellipticities.

\begin{figure*}
  \includegraphics[height=8.7cm,bb=40 100 544 740, clip==,angle=-90]{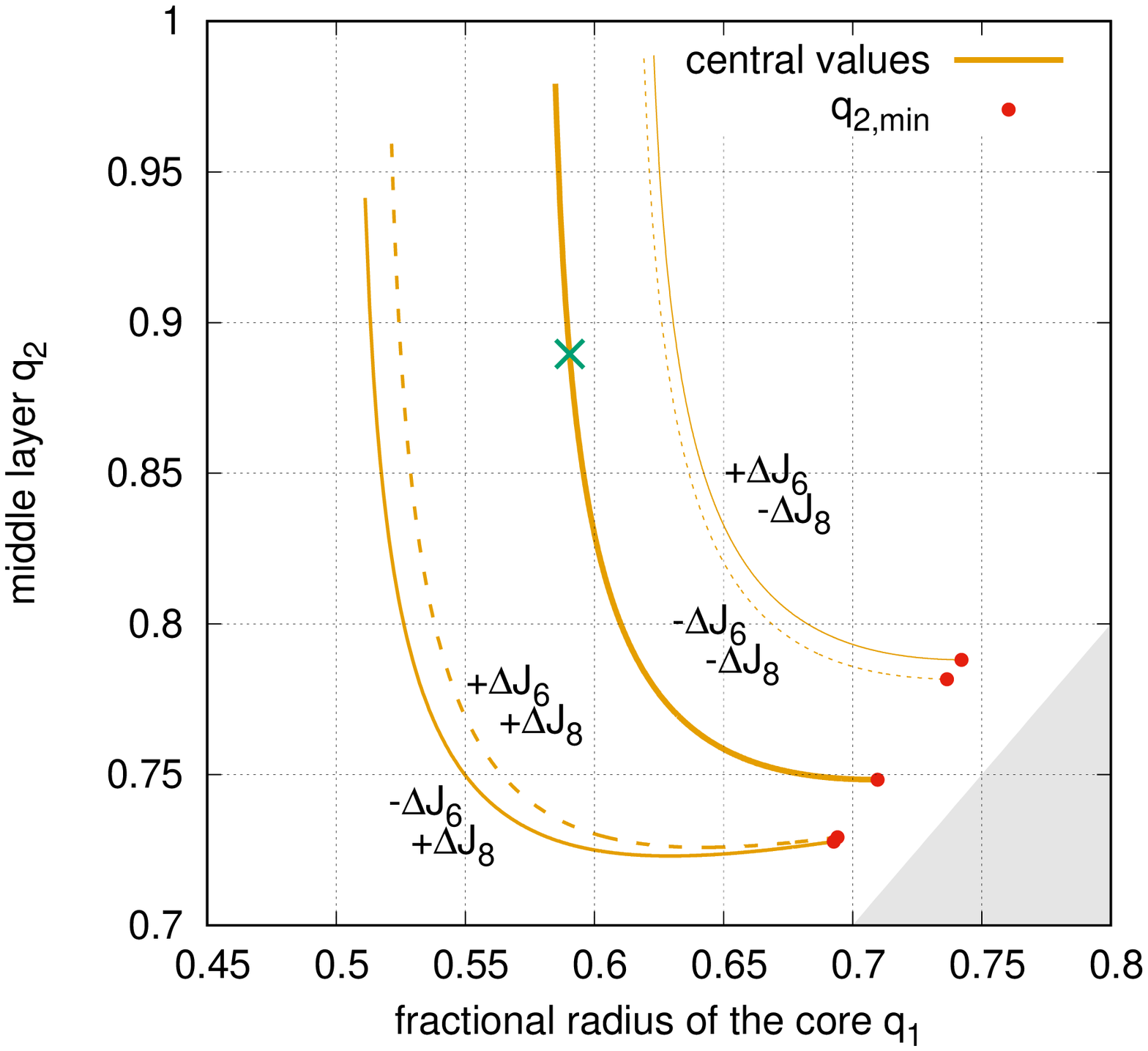}
  \includegraphics[height=8.7cm,bb=40 84 544 724, clip==,angle=-90]{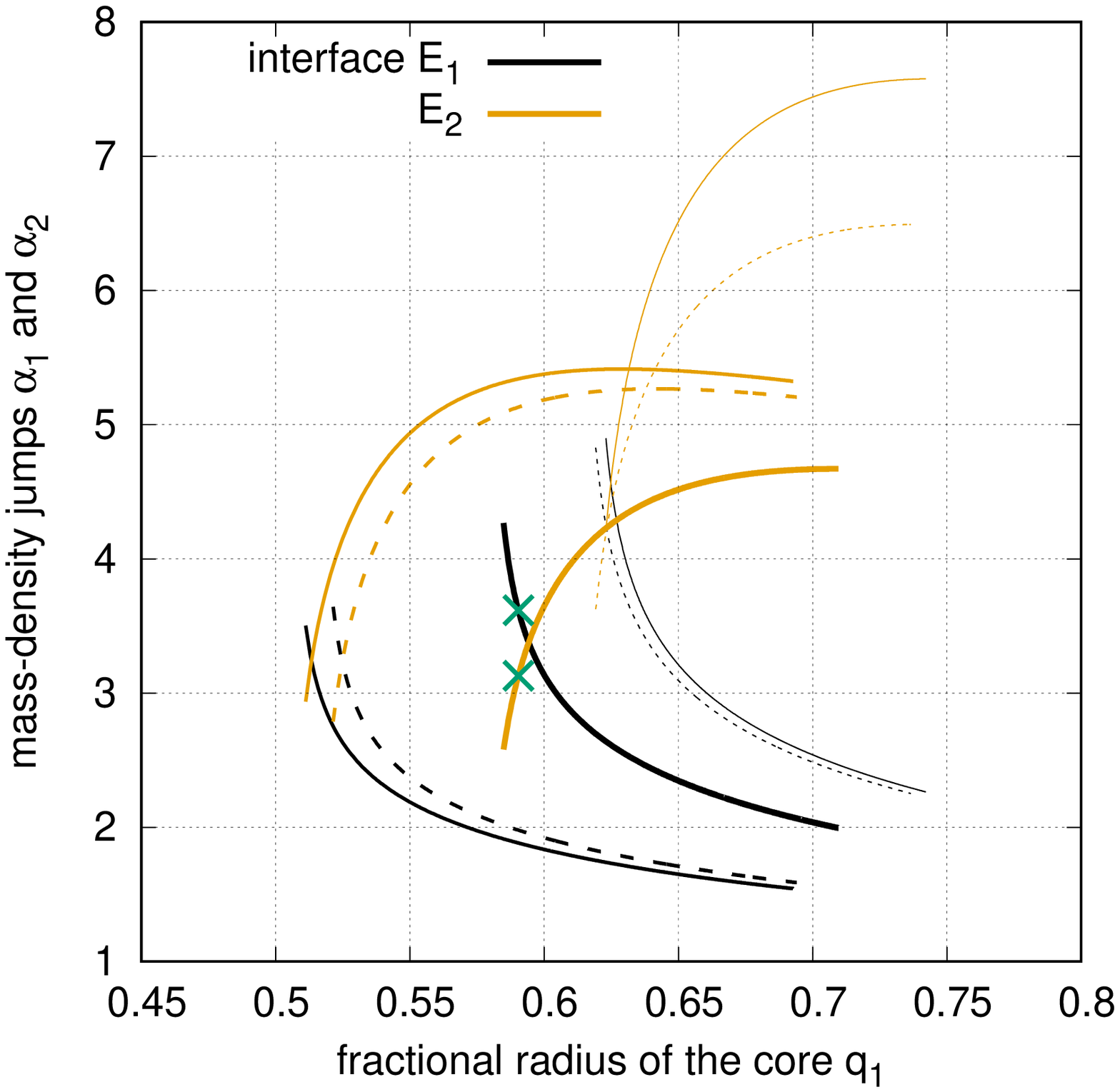}\\
  \includegraphics[height=8.7cm,bb=40 100 544 740, clip==,angle=-90]{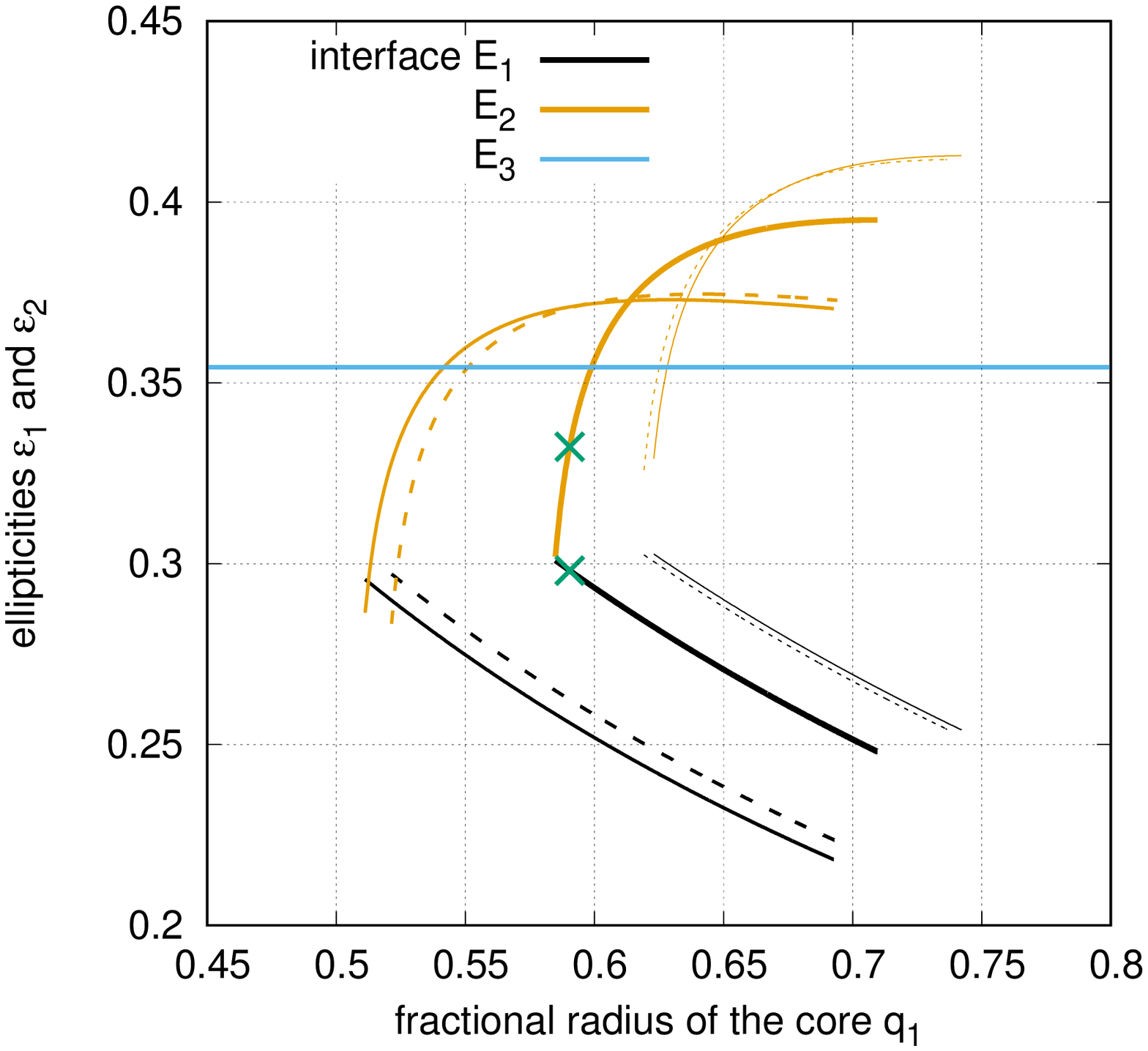}
  \includegraphics[height=8.7cm,bb=40 100 544 740, clip==,angle=-90]{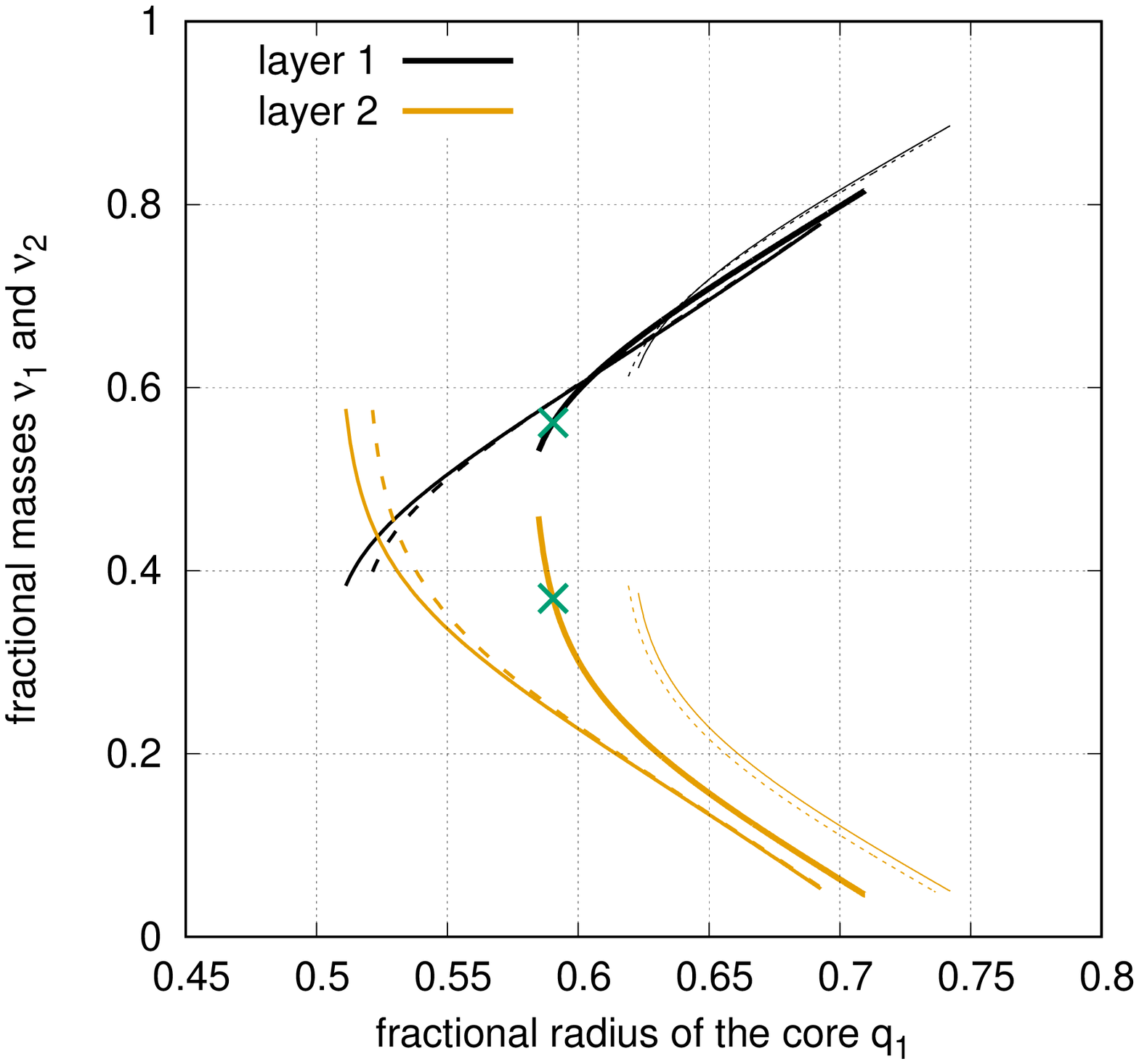}\\
  \includegraphics[height=8.7cm,bb=40 94 544 734, clip==,angle=-90]{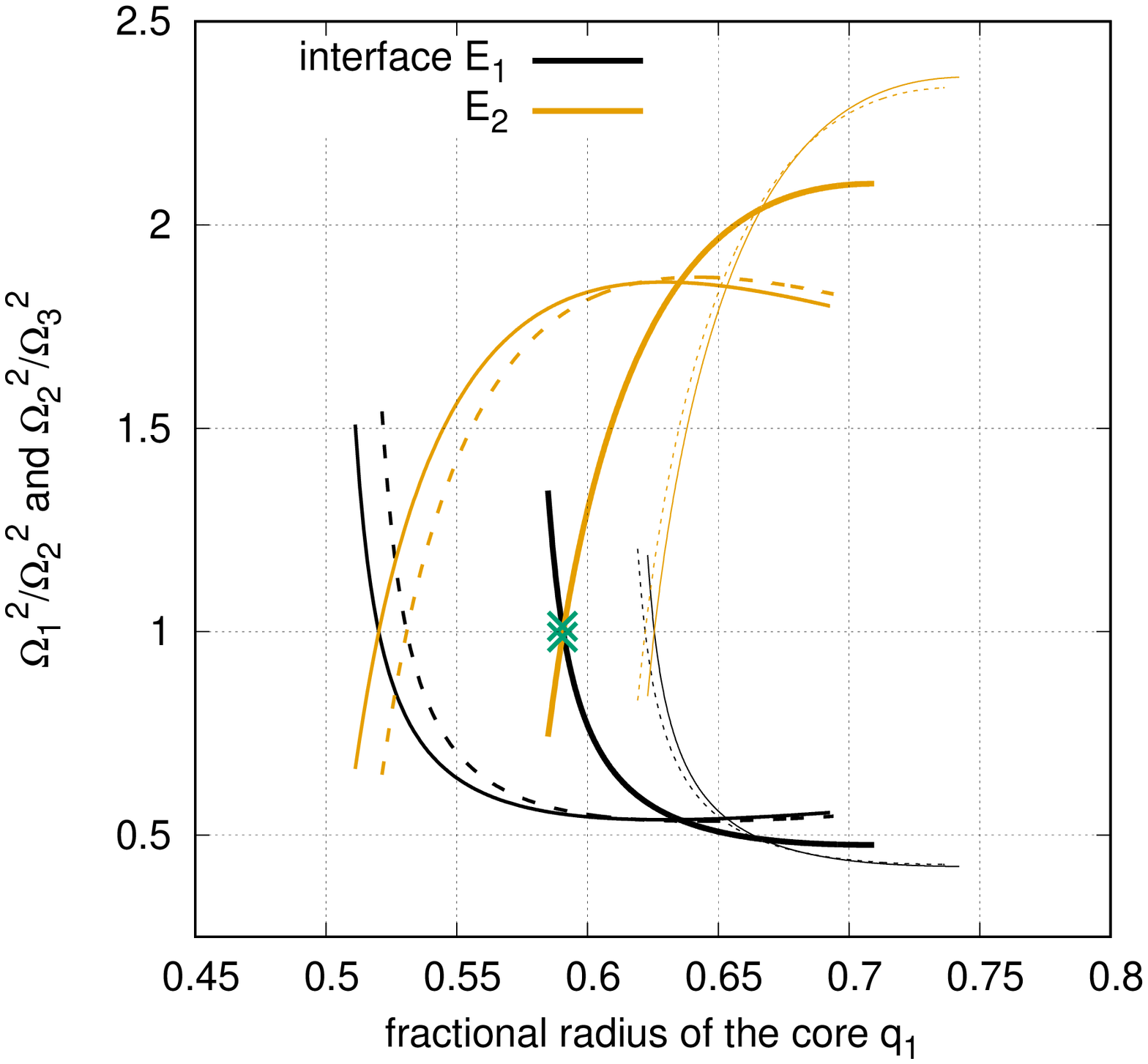}
  \includegraphics[height=8.7cm,bb=40 100 544 740, clip==,angle=-90]{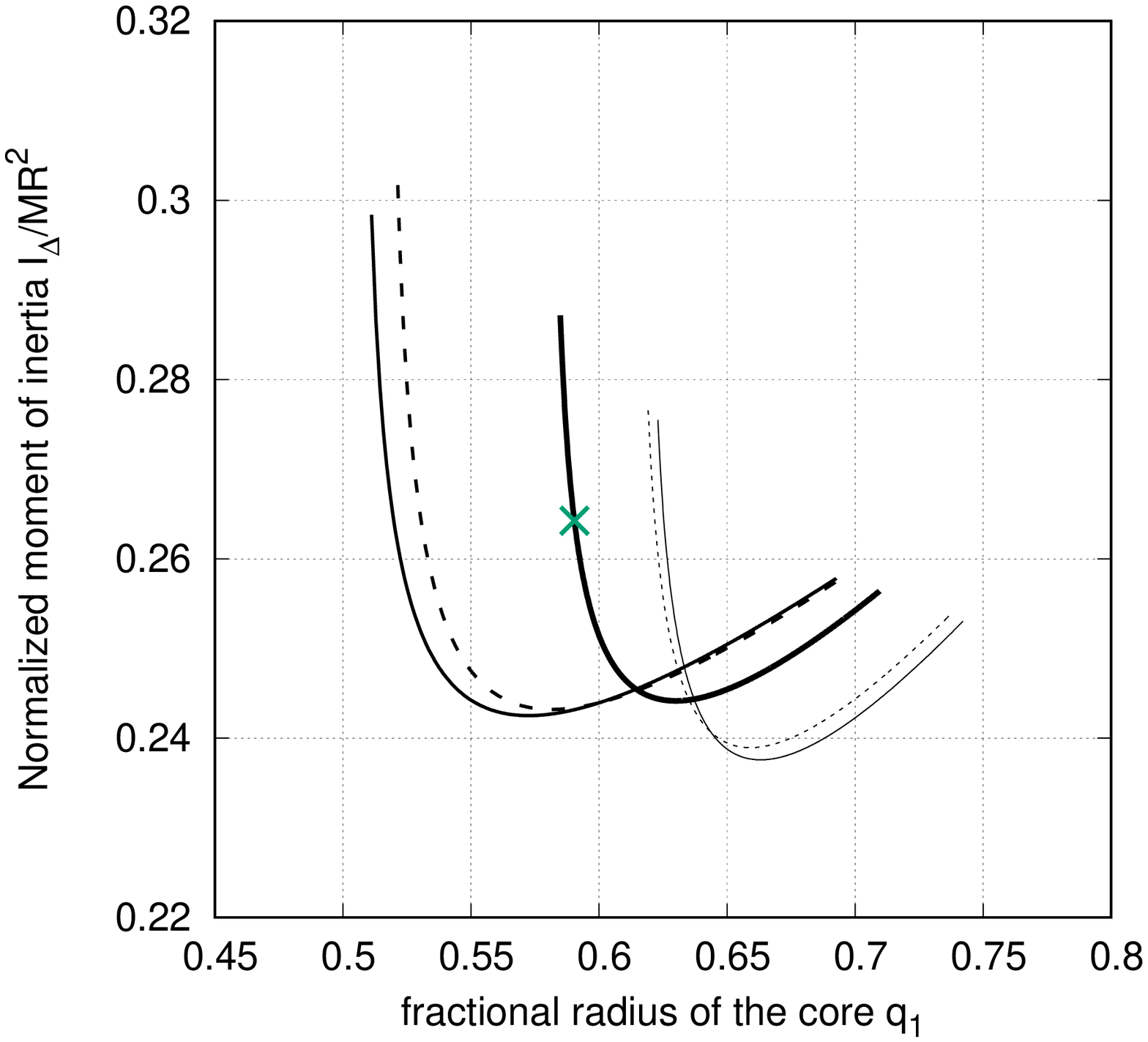}
 \caption{Equilibrium configurations for the canonical solution $S_{1,2}$: the fractional radius $q_2$ ({\it top left panel}), the mass-density jumps at the interfaces ({\it top right panel}), the ellipticities of $E_1$ and $E_2$ ({\it middle left panel}), the mass fractions for lyers $1$ and $2$ ({\it middle right panel}), the ratio of the rotation rates squared ({\it bottom left panel}), and normalized moment of inertia ({\it bottom right panel}) as a function of the fractional radius of the core $q_1$ for the $3$-layer model applied to Jupiter. Curves are obtained for $5$ values of the gravitational moments: for central values of $J_6$ and $J_8$ ({\it thick lines}), for $J_6 \pm \Delta J_6$ ({\it thin lines}) and for $J_8 \pm \Delta J_8$ ({\it dotted lines}). The internal structure corresponding to $q_1 \approx 0.59038$ ({\it green cross}) and computed with the {\tt DROP}-code is detailed in Fig. \ref{fig:drop3layers_perm12.ps}; see also Tab. \ref{tab:drop3layers12}.}
\label{fig:q1q2_perm12.ps}
\end{figure*}

The results of the full scan are displayed in Fig. \ref{fig:q1q2_perm12.ps} for the canonical set. We have used the ``central values'' for $J_6$ and $J_8$ and the values at the four vertex of the error rectangle defined by $J_6 \pm \Delta J_6$ and $J_8 \pm \Delta J_8$. Actually, by varying the moment, the coefficients $A_i$ change, which produces new canonical roots $(Y_1,Y_2)$ for the $y_i$-problem, and subsequently modifies the range of equilibria available. The graphs show $q_2$, $\epsilon_1$, $\epsilon_2$, $\alpha_1$, $\alpha_2$, $\nu_1$, $\nu_2$, $\Omega_2^2/\Omega_1^2$, $\Omega_3^2/\Omega_2^2$, and $I_\Delta$ versus $q_1$. For this set, the fractional radius of the core stands roughly in the range $[0.5,0.75]$, while we have $q_2 \gtrsim 0.72$. The mass density jumps at the interfaces $E_1$ and $E_2$ are of similar magnitude, with $\alpha_2 \gtrsim \alpha_1$ in general. The core is roughly more massive than the two layers located above, except when $q_1 \lesssim 0.52$. The ellipticity of the core is always smaller than that of the planet, while the ellipticity of the surface $E_2$ bounding layers 2 and 3 can be lower or higher than for the free boundary. Globally, the intermediate layer $2$ rotates faster than the core, except when $q_2$ is close to unity. The moment of inertia stands in the range $[0.24,0.32]$ typically. We notice that the uncertainties in $J_6$ and especially in $J_8$ have significant effects in the solutions. Configurations obtained for $q_2 \gtrsim 0.96$ are to be taken with caution, as the transition occurs within the domain where zonal flows are observed. These correspond to the smallest cores. 

\begin{figure*}
  \includegraphics[height=8.7cm,bb=40 100 544 740, clip==,angle=-90]{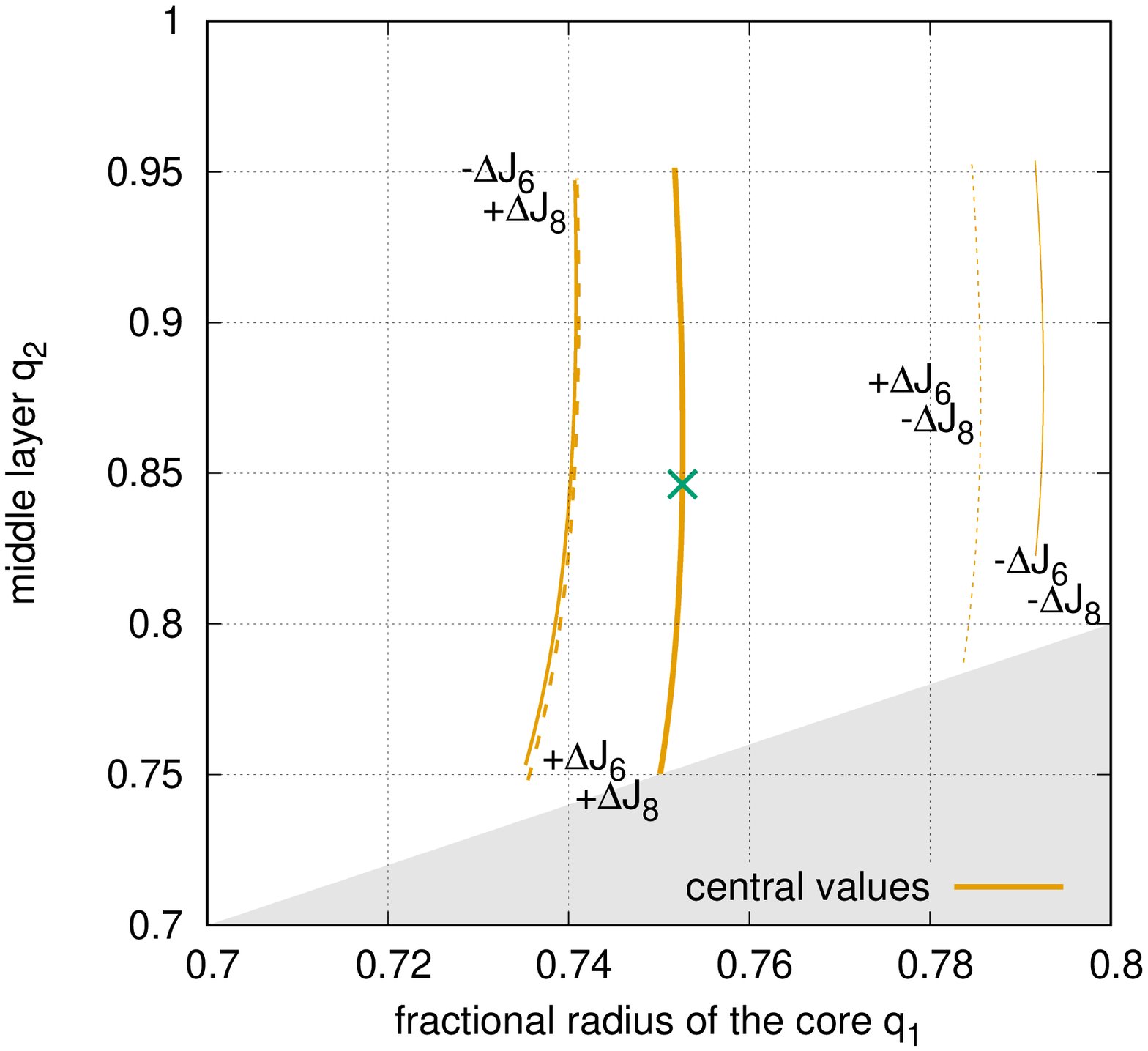}
  \includegraphics[height=8.7cm,bb=40 84 544 724, clip==,angle=-90]{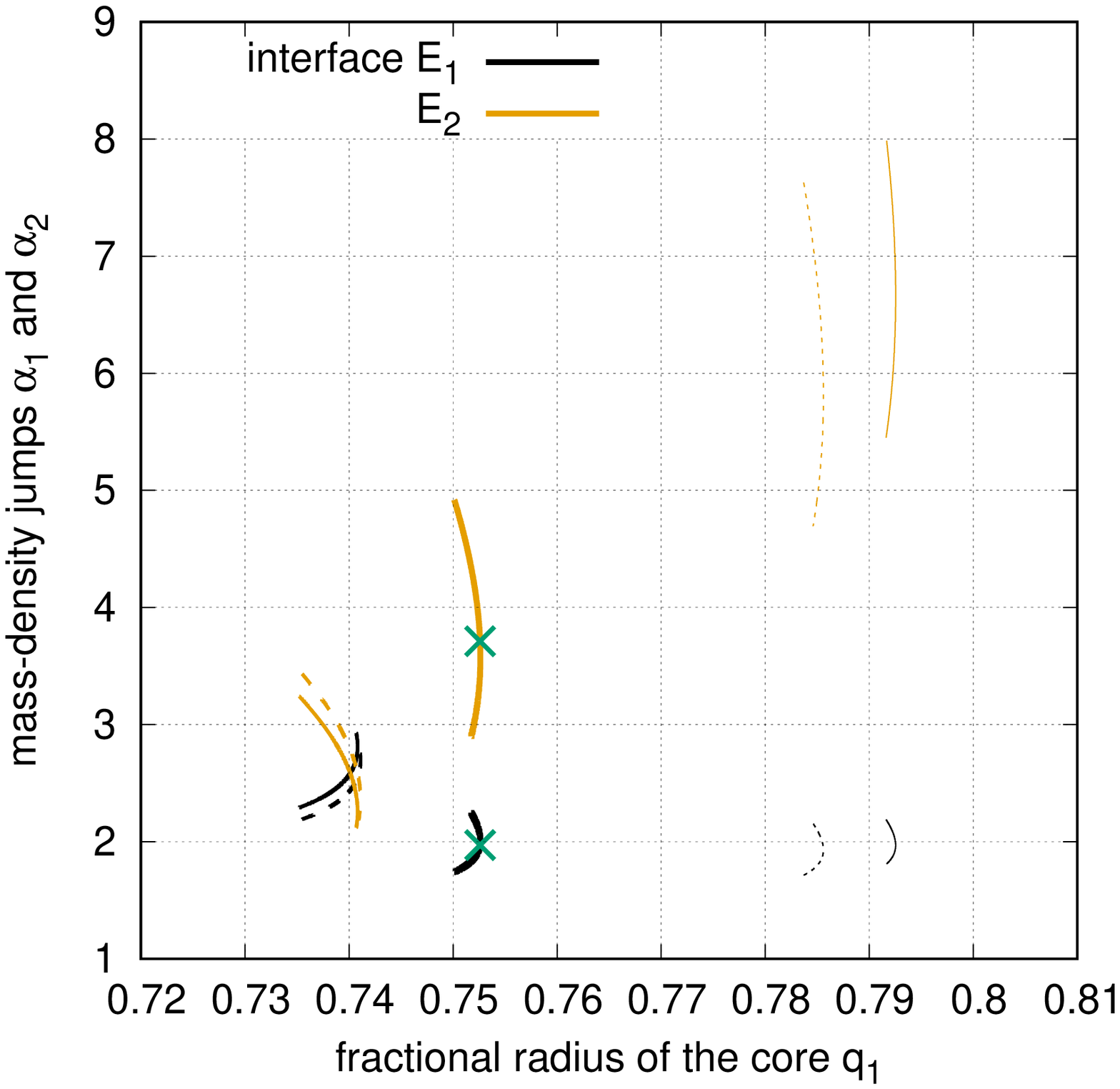}\\
  \includegraphics[height=8.7cm,bb=40 100 544 740, clip==,angle=-90]{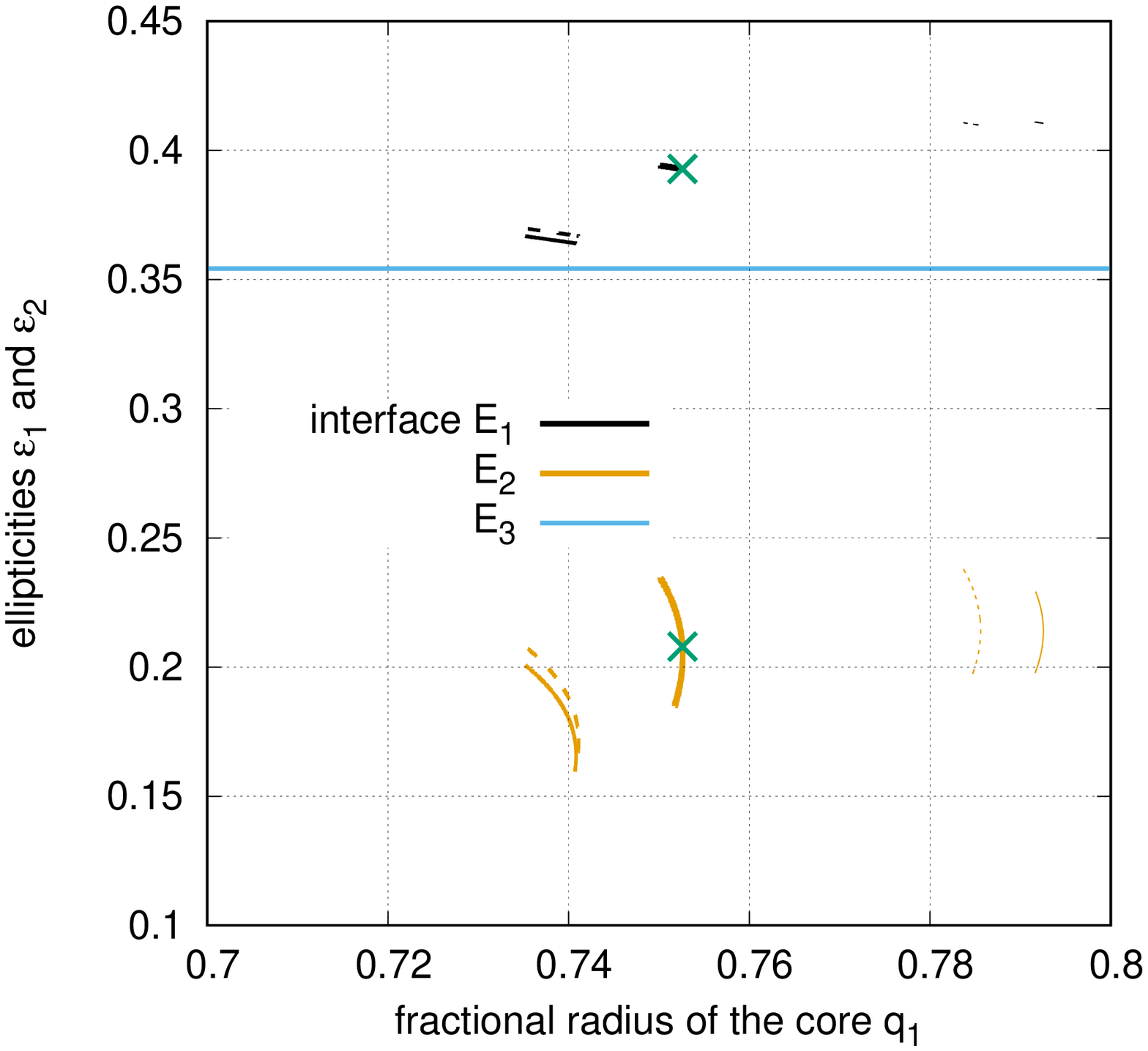}
  \includegraphics[height=8.7cm,bb=40 100 544 740, clip==,angle=-90]{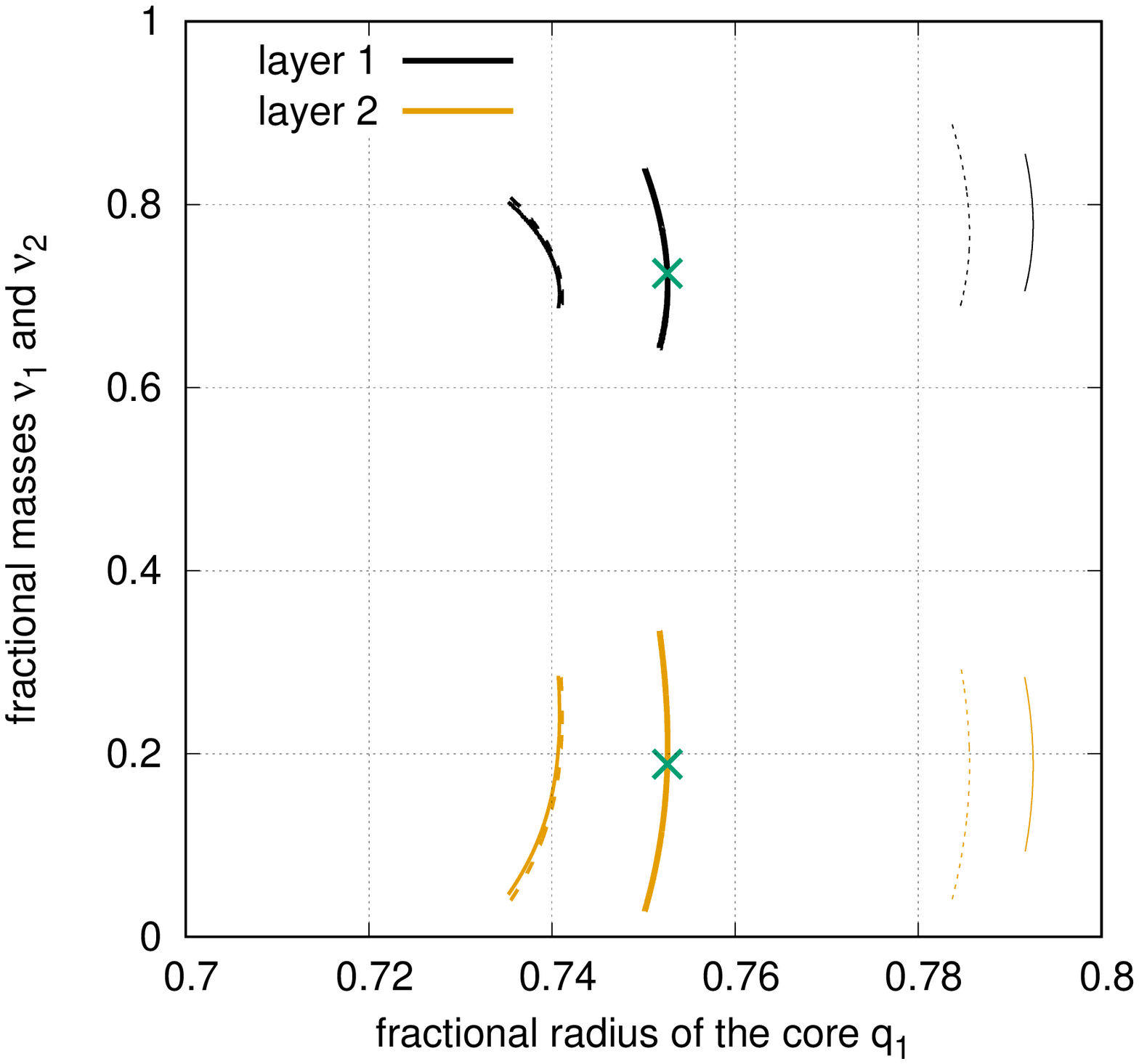}\\
  \includegraphics[height=8.7cm,bb=40 94 544 734, clip==,angle=-90]{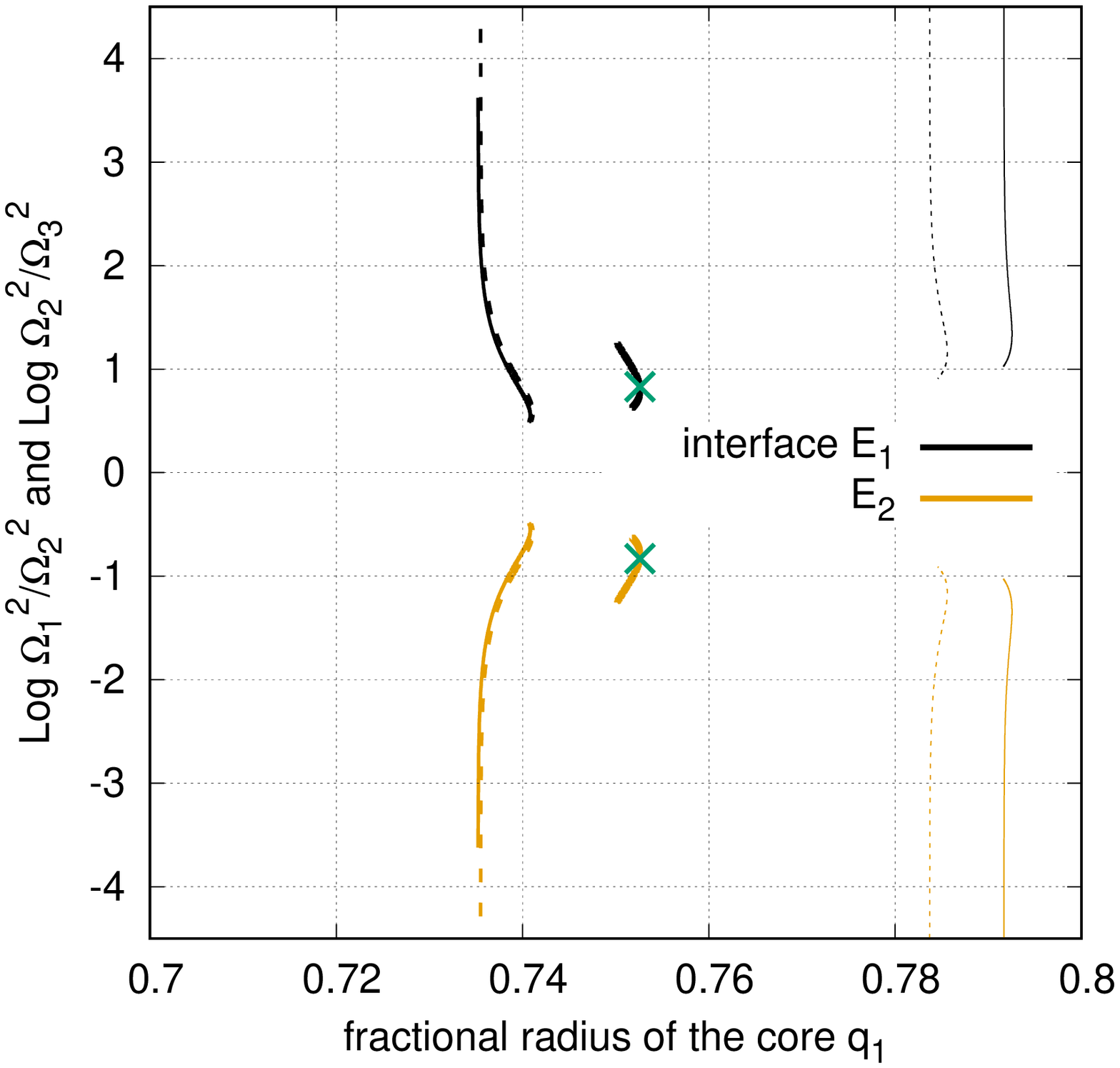}
  \includegraphics[height=8.7cm,bb=40 100 544 740, clip==,angle=-90]{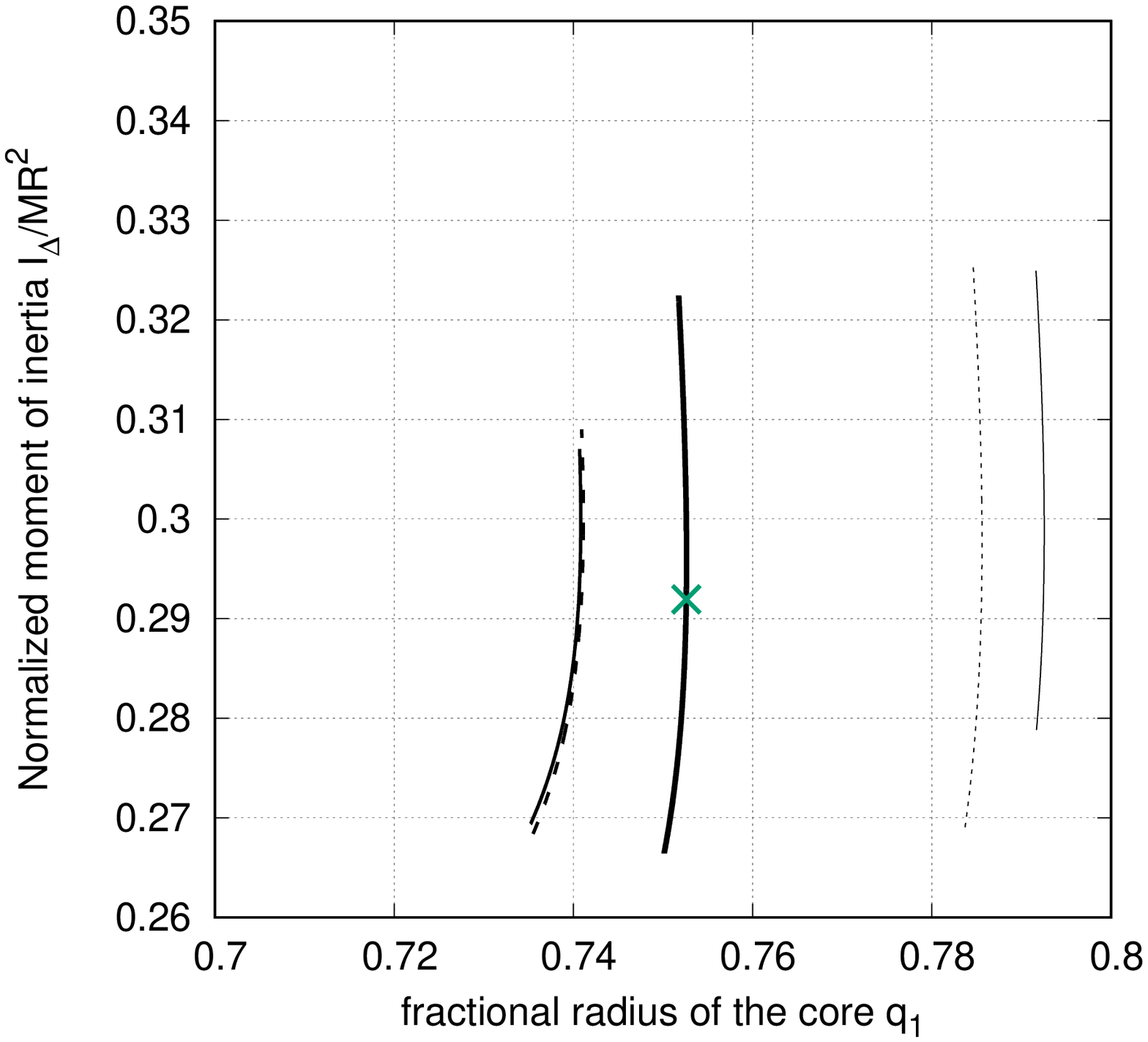}
 \caption{Same legend as for Fig. \ref{fig:q1q2_perm12.ps} but for the solution $S_{2,1}$. The internal structure corresponding to $q_1 = 0.75262$  ({\it green cross}) and computed with the {\tt DROP}-code is shown in Fig. \ref{fig:drop3layers_perm21.ps}; see also Tab. \ref{tab:drop3layers21}.}
\label{fig:q1q2_perm21.ps}
\end{figure*}

We show in Fig. \ref{fig:q1q2_perm21.ps}  the results obtained for the second set $S_{2,1}=(Y_2,Y_1)$. The fractional radius of the core is about $0.75$ while $0.75 \lesssim q_2 \lesssim 0.95$ typically. The mass-density jump at $E_1$ is about $2$, and about $4$ at the next interface $E_2$. The core is roughly $4$ times more massive than the two layers located above. It is more oblate than the surface of the planet, while $E_2$ is almost spherical, and it rotates $10$ times faster than the middle layer, which results in the significant flatening. The normalized moment of inertia stands in the range $[0.27,0.32]$ typically. The results are very sensitive to $J_6$ and especially to $J_8$, and especially regarding the mgnitude of the rotation rates. As for the first set, increasing $J_8$ shifts the curves towards short radii. For these configurations, the outer layer is significantly larger than the depth of observed zonal flows.

\begin{figure*}
  \includegraphics[height=17.9cm,bb=60 60 230 730,clip==, angle=-90]{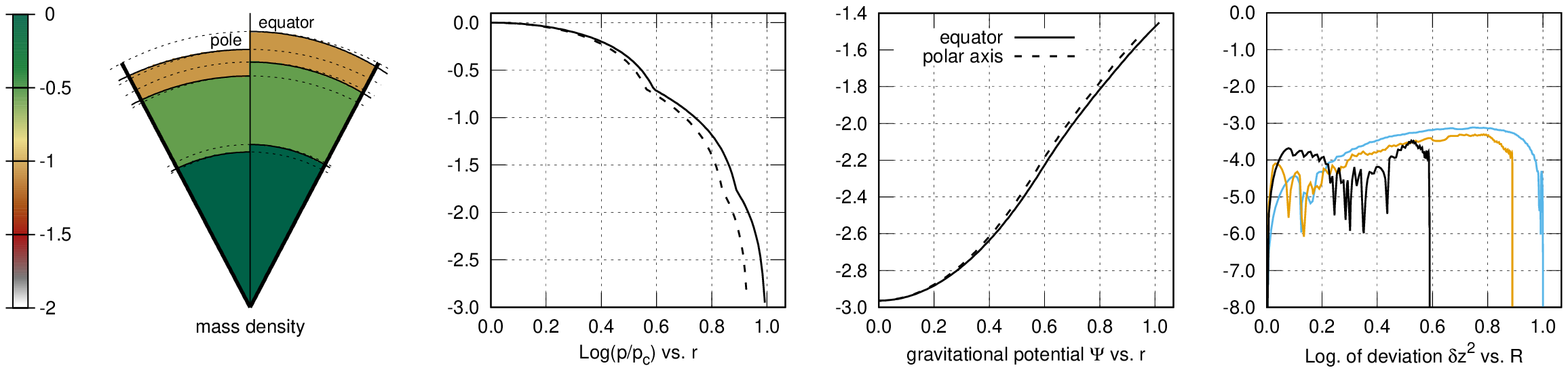}
  \caption{Same legend as for Fig. \ref{fig:drop2layers.ps} but for the $3$-layer model. The structure corresponds to $q_1 = 0.59038$ as one of the solutions $S_{1,2}$ displayed in Fig. \ref{fig:q1q2_perm12.ps}; see Tabs. \ref{tab:jup3layers} and \ref{tab:drop3layers12}.}
  \label{fig:drop3layers_perm12.ps}
\end{figure*}

\begin{figure*}
  \includegraphics[height=17.9cm,bb=60 60 230 730,clip==, angle=-90]{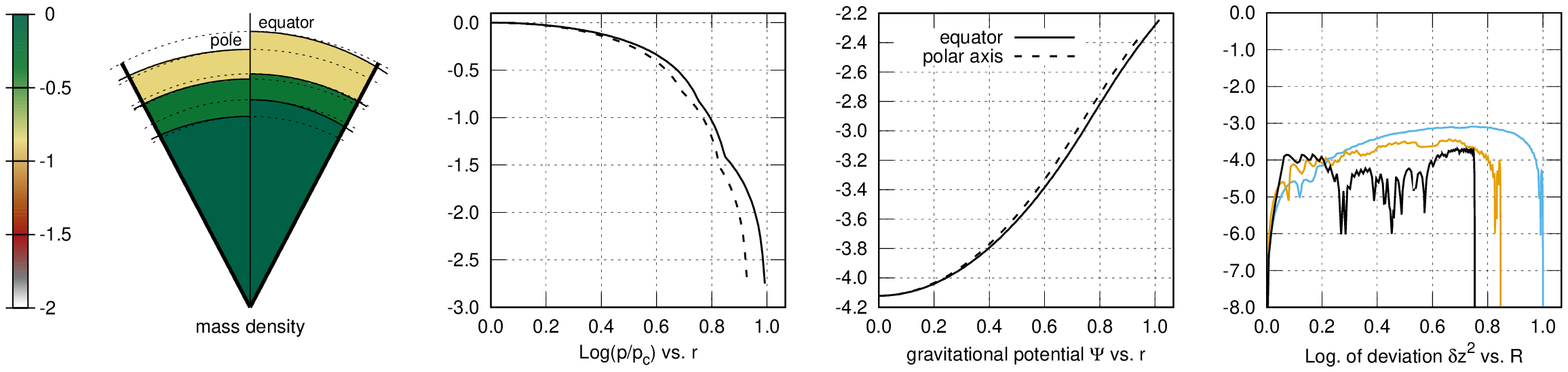}
  \caption{Same legend as for Fig. \ref{fig:drop2layers.ps} but for the $3$-layer model. The structure corresponds to $q_1 = 0.75262$ as one of the solutions $S_{2,1}$ displayed in Fig. \ref{fig:q1q2_perm21.ps}; see Tabs. \ref{tab:jup3layers} and \ref{tab:drop3layers21}.}
  \label{fig:drop3layers_perm21.ps}
\end{figure*}

\subsection{Examples of internal structure}

We can produce a particular configuration by setting, for instance, the fractional radius $q_1$ of the core. We give in Tab. \ref{tab:drop3layers12} (left, column 2) the results obtained for $q_1= 0.59048$, accessible from the canonical set $S_{1,2}$; see Figs. \ref{fig:chi3.ps} and \ref{fig:q1q2_perm12.ps}. For this case, the moment of inertia stands in the range of plausible values discussed by \cite{Neuenschwander21}, and we are close to global rotation within a few percents. There is a factor about $10$ between the mass density of the core and the mass density of the outermost layer. The mass fraction of the core is $56 \%$. The gradient of ellipticities is positive (the layers are more and more oblate from the center to the surface). Layer $2$ appears to rotate a little bit slower than layers $1$ and $3$. The configuration computed from the {\tt DROP}-code in the same conditions as for ${\cal L}=2$ is displayed in Fig. \ref{fig:drop3layers_perm12.ps}; see note \ref{note:drop}. The axis ratios $b_1/a_3$ and $b_2/a_3$ are varied until the $q_1$ and $q_2$ on output roughly coincide (given the low resolution). The configuration retained corresponds to $q_1 = 0.59038$, which is very close to the analytical analogous. The convergence occurs after $15$ SCF-cycles, and the relative virial parameter is $1.07 \times 10^{-4}$. The results are given in Tab. \ref{tab:drop3layers12} (left, column 3). We see that both approaches are in very good agreement (the confocal parameters are small, in absolute). The interfaces are very close to ellipses. The deviations are of a few $10^{-3}$ in relative for most quantities, in particular for the mass fractions and for the ellipticities.

\begin{table}
  \centering
   \begin{tabular}{rrrr}
              & this work$^\dagger$  & {\tt DROP}-code$^\star$\\ \hline
     $c_{1,3}$ & $-0.09456$ & $-0.09403$ \\
     $c_{2,3}$ & $-0.03813$ & $-0.03792$ \\
     $q_1$ & $0.59049$ & $0.59038$ \\
     $q_2$ & $0.88829$ & $0.88956$ \\
     $\epsilon_1$ & $0.29805$ & $0.30063$ \\
     $\epsilon_2$ & $0.33282$ & $0.33275$ \\
     $b_1/a_3=q_1 \bar{\epsilon}_1$ & $0.56362$ &  $0.56307$\\
     $b_2/a_3=q_2 \bar{\epsilon}_2$ & $0.83764$ &  $0.83887$\\
     $\alpha_1$  & \multicolumn{2}{c}{$3.61692$}\\
     $\alpha_2$  & \multicolumn{2}{c}{$3.12887$} \\
     $\Omega_1^2/\Omega_2^2$  & \multicolumn{2}{c}{$1.01326$} \\
      $\Omega_2^2/\Omega_3^2$  & \multicolumn{2}{c}{$0.98702$} \\
     $\nu_1$ & $0.56234$ & $0.56158$ \\
     $\nu_2$ & $0.36836$ & $0.36983$ \\    
     $I_\Delta/MR_e^2$ & $0.26397$ & $0.26419$ \\\hline
   \end{tabular}\\
   \raggedright
   $^\dagger$input data: $a_3,\epsilon_3,M,J_2,J_4,,J_6,J_8,\omobs$\\
   $^\star$input data: $a_3,\epsilon_3,M,b_1/a_3,b_2/a_3,\alpha_1,\alpha_2,\Omega_2^2/\Omega_1^2,\Omega_3^2/\Omega_2^2$\\
   \caption{Comparison between this approach (column 2) and the numerical SCF-method (column 3; see note \ref{note:drop}) for the $3$-layer model for $q_1 = 0.59038$ accessible for the canonical set $S_{1,2}=(Y_1,Y_2)$. See also Figs. \ref{fig:chi3.ps} and \ref{fig:q1q2_perm12.ps} and Tabs. \ref{tab:jupdata} and \ref{tab:jup3layers}.}
  \label{tab:drop3layers12}
\end{table}

\begin{table}
  \centering
  \begin{tabular}{rrrr}
     & this work$^\dagger$  & {\tt DROP}-code$^\star$\\ \hline
     $c_{1,3}$ & $-0.03813$ & $-0.03848$ \\
     $c_{2,3}$ & $-0.09456$ & $-0.09474$ \\
     $q_1$ & $0.75259$ & $0.75262$ \\
     $q_2$ & $0.84640$ & $0.84635$ \\
     $\epsilon_1$ & $0.39283$ & $0.39202$ \\
     $\epsilon_2$ & $0.20794$ & $0.20733$ \\
     $b_1/a_3=q_1 \bar{\epsilon}_1$ & $0.69209$ &  $0.69238$\\
     $b_2/a_3=q_2 \bar{\epsilon}_2$ & $0.82789$ &  $0.82796$\\
     $\alpha_1$  & \multicolumn{2}{c}{$1.97147$} \\
     $\alpha_2$  & \multicolumn{2}{c}{$3.71150$} \\
     $\Omega_1^2/\Omega_2^2$  & \multicolumn{2}{c}{$6.77728$} \\
      $\Omega_2^2/\Omega_3^2$  & \multicolumn{2}{c}{$0.14756$} \\
     $\nu_1$ & $0.72491$ & $0.72551$ \\
     $\nu_2$ & $0.18864$ & $0.18816$ \\    
     $I_\Delta/MR_e^2$ & $0.29193$ & $0.29188$ \\ \hline
  \end{tabular}\\
   \raggedright
   $^\dagger$input data: $a_3,\epsilon_3,M,J_2,J_4,,J_6,J_8,\omobs$\\
   $^\star$input data: $a_3,\epsilon_3,M,b_1/a_3,b_2/a_3,\alpha_1,\alpha_2,\Omega_2^2/\Omega_1^2,\Omega_3^2/\Omega_2^2$\\
  \caption{Same caption as Tab. \ref{tab:drop3layers12} but for $q_1 \approx 0.75259$  accessible for the second set $S_{2,1}=(Y_2,Y_1)$. See also Figs. \ref{fig:chi3.ps} and \ref{fig:q1q2_perm21.ps}.}
  \label{tab:drop3layers21}
\end{table}

 We give in Tab. \ref{tab:drop3layers21} (right, column 2) the results obtained for $q_1 = 0.75259$, which is reachable from $S_{2,1}$; see Figs. \ref{fig:chi3.ps} and \ref{fig:q1q2_perm12.ps}. In contrast to the first example, the intermediate layer is rotating slowly compared to layers $1$ and $3$. The core is more massive, although the mass density jump is lower by a factor $3$. There is an ellipticity reversal (layer $2$ is more oblate than the others). The configuration computed from the {\tt DROP}-code in the same conditions is displayed in Fig. \ref{fig:drop3layers_perm21.ps}. The nominal configuration is obtained for $q_1 = 0.75262$, which is very close to the analytical analogous. The SCF-cycles convergence after $14$ SCF-cycles, and the relative virial parameter is $5 \times 10^{-6}$. The results of the simulation are gathered in Tab. \ref{tab:drop3layers21} (right, column 3). Again, due to the small confocal parameters (less than $0.01$ is absolute), there is a good agreement between the two methods.

 \section{The four-layer case}
\label{sec:4layers}

\subsection{The cubic equation for the $y_i$-problem}

As any supplementary layer brings two new parameters, we can fix $2$ to additionnal gravitational moments with four layers, namely $J_{10}$ and $J_{12}$. 
For ${\cal L}=4$, (\ref{eq:mj2nmassa}) and (\ref{eq:mj2nmassb}) yields $7$ coupled equations in total. In dimensionless form, we have
\begin{subnumcases}{}
\bar{\rho} =\rho_4  (1+ C),\label{eq:j2toj12massa}\\
\eta_2 (1+C) = 1+C_1y_1+ C_2 y_2 + C_3y_3,\label{eq:j2toj12massb}\\
\eta_4 (1+C)= 1+C_1y_1^2 + C_2 y_2^2+ C_3 y_3^2,\label{eq:j2toj12massc}\\
\eta_6 (1+C)= 1 +C_1y_1^3+ C_2 y_2^3+ C_3 y_3^3\label{eq:j2toj12massd}, \\
\eta_8 (1+C)= 1 +C_1y_1^4 + C_2 y_2^4+ C_3 y_3^4\label{eq:j2toj12masse},\\
\eta_{10} (1+C)= 1 +C_1y_1^5 + C_2 y_2^5+ C_3 y_3^5\label{eq:j2toj12massf},\\
\eta_{12} (1+C)= 1 +C_1y_1^6 + C_2 y_2^6+ C_3 y_3^6\label{eq:j2toj12massg},
\end{subnumcases}
where $C=C_1+ C_2+C_3$ is set for convenience (not the same as for ${\cal L}=3$), $y_1 \epsilon_4^2=q_1 \epsilon_1^2$, $y_2 \epsilon_4^2=q_2\epsilon_2^2$, $y_3 \epsilon_4^2=q_3 \epsilon_3^2$, $q_1=\frac{a_1}{a_4}$, $q_2=\frac{a_2}{a_4}$, $q_3=\frac{a_3}{a_4}$, $\alpha_1=\frac{\rho_1}{\rho_2}$, $\alpha_2=\frac{\rho_2}{\rho_3}$, $\alpha_3=\frac{\rho_3}{\rho_4}$, and

\begin{subnumcases}{}
  C_1 = (\alpha_1-1) q_1^3 \frac{\bar{\epsilon}_1}{\bar{\epsilon}_4} \alpha_3 \alpha_2,\\
  C_2 = (\alpha_2-1) q_2^3 \frac{\bar{\epsilon}_2}{\bar{\epsilon}_4 } \alpha_3 .\\
  C_3 = (\alpha_3-1) q_3^3 \frac{\bar{\epsilon}_3}{\bar{\epsilon}_4 }.
\end{subnumcases}

At any order $n \ge 0$, we have from (\ref{eq:mj2nmassb})
\begin{flalign}
\eta_{2n} (1+C)= 1 +C_1y_1^{n} + C_2 y_2^{n}+ C_3 y_3^{n},
\end{flalign}
with the consequence that the three quantities $\eta_{2n-4}-\eta_{2n-6}$, $\eta_{2n-2}-\eta_{2n-4}$ and $\eta_{2n}-\eta_{2n-2}$ are linked by
\begin{flalign}
\label{eq:j2n4layers}
  & \left(\eta_{2n-4}-\eta_{2n-6}\right)y_1 y_2 y_3\\
  \nonumber
 &\quad-\left(\eta_{2n-2}-\eta_{2n-4}\right)(y_1y_2+y_1y_3+y_2y_3)\\
  \nonumber
   &\quad\quad +(\eta_{2n}-\eta_{2n-2})(y_1+y_2+y_3)-(\eta_{2n+2}- \eta_{2n}) =0,
\end{flalign}
valid for $n \ge 3$. In particular, by setting successively $n=3$, $n=4$ and $n=5$ in this recurrence relationship, we find the analogous of (\ref{eq:solution3}) and (\ref{eq:solution3bis}). In matrix form, this is
\begin{flalign}
	\begin{pmatrix} 
		1 & A_1 & A_2\\ 
		1 & A_4 & A_5 \\
                1 & A_7 & A_8
	\end{pmatrix}
        \begin{pmatrix}
                y_1 y_2 y_3\\
		y_1y_2 + y_1y_3+y_2y_3\\
                y_1+y_2+y_3
	\end{pmatrix}
        =     -   \begin{pmatrix} 
		A_3\\
                A_6\\
                A_9
	\end{pmatrix}.
        \label{eq:y1y2y3matrix}
\end{flalign}

where
\begin{flalign}
\begin{cases}
  A_1=\frac{\eta_2-\eta_4}{\eta_2-1}, \quad  A_2=\frac{\eta_6-\eta_4}{\eta_2-1},  \quad A_3=\frac{\eta_6-\eta_8}{\eta_2-1},\\
  A_4=\frac{\eta_4-\eta_6}{\eta_4-\eta_2}, \quad  A_5=\frac{\eta_8-\eta_6}{\eta_4-\eta_2},  \quad A_6=\frac{\eta_8-\eta_{10}}{\eta_4-\eta_2},\\
  A_7=\frac{\eta_6-\eta_8}{\eta_6-\eta_4},  \quad A_8=\frac{\eta_{10}-\eta_8}{\eta_6-\eta_4}, \quad A_8=\frac{\eta_{10}-\eta_{12}}{\eta_6-\eta_4}.\\  
\end{cases}
\label{eq:a1a9}
\end{flalign}

We can easily isolate $s=y_1+y_2+y_3$, $d=y_1 y_2+y_2 y_3+y_1y_3$ and $p=y_1y_2y_3$, and solve (\ref{eq:y1y2y3matrix}) for $y_1$, $y_2$ and $y_3$. The expressions for $s$, $d$ and $p$, which are combinations of the $A_i$'s, are reproduced in the Appendix \ref{app:4layer}. As $y_2 y_3$ and $y_2+y_3$ can be expressed as a function of $y_1$, $s$, $d$ and $p$ only, the problem of solving (\ref{eq:y1y2y3matrix}) is equivalent to finding the roots of the cubic equation
\begin{flalign}
  Y^3-sY^2+dY-p=0,
  \label{eq:p3y}
\end{flalign}
where $Y$ stands for either $y_1$, or $y_2$ or $y_3$. As we are seeking for solutions compatible with (\ref{eq:immersionconditionsc}) and (\ref{eq:maxyl}), complex roots or negatives real roots are ruled out. The relevant roots can be expressed for instance from Cardano's formula
\begin{flalign}
\label{eq:solution4y1y2y3full}
Y_k =\frac{1}{3}s +2\sqrt{-Q} \cos \left[\theta_0+(k-1)\frac{2\upi}{3}\right],
\end{flalign}
with $k=\{1,2,3\}$ and
\begin{flalign}
  \begin{cases}
    Q=\frac{1}{9}(3d-s^2),\\
    R=\frac{1}{54}(27p-9sd+2s^3)\\
    3\theta_0=\arccos \frac{R}{\sqrt{-Q^3}}, \quad \text{with } Q^3+R^2 \le 0.
  \end{cases}
\end{flalign}

\begin{table}
  \begin{tabular}{rrrrrr}
      \multicolumn{6}{l}{the $y_i$-problem$^\dagger$} \\ \hline
      $A_1$ & $-0.36437$ &$A_2$ & $+0.17180$ & $A_3$ & $-0.09407$\\
      $A_4$ & $-0.47149$ & $A_5$ & $+0.27281$ &  $A_6$ & $-0.29847$\\
      $A_7$ & $-0.57861$ & $A_8$ & $+0.63303$ & $A_9$ & $-18.33437$\\
    $s$   & $+68.81435$\\
    $d$   & $+63.03239$\\
    $p$   & $+11.24436$\\
    $Y_1$ & $+0.24221$\\
    $Y_2$ & $+0.68381$ \\
    $Y_3$ & \textcolor{red}{$+67.88832$}\\
    $C_1$ & $+1.63992$\\
    $C_2$ & $+1.50302$\\
    $C_3 (\times 10^{11}$) & $-5.53540$\\
    $\rho_4/\bar{\rho}$ & $0.24137$\\
    $J_{14} \; (\times 10^9)$ & $-511$ \\ \hline
  \end{tabular}\\
   \raggedright
   $^\dagger$input data: $a_4,\epsilon_4,M,J_2,J_4,J_6,J_8,J_{10},J_{12}$\\
  \caption{The canonical solution of the $y_i$-problem for the $4$-layer problem applied to Jupiter's data; $J_{12}$ is computed from (\ref{eq:j2n4layers}). See Tab. \ref{tab:jupdata} for the reference data.}
  \label{tab:jup4layers_notworking}
\end{table}

\begin{figure}
  \includegraphics[height=9.2cm,bb=40 90 544 730, clip==,angle=-90]{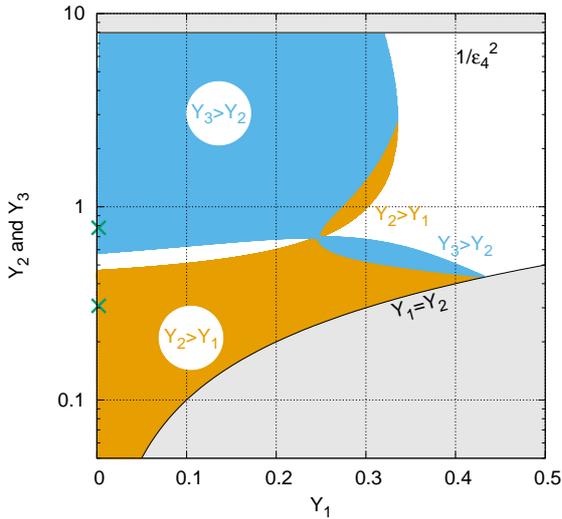}
  \caption{The canonical solution $Y_2 > Y_1$ ({\it orange}) and $Y_3 > Y_2$ ({\it blue}) of the $y_i$-problem for the $4$-layer problem applied to Jupiter, and for $J_{10}$ and $J_{12}$ within the error bars. See Fig. \ref{fig:4layers_tworootsq1q2q3.ps} for the equilibrium configuration obtained for the canonical set of Tab. \ref{tab:jup4layers_workinga} ({\it green cross}). See Tab. \ref{tab:jupdata} for the reference data.}
\label{fig:4layersy1y2y3.ps}
\end{figure}

There are at most $6$ triplets $(y_1,y_2,y_3)$ that can be built from the canonical set, and this depends strongly on the $A_i'$s.  Note that, contrary to (\ref{eq:solution3y1y2}), the $Y_k$'s as given by (\ref{eq:solution4y1y2y3full}) are not sorted in ascending order.


\begin{figure}
  \includegraphics[height=8.7cm,bb=40 90 544 730, clip==,angle=-90]{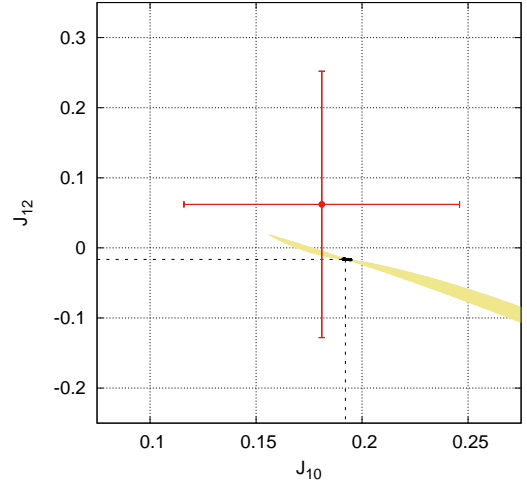}\\
  \includegraphics[height=8.4cm,bb=100 50 484 765, clip==,angle=-90]{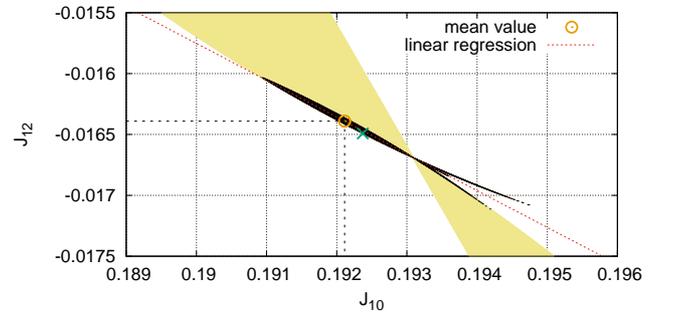}
  \caption{Values of $J_{10}$ and $J_{12}$ (in units of $10^{-6}$) associated with all the solutions of the $y_i$-problem for the $4$-layer model and fulfilling (\ref{eq:maxyl}) ({\it yellow}), with a zoom around mean values ({\it bottom panel}). Values corresponding to $y_i \le 1$ have been separated ({\it black}). See Fig. \ref{fig:4layers_tworootsq1q2q3.ps} for an example of internal structure computed with the set of Tab. \ref{tab:jup4layers_workinga} ({\it green cross}).}
\label{fig:layer4deltaj10j12.ps}
\end{figure}

\begin{figure}
  \includegraphics[height=8.4cm,bb=100 50 484 765, clip==,angle=-90]{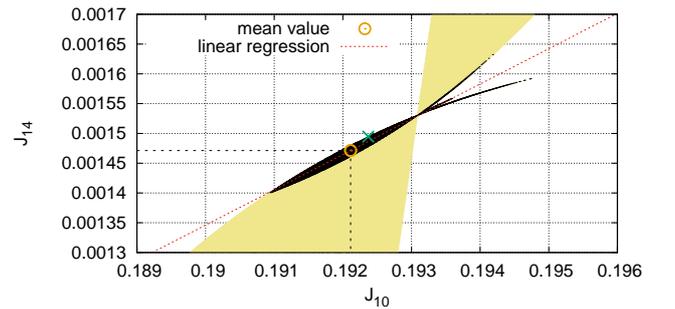}\\
 \caption{Same caption as for Fig. \ref{fig:layer4deltaj10j12.ps} but for $J_{10}$ and $J_{14}$.}
\label{fig:layer4j14.ps}
\end{figure}

\subsection{Results with Jupiter's data. Rejection of the canonical set for central values. Error bars}

Based on Jupiter's data, we can determine the nine coefficients $A_1$ to $A_9$, then $s$, $d$ and $p$, the $Y_k$'s. The results are given in Tab. \ref{tab:jup4layers_notworking}. We see that one of the roots exceeds the limit given by (\ref{eq:maxyl}). This solution, based on central values of the $J_{2n}$'s, is therefore to be rejected. But we can consider the error bars in $J_{10}$ and in $J_{12}$. In pratice, we find easier to vary $Y_1$ and $Y_2>Y_1$ (for instance), and to retain only the canonical solutions $(Y_1,Y_2,Y_3>Y_2)$ compatible with $J_{10}\pm \Delta J_{10}$ and $J_{12} \pm \Delta J_{12}$, and fulfilling (\ref{eq:maxyl}). The reason for this procedure is that, in contrast with the case ${\cal L}=3$, each point inside the error box $(J_{10}\pm \Delta J_{10}) \times (J_{12}\pm \Delta J_{12})$ does not lead to acceptable $Y_k$'s. The result of the scan is presented in Fig. \ref{fig:4layersy1y2y3.ps}. Figure \ref{fig:layer4deltaj10j12.ps} shows the corresponding domain in the $(J_{10},J_{12})$-plane. It happens that this domain is extremely reduced in size, off-centered with respect to central values. There is a remarkable correlation between these two quantities, especially if we limit the data to $Y_k \le 1$ (see Sect. \ref{subsec:yellproblem}), in which case we find $J_{10} \approx +0.19211 \times 10^{-6}$ and $J_{12} \approx -0.01639 \times 10^{-6}$ as mean values, and $J_{12}  \approx -0.30217 J_{10} + 0.04166 \times 10^{-6}$ from a linear regression. We show in Fig. \ref{fig:layer4j14.ps} values of $J_{14}$ deduced by (\ref{eq:j2n4layers}) from the same sample. The mean value is about $+1.47 \times 10^{-9}$, which agrees with the prediction of the $3$-layer problem. A linear regression yields $J_{14} \approx 0.059234 J_{10} - 0.009908 \times 10^{-6}$. Note that the $J_{2n}$'s are not affected by the set $(y_1,y_2,y_3)$ formed by permutation of the $Y_k$'s, according to (\ref{eq:p3y}). Interestingly enough, the incertainty in $J_8$, when accounted for, does not produce big changes in the solutions of the $y_i$-problem, and Fig. \ref{fig:4layersy1y2y3.ps} is slightly stretched but globally conserved. The impact on the moments $J_{10}$ to $J_{14}$ is also weak, with shifts of a few purcents (which is the magnitude of $\Delta J_8/J_8$), as follows (see Tab. \ref{tab:jupdata}):
\begin{flalign}
  \Delta J_8 = + 0.021  \times 10^{-6} \rightarrow
  \begin{cases}
   \Delta J_{10} \approx + 0.0050 \times 10^{-6},\\
   \Delta J_{12} \approx - 0.0008 \times 10^{-6},\\
   \Delta J_{14}\approx + 0.11 \times 10^{-9}.
  \end{cases}
\end{flalign} 

\subsection{The key-function $\chi_4(q_1,q_2,q_3)$. Result for a given canonical set. Example of structure}

The rotation rates $\Omega_1$ to $\Omega_4$ are still expressed from (32) and (33) of Paper II. Each $\Omega_i$ depends on $3{\cal L}-1=11$ variables, namely
\begin{flalign}
\Omega_i \equiv \Omega_i(\rho_4,q_1,q_2,q_3,\epsilon_1,\epsilon_2,\epsilon_3,\epsilon_4,\alpha_1,\alpha_2,\alpha_3),
\end{flalign}
for $i =\{1,2,3,4\}$, where $\epsilon_4$ is the ellipticity of the planet. As $y_1$ to $y_3$, $C_1$ to $C_3$ and $\rho_4$ are already known (see above), we can eliminate $8$ variables, for instance the $\epsilon_i$'s and the $\alpha_i$'s. It follows that the $\Omega_i$'s depend intrinsically on the three variables $q_1$, $q_2$ and $q_3$. Then, (\ref{eq:omobs}) holds, where the $\chi$-function for the four-layer problem is of the form
\begin{equation}
  \chi_4(q_1,q_2,q_3) \equiv \frac{\Omega_1^2(q_1,q_2,q_3)}{\omobs^2}-1,
  \label{eq:chi4}
\end{equation}
where we assume, as for the $2$-layer and $3$-layer problems, that the rotation of the core coincides with $\omobs$. In principle, the roots of the equation $\chi_4=0$ are easily found numerically, by varying the $q_i$'s in the allowed ranges. However, this is a tedious task because of the number of dimensions involved, which now amounts to $6$. Actually, we have two probe the $(q_1,q_2,q_3)$-space for each set of $\{y_i\}$, and its $6$ possible permutations. As we have checked, not only the computing time is rapidly prohibitive, but the information on output is difficult to summarize. We find more illustrative to select one canonical set among the triplets $(Y_1,Y_2,Y_3)$ shown in Fig. \ref{fig:4layersy1y2y3.ps}. We have chosen a set with a very low value of $Y_1$, which enables to catch small cores that are not permitted for ${\cal L} \le 3$. The data for this specific set are given in Tab. \ref{tab:jup4layers_workinga}. The table also contains the associated moments $J_{10}$ and $J_{12}$, and the prediction for $J_{14}$ computed from (\ref{eq:j2n4layers}); see also Figs. \ref{fig:layer4deltaj10j12.ps} and \ref{fig:layer4j14.ps}. The physical equilibria are then determined by solving (\ref{eq:chi4}). In this experiment, we have used a moderate resolution of $0.01$ for both $q_1 \in [0,1]$ and 
\begin{flalign}
  q_2 \ge \sqrt{\max\{q_1^2,q_1^2 + (y_2-y_1) \epsilon_4^2\}} \equiv q_{2,\rm min},
  \end{flalign}
to scan the $(q_1,q_2)$-plane. For given $q_1$ and $q_2$, the root $q_3$ in (\ref{eq:chi4}) is then searched for in the range $[q_{3,\rm min},1]$ where 
\begin{flalign}
  q_3 \ge \sqrt{\max\{q_2^2,q_2^2 + (y_3-y_2) \epsilon_4^2\}} \equiv q_{3,\rm min},
  \end{flalign}
according to the immersion conditions, again.

\begin{table}
  \begin{tabular}{rrr}
    & the $y_i$-problem \\ \hline
    $A_8$ & $+0.38903$\\
    $A_9$ & $-0.28398$\\
    $s$   & $+1.08772$\\
    $d$   & $+0.24135$\\
    $p$   & $+4.78269 \times 10^{-4}$\\
    $Y_1$ & $+1.99959 \times 10^{-3}$\\
    $Y_2$ & $+0.30724$\\
    $Y_3$ & $+0.77847$\\
    $C_1$ & $+0.13120$\\
    $C_2$ & $+2.24911$\\
    $C_3$ & $+1.48762$\\
    $\rho_4/\bar{\rho}$ & $+0.20542$\\
    $J_{14} \; (\times 10^9)$ & $+1.49$ \\ \hline
  \end{tabular}\\
  $^\dagger$input data : $a_2,\epsilon_2,M,J_2,J_4,J_6,J_8,\omobs$ from Tab. \ref{tab:jupdata}\\ \hspace*{1.7cm} $J_{10} = \mathbf{+0.1923711\dots}$, $J_{12} = \mathbf{-0.0164917\dots}$
  \caption{A canonial solution $\{Y_k\}$ of the $y_i$-problem in the $4$-layer case; see also Figs. \ref{fig:4layersy1y2y3.ps}, \ref{fig:layer4deltaj10j12.ps} and \ref{fig:layer4j14.ps}. Values for $A_1$ to $A_7$, which do not depend on $J_{10}$ and $J_{12}$, are unchanged; see Tab. \ref{tab:jup4layers_notworking}.}
   \label{tab:jup4layers_workinga}
\end{table}

\begin{figure*}
  \includegraphics[height=8.7cm,bb=40 90 544 730, clip==,angle=-90]{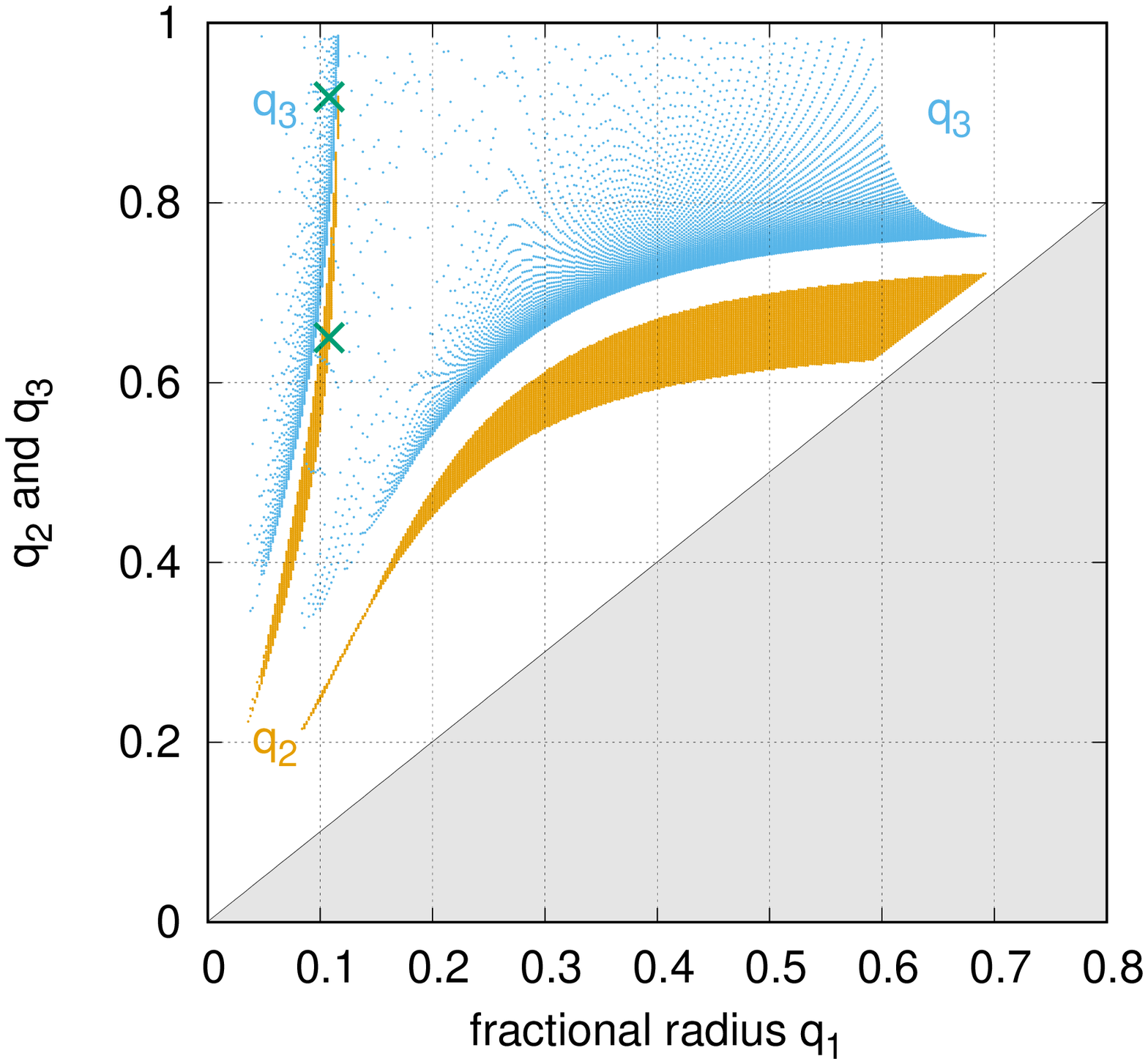}
  \includegraphics[height=8.7cm,bb=40 90 544 730, clip==,angle=-90]{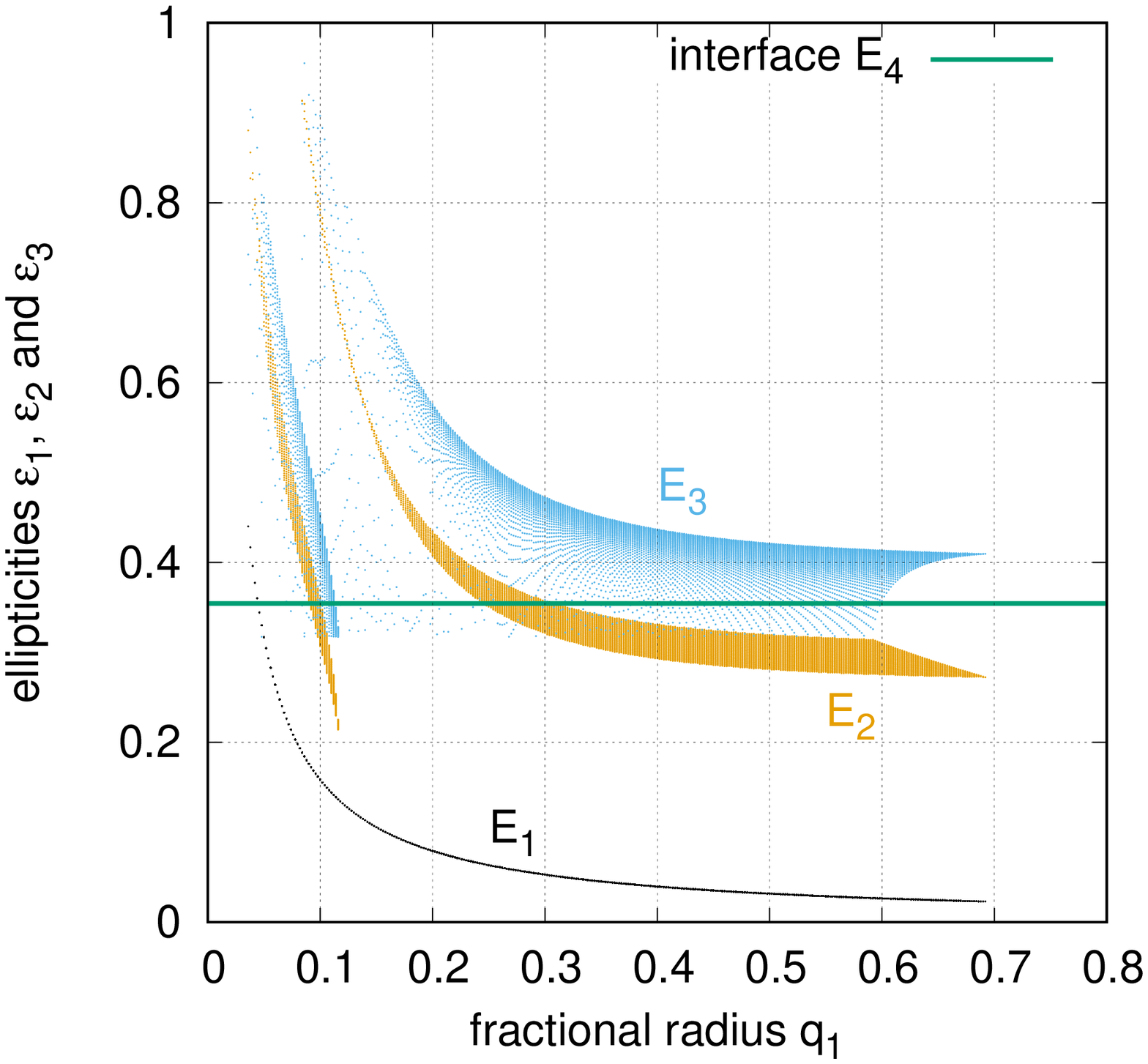}\\
  \includegraphics[height=8.7cm,bb=40 90 544 730, clip==,angle=-90]{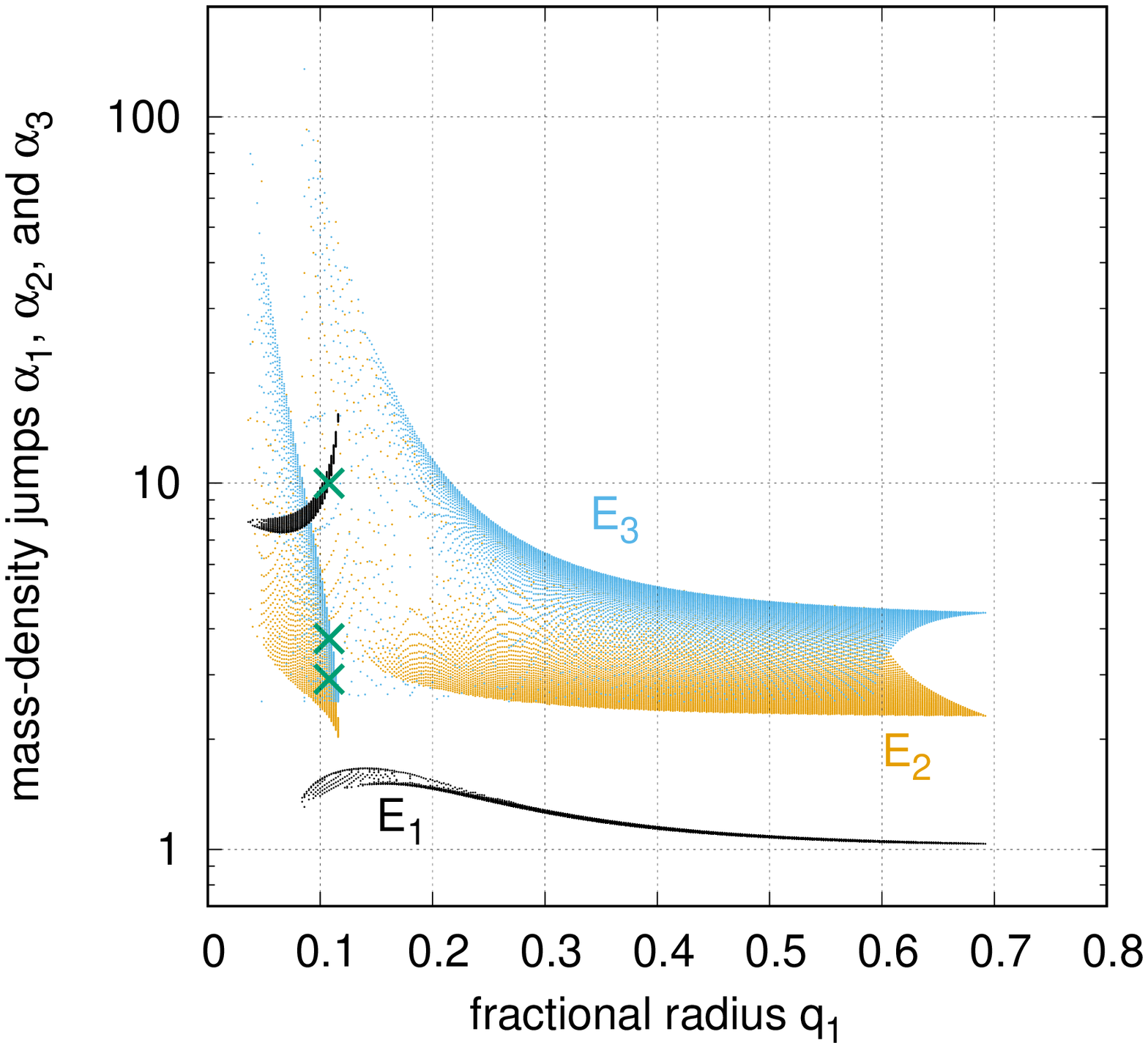}
  \includegraphics[height=8.7cm,bb=40 90 544 730, clip==,angle=-90]{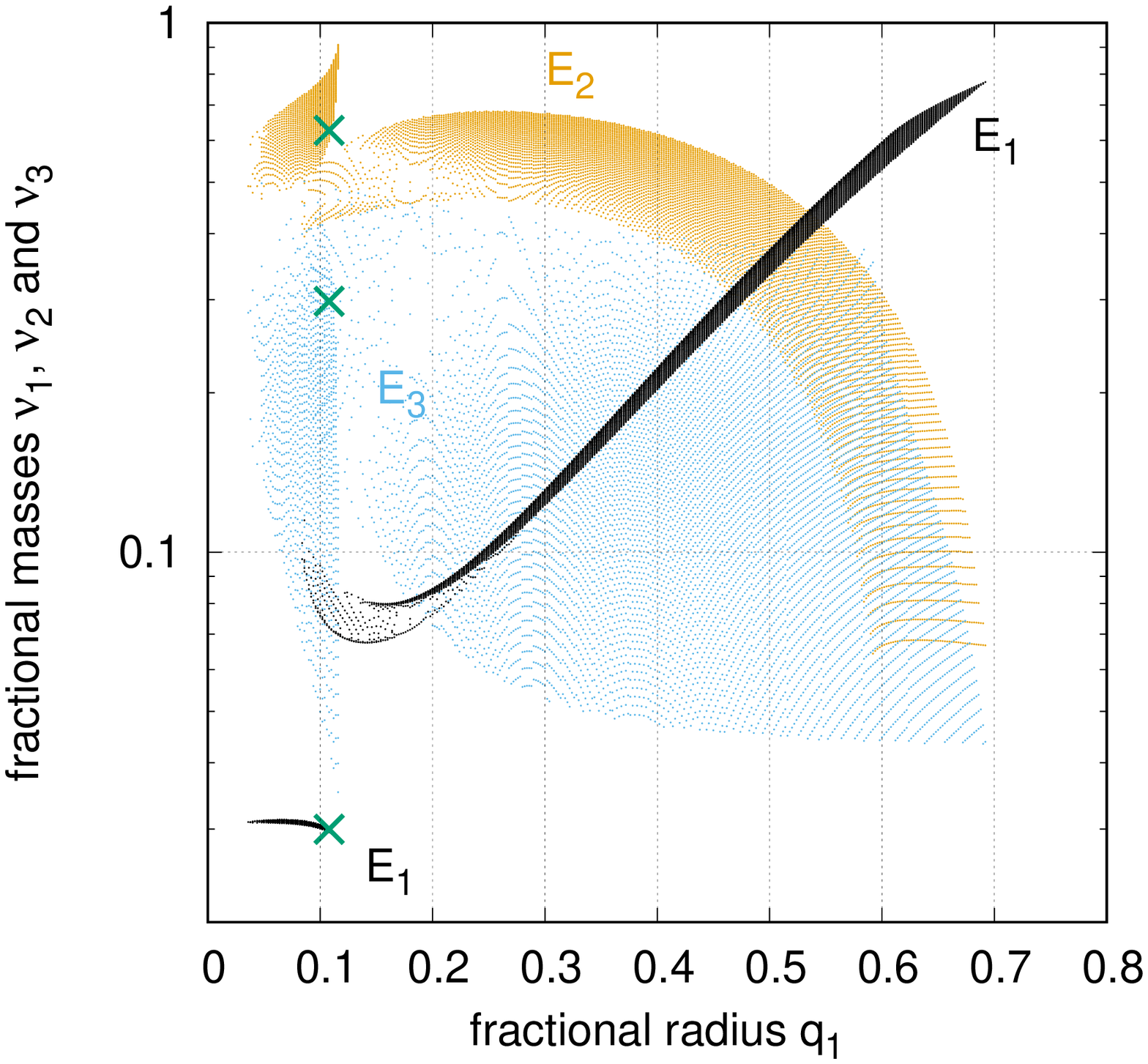}\\
  \includegraphics[height=8.7cm,bb=40 90 544 730, clip==,angle=-90]{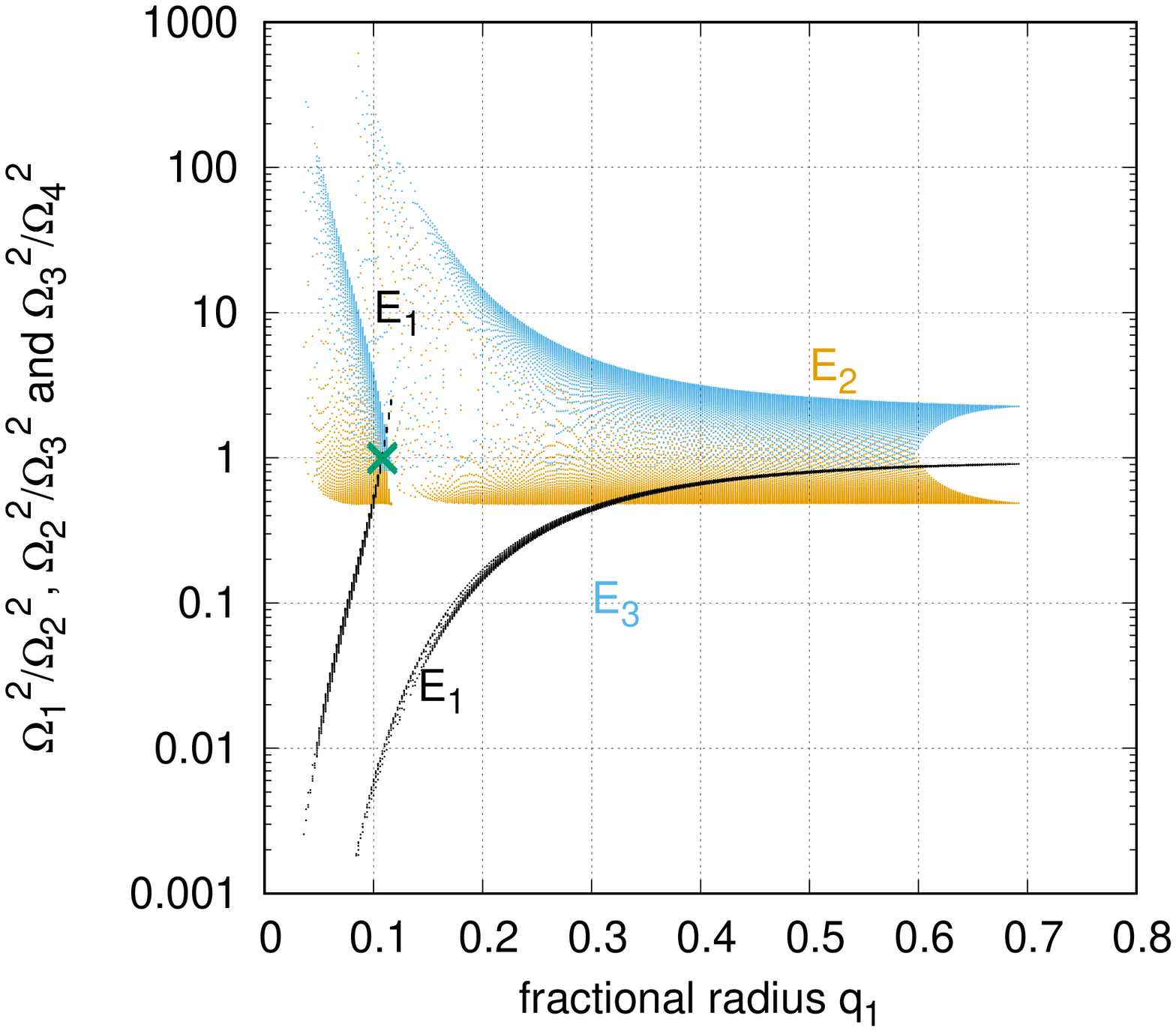}
 \includegraphics[height=8.7cm,bb=40 90 544 730, clip==,angle=-90]{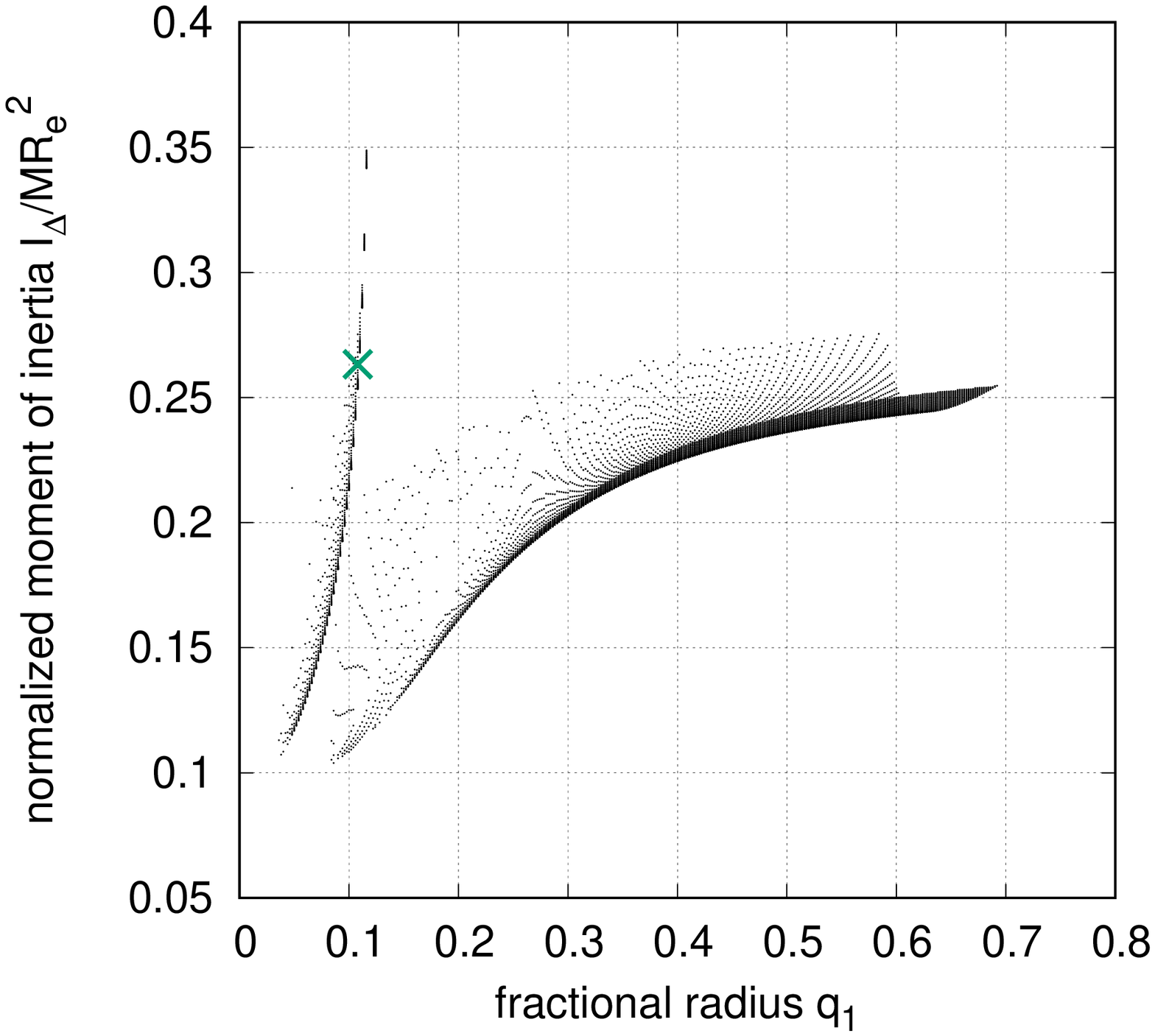}
 \caption{Results for the canonical set of Tab. \ref{tab:jup4layers_workinga} (each dot is an equilibrium). The graphs shows the fractional radii $q_1$, $q_2$ and $q_3$ ({\it top left panel}), mass-density jumps $\alpha_1$, $\alpha_2$ and $\alpha_3$ at the interfaces ({\it top right panel}) ({\it middle panel}), the ellipticities $\epsilon_1$, $\epsilon_2$ and $\epsilon_3$ ({\it middle left panel}), the mass fractions $\nu_1$, $\nu_2$ and $\nu_3$ ({\it middle right panel}), the ratio of rotation rates squared $\Omega_2^2/\Omega_1^2$, $\Omega_3^2/\Omega_2^2$ and $\Omega_4^2/\Omega_1^2$ ({\it bottom left panel}), and the normalized moment of inertia ({\it bottom right panel}) as a function of the fractional radius of the core $q_1$.}
\label{fig:4layers_tworootsq1q2q3.ps}
\end{figure*}

\begin{figure*}
  \includegraphics[height=17.9cm,bb=60 60 230 730,clip==, angle=-90]{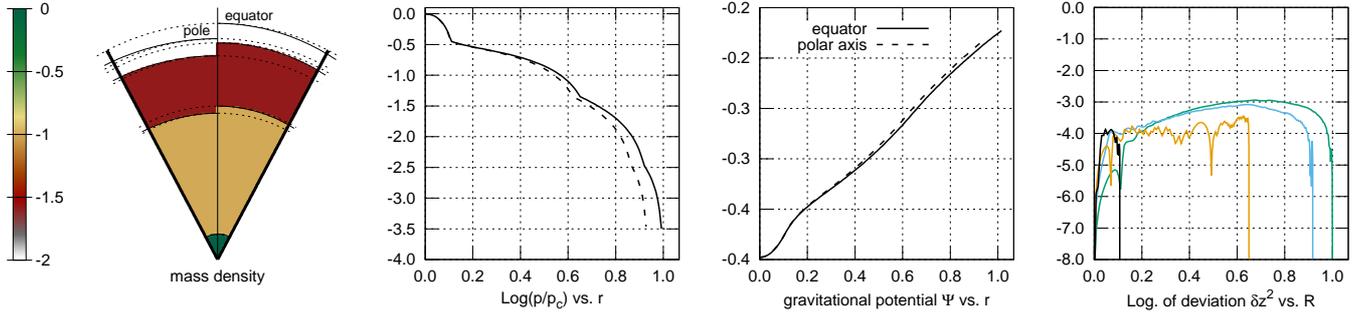}
  \caption{Same legend as for Fig. \ref{fig:drop3layers_perm12.ps} but the $4$-layer model applied to Jupiter, with the parameters of Tabs. \ref{tab:jup4layers_workinga} (a canonical solution) and \ref{tab:drop4layers}; see also Fig. \ref{fig:4layers_tworootsq1q2q3.ps}.}
  \label{fig:drop4layers_perm12.ps}
  \end{figure*}

  \begin{table}
 \centering
  \begin{tabular}{rrr}
    & this work  & {\tt DROP}-code\\ \hline
     $c_{14}$ & $-0.12528$ & $-0.12533$ \\
     $c_{24}$ & $-0.08696$ & $-0.08766$ \\
     $c_{34}$ & $-0.02780$ & $-0.02800$ \\
    $q_1$ & $0.10800$ & $0.10800$ \\
    $q_{2,\rm min}$ & $0.22357$ \\
     $q_2$ & $0.64980$ & $0.64926$ \\
    $q_{3,\rm min}$ & $0.69383$ \\
     $q_3$ & $0.91772$ & $0.91762$ \\
     $\epsilon_1$ & $0.14670$ & $0.13305$ \\
     $\epsilon_2$ & $0.30223$ & $0.29973$ \\
     $\epsilon_3$ & $0.34064$ & $0.34035$ \\
    $b_1/a_4=q_1 \bar{\epsilon}_1$ & $0.10683$ & $0.10704$\\
    $b_2/a_4=q_2 \bar{\epsilon}_2$ & \multicolumn{2}{c}{$0.61941$}\\
    $b_3/a_4=q_3 \bar{\epsilon}_3$ & \multicolumn{2}{c}{$0.86283$} \\
   $\alpha_1$  & \multicolumn{2}{c}{$9.98762$}  \\
   $\alpha_2$  & \multicolumn{2}{c}{$3.75928$} \\
   $\alpha_3$  & \multicolumn{2}{c}{$2.91428$} \\
   $\Omega_1^2/\Omega_2^2$  & \multicolumn{2}{c}{$1.01294$} \\
   $\Omega_2^2/\Omega_3^2$  & \multicolumn{2}{c}{$1.04399$} \\
   $\Omega_3^2/\Omega_4^2$  & \multicolumn{2}{c}{$1.01262$} \\
    $\nu_1$ & $0.02995$ & $0.02986$\\
     $\nu_2$ & $0.62646$ & $0.62632$\\
    $\nu_3$ & $0.29779$ & $0.29799$ \\
      $I_\Delta/MR_e^2$ & $0.26328$ & $0.26313$ \\ \hline
  \end{tabular}
  \caption{Results for the configuration with $q_1=0.108$ obtained for the parameters listed in Tab. \ref{tab:jup4layers_workinga}; see also Fig. \ref{fig:drop4layers_perm12.ps}.}
  \label{tab:drop4layers}
  \end{table}

We find that all permutations are relevant. We show in Fig. \ref{fig:4layers_tworootsq1q2q3.ps} the physical configurations accessible from the canonical set $S_{1,2,3}$. There are two groups of configurations. The first group consists in small cores with $q_1 \lesssim 0.12$. The ellipticities are rather large for the smallest values of $q_1$. The mass density jumps are the largest, with $\alpha_3$ of a few hundreds at very small $q_1$. This means a contrast between the centre and the surface of several thousands. The fractional mass of the core is small, of the order of $0.02$, while the most massive layer is layer $2$. The rotation rates are very different (by up to 2 orders of magnitudes at very small $q_1$). Layer $2$ is rotating very fast, while surface layers are under-rotating in relative. This second group is comparable to what is obtained from $S_{1,2}$ of the three-layer problem (although smaller values of $q_1$ are obtained here). We have $q_1 \gtrsim 0.09$ and $q_2$ does not exceed about $0.73$. The mass density jumps are mostly of the order of $4$, except at $E_1$ where $\alpha_1$ is close to unity. The core is highly spherical and its fractional mass is as low as $7 \%$ at small $q_1$ but reaches $80\%$ at large $q_1$. The two groups of equilibria overlap a little bit at $q_1 \approx 0.1$, with the possibility of two distinct roots for $\chi_4$ for a given pair $(q_2,q_3)$. For reasons evoked already for the $3$-layer problem, the configurations with $q_3 \gtrsim 0.96$ correspond to a transition inside the upper atmosphere where rotation is latitude-dependent. This is a negligible part of allowed configurations. 


We give in Tab. \ref{tab:drop4layers} (column 2) the results obtained for a small core with fractional radius $q_1=0.10800$, which is marked in Fig. \ref{fig:4layers_tworootsq1q2q3.ps}. The core is close to spherical ($\epsilon_1 \approx 0.146$), with mass density about $10$ times larger than layer $2$, but a small mass-fraction of about $3\%$. The most massive layer is layer $2$ (sat on the core). The gradient of ellipticity is positive from the center to the surface. The structure is close to global rotation within a few purcents. The internal structure computed from the {\tt DROP}-code is displayed in Fig. \ref{fig:drop4layers_perm12.ps} and the associated output data are listed in Tab. \ref{tab:drop4layers} (column 3). As expected from the confocal parameter, the two approaches agree globally within a few $10^{-4}$ in relative (and $10^{-3}$ for the ellipticities).

\begin{figure*}
  \includegraphics[height=8.7cm,bb=40 90 544 730, clip==,angle=-90]{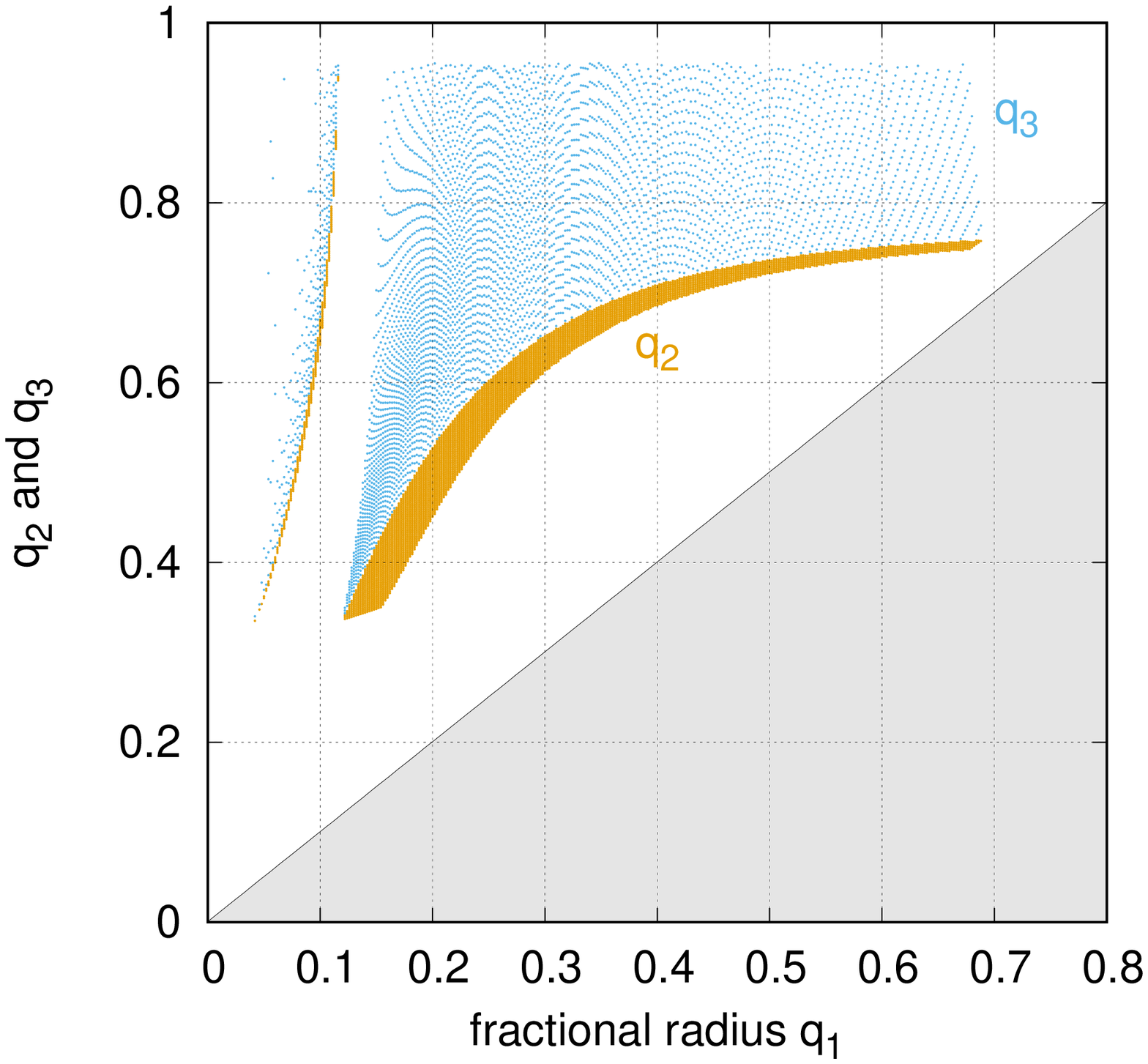}
  \includegraphics[height=8.7cm,bb=40 90 544 730, clip==,angle=-90]{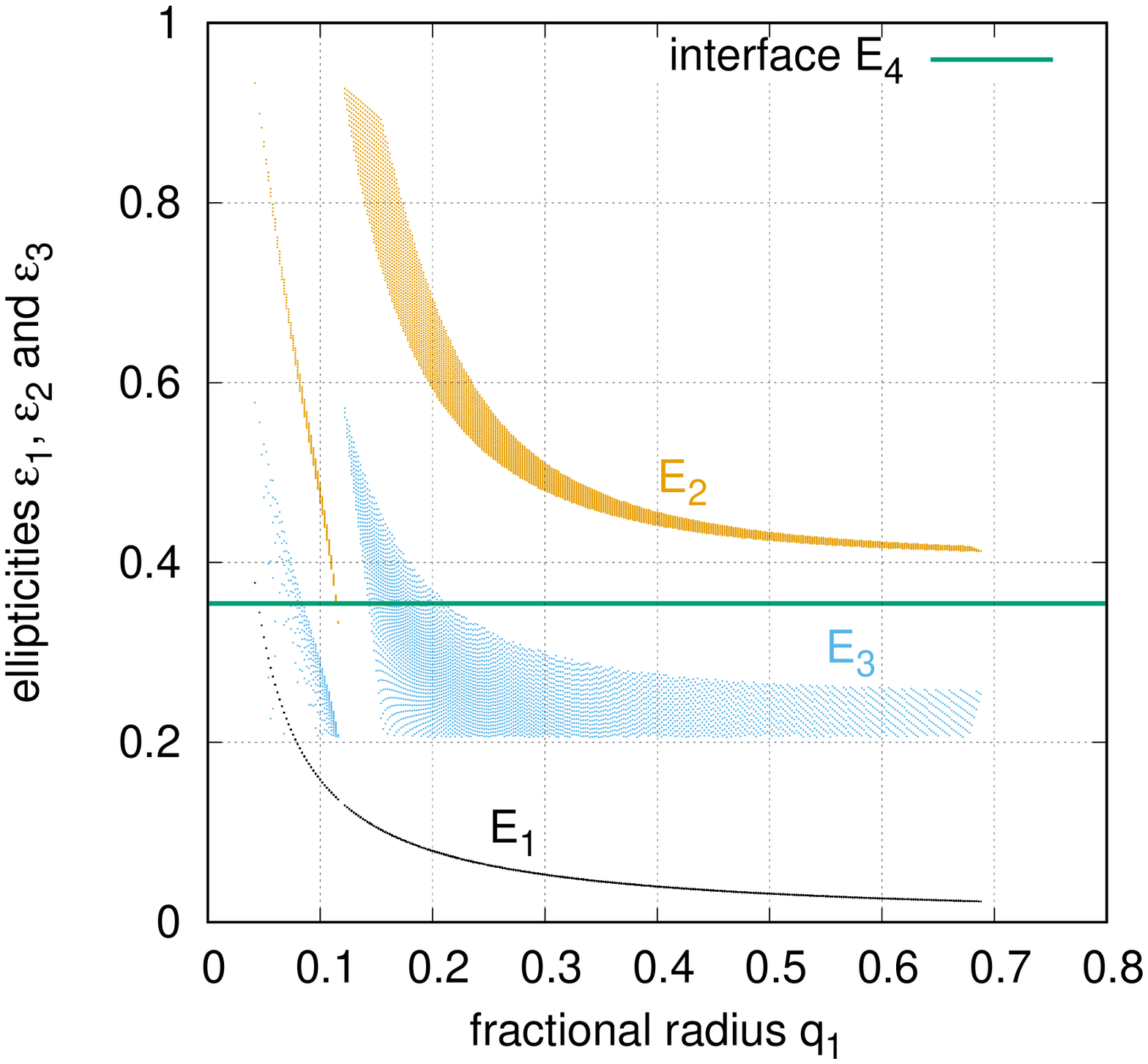}\\
  \includegraphics[height=8.7cm,bb=40 90 544 730, clip==,angle=-90]{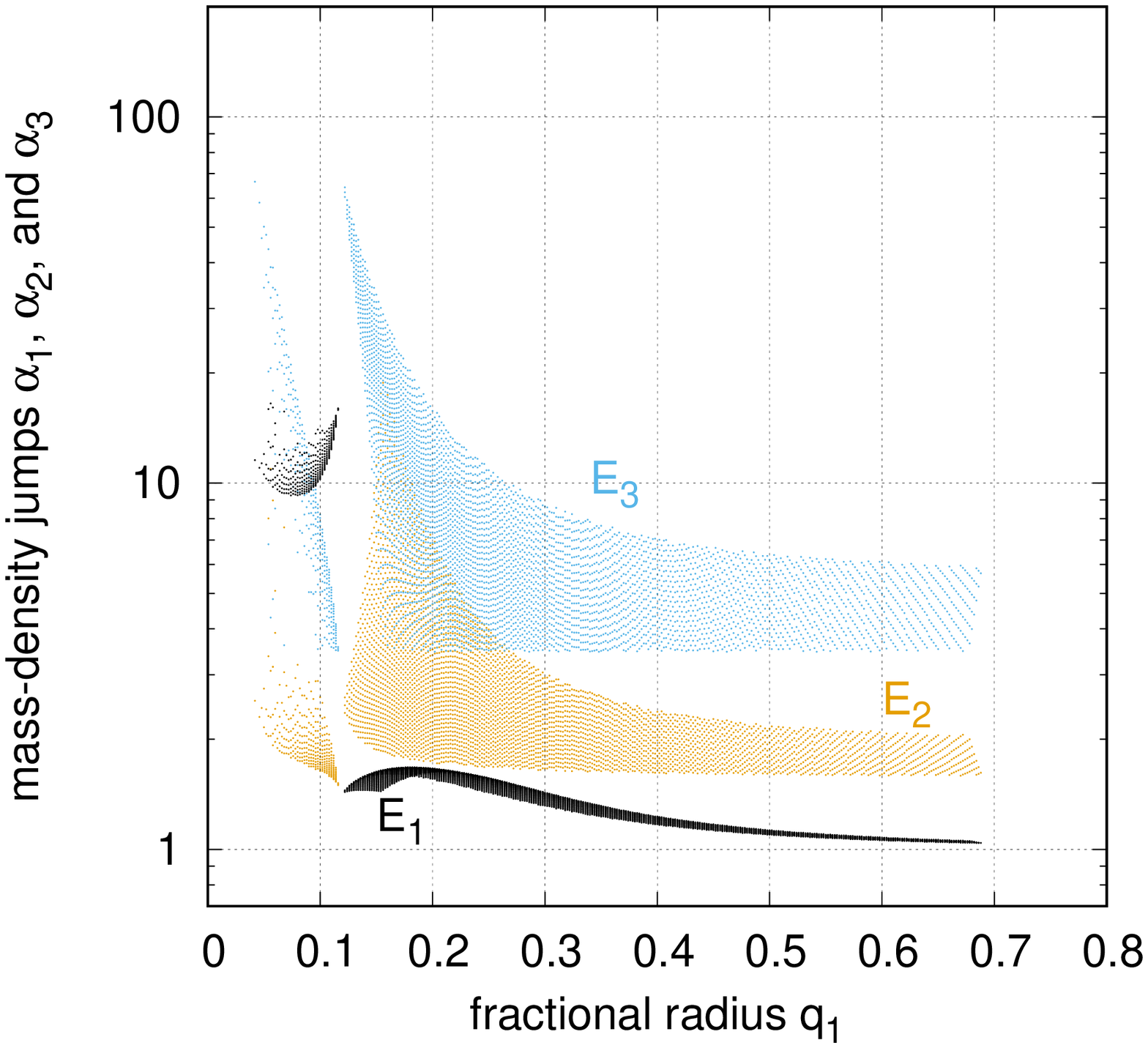}
  \includegraphics[height=8.7cm,bb=40 90 544 730, clip==,angle=-90]{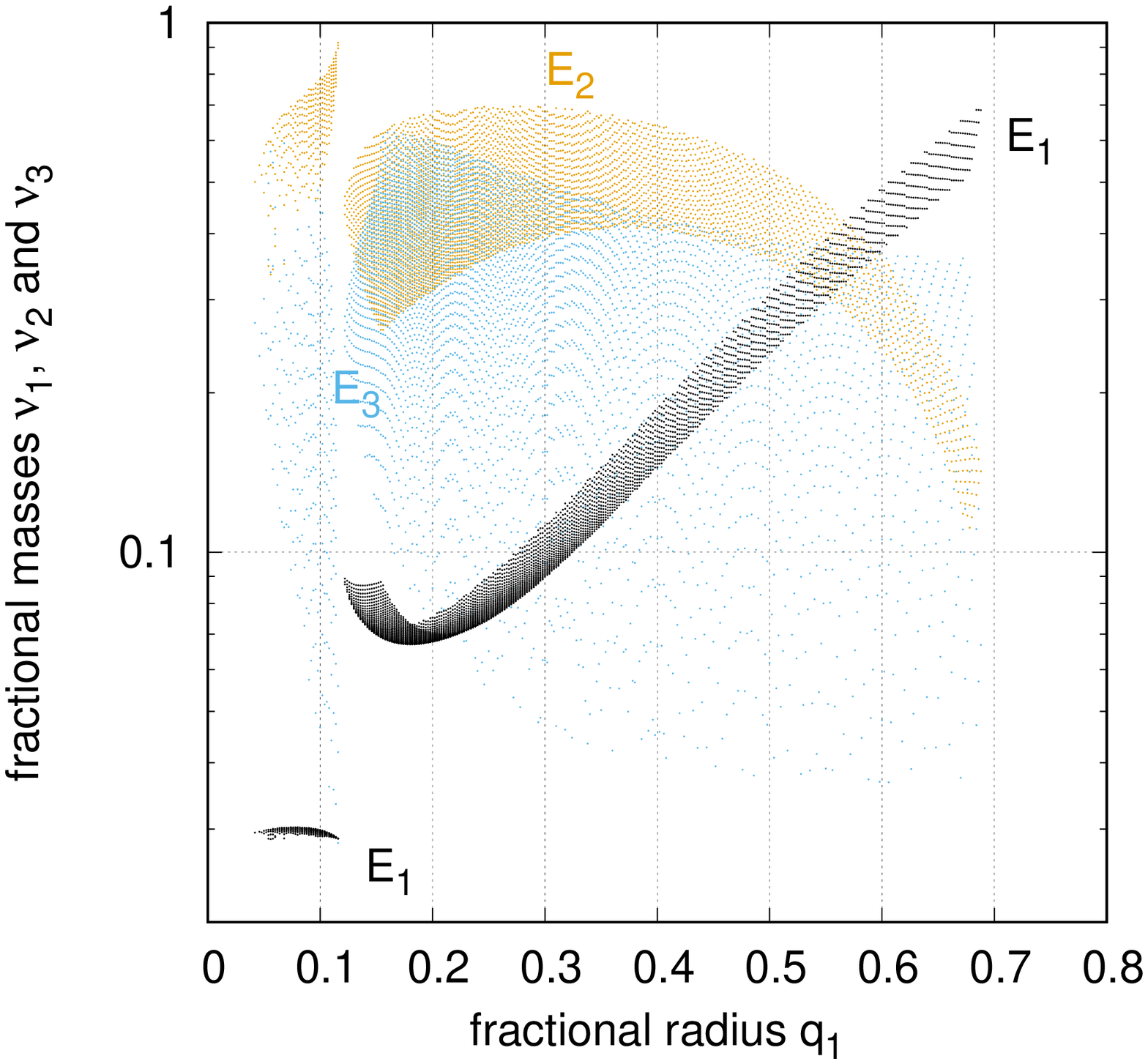}\\
  \includegraphics[height=8.7cm,bb=40 90 544 730, clip==,angle=-90]{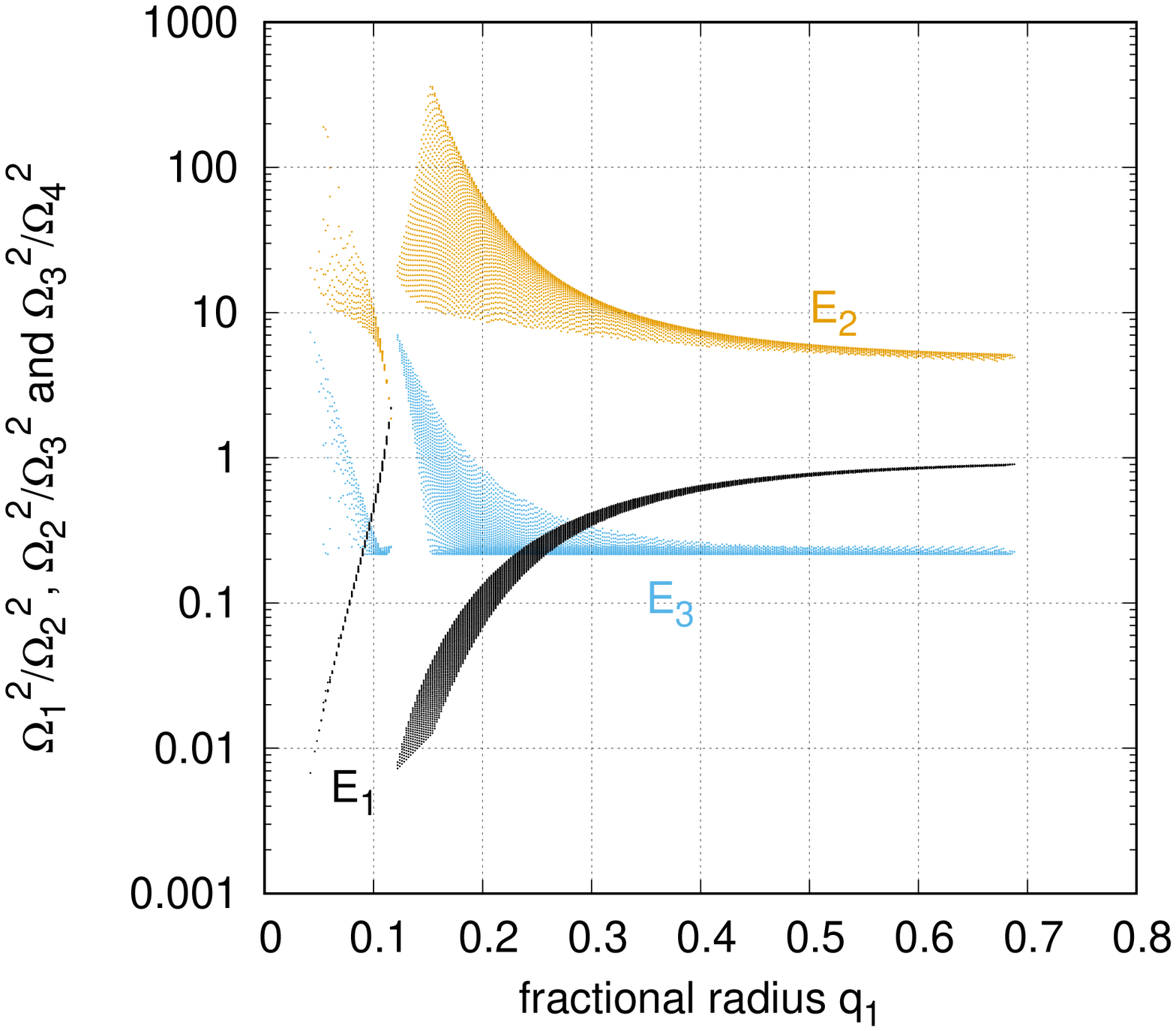}
 \includegraphics[height=8.7cm,bb=40 90 544 730, clip==,angle=-90]{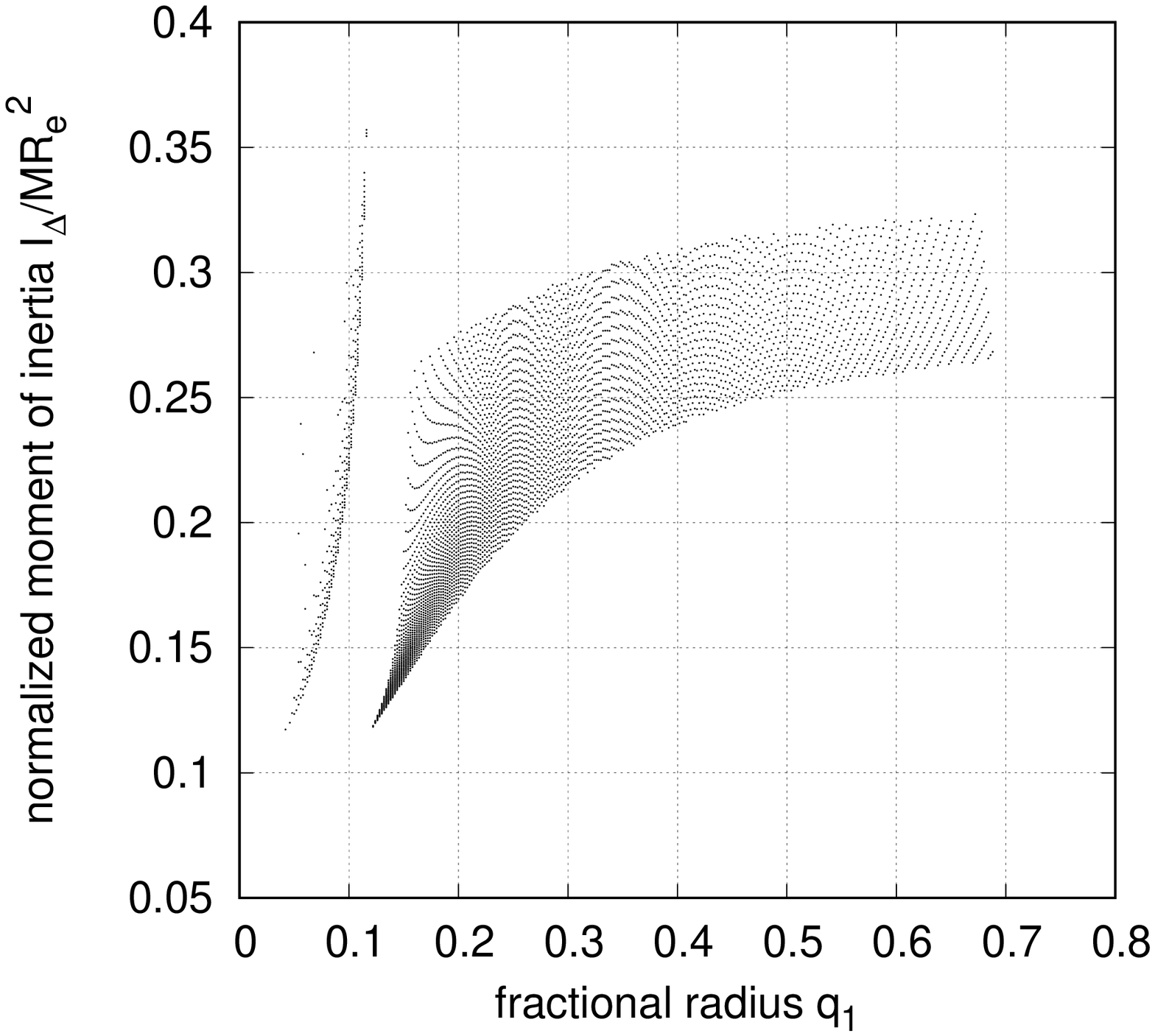}
 \caption{Same caption as for Fig. \ref{fig:4layers_tworootsq1q2q3.ps} but for the set $S_{1,3,2}$ ($y_3$ and $y_2$ are permuted with respect to $S_{1,2,3}$); set Tab. \ref{tab:jup4layers_workinga}.}
\label{fig:4layers_S132_q1q2q3.ps}
\end{figure*}

\subsection{Results for a given canonical set (con't)}

The results obtained for $S_{1,3,2}$, shown in Fig. \ref{fig:4layers_S132_q1q2q3.ps}, are very similar. By permuting $y_2$ and $y_3$, the thresholds $q_{2,\rm min}$ and $q_{3,\rm min}$ are modified. Accordingly, higher values of $q_2$ are required and lower values of $q_3$ are possible. The two groups of configurations are still present, but well separated without any overlap. The main differences concern the distribution of ellipticities (there is a reversal between $\epsilon_2$ and $\epsilon_3$, as a direct consequence of the permutations $y_2 \leftrightarrow y_3$) and the rotation rates (there is, again, a reversal between the relative motion of layers $2$).

\begin{figure*}
  \includegraphics[height=8.7cm,bb=40 90 544 730, clip==,angle=-90]{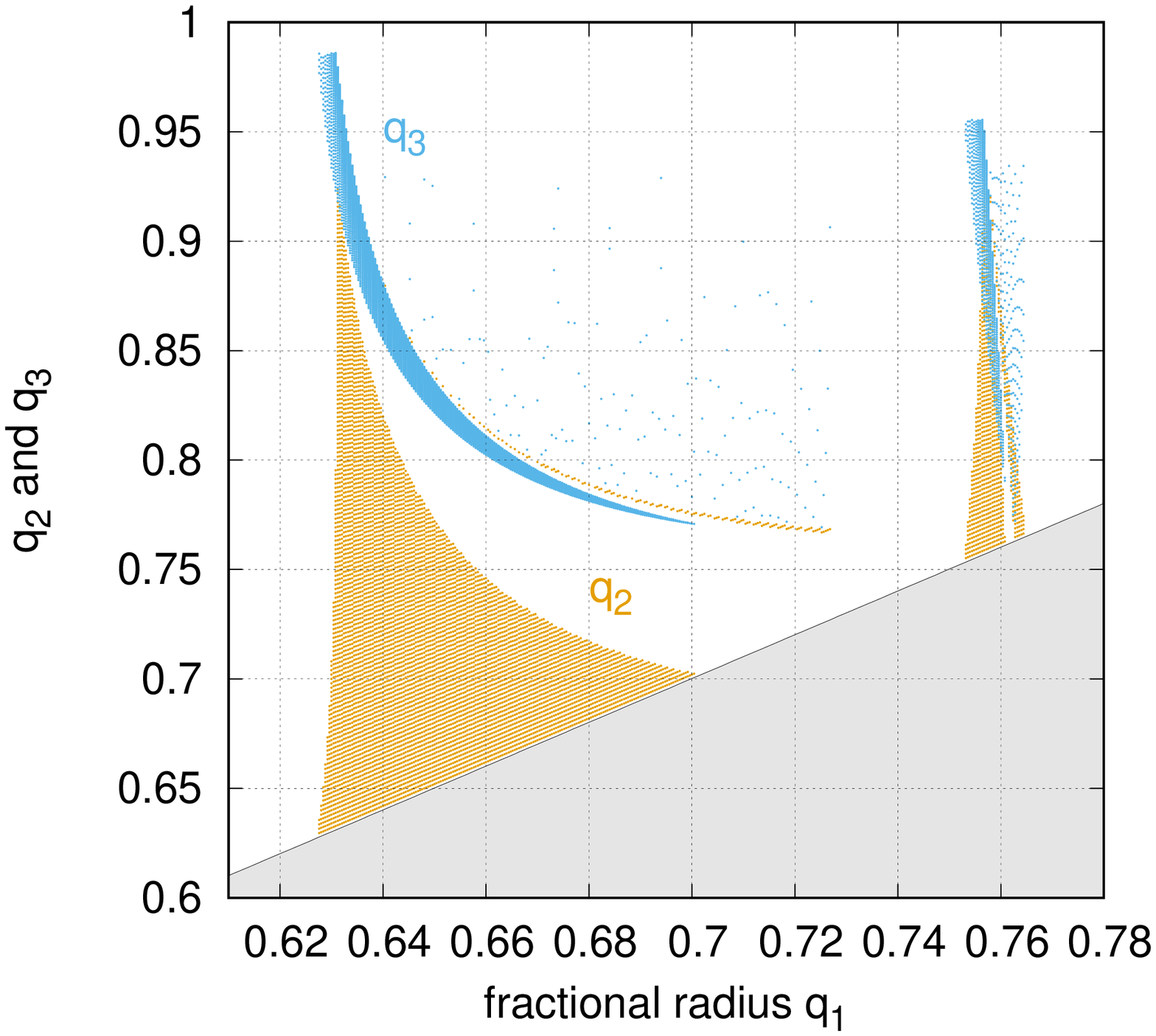}
  \includegraphics[height=8.7cm,bb=40 90 544 730, clip==,angle=-90]{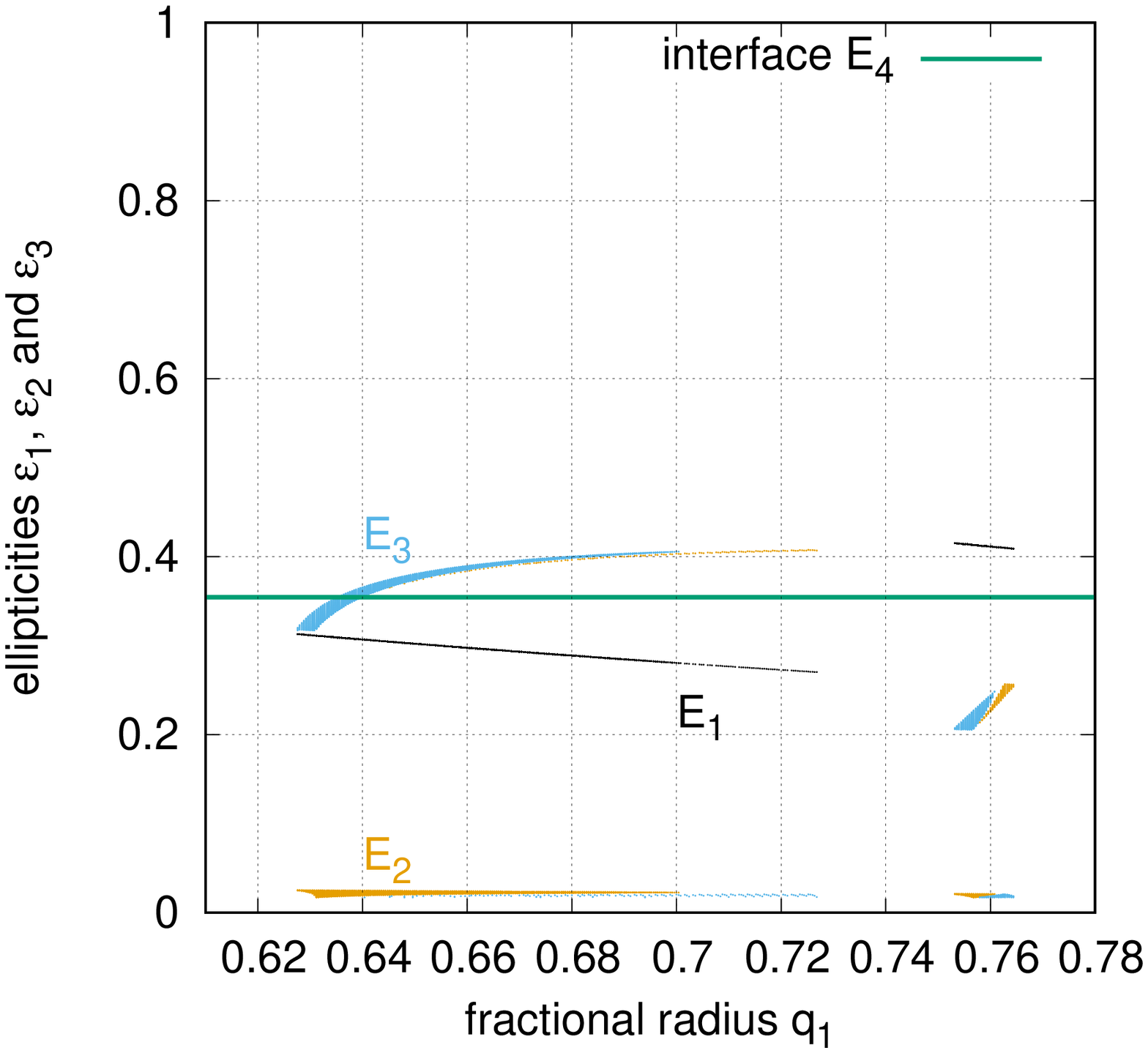}\\
  \includegraphics[height=8.7cm,bb=40 90 544 730, clip==,angle=-90]{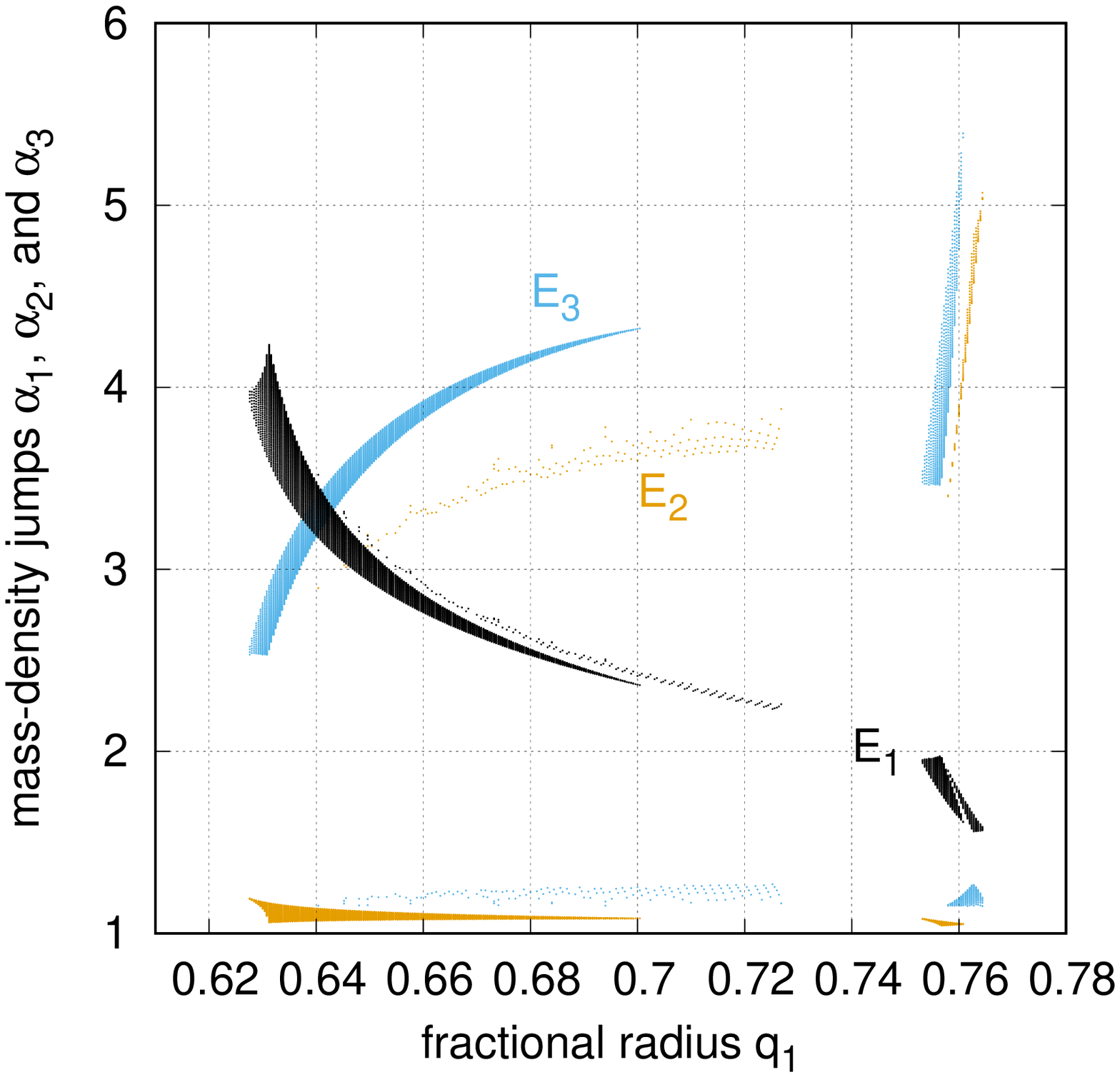}
  \includegraphics[height=8.7cm,bb=40 90 544 730, clip==,angle=-90]{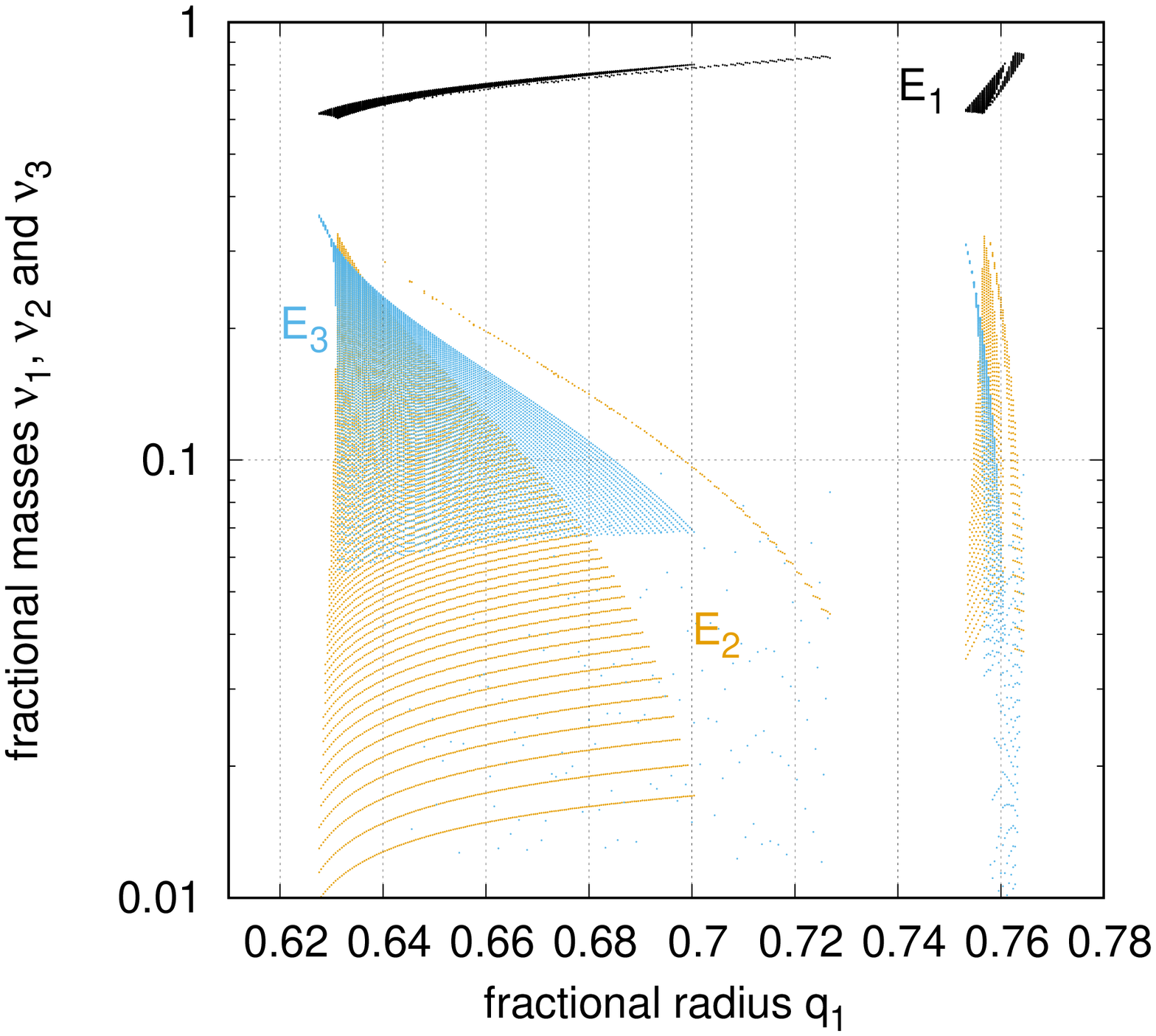}\\
  \includegraphics[height=8.7cm,bb=40 90 544 730, clip==,angle=-90]{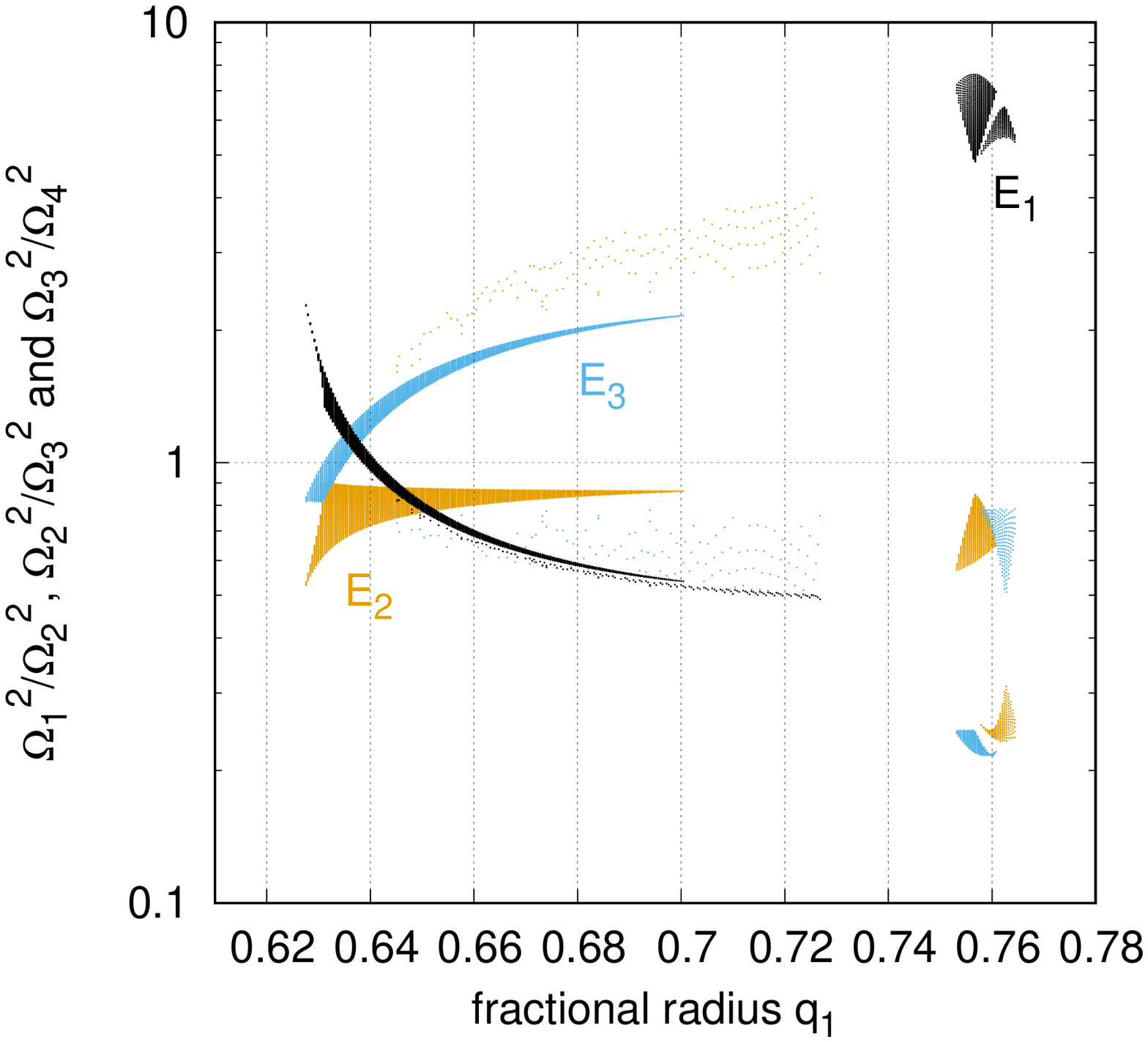}
 \includegraphics[height=8.7cm,bb=40 90 544 730, clip==,angle=-90]{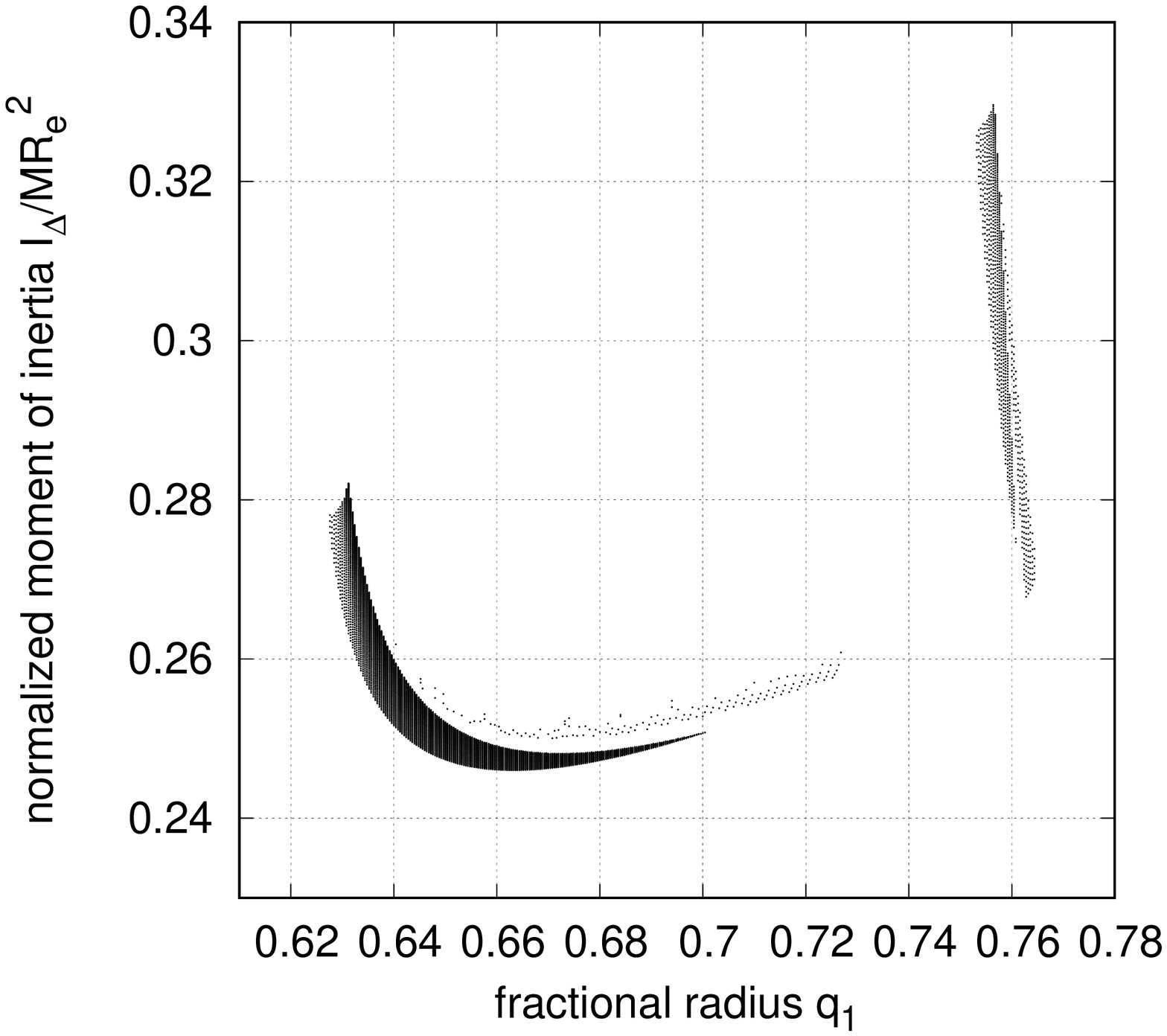}
 \caption{Same caption as for Fig. \ref{fig:4layers_tworootsq1q2q3.ps} but for the four sets $S_{2,1,3}$, $S_{2,3,1}$, $S_{3,1,2}$ and $S_{3,3,1}$; set Tab. \ref{tab:jup4layers_workinga}.}
\label{fig:4layers_Snext_q1q2q3.ps}
\end{figure*}

The results for the four remaing sets are all gathered in Fig. \ref{fig:4layers_Snext_q1q2q3.ps}. With much larger and more massive cores are involved, the configurations resemble to what is obtained from $S_{2,1}$ in the three-layer problem, namely
\begin{itemize}
\item for $S_{2,1,3}$, we have $0.63 \lesssim q_1 \lesssim 0.7$, $\epsilon_2 \ll  \epsilon_1 \lesssim \epsilon_4 \approx \epsilon_3$ ($E_2$ is quasi-spherical), $\alpha_1 \approx 1$ and $\alpha_1  <  \alpha_1\approx \alpha_3 \in [2.5,4.5]$ typically $\nu_1 \gtrsim 0.6 $ and $\nu_2 \lesssim \nu_2$. The rotational discontinuities are in the range $0.5-2$ typically, with no case close to synchronization. The moment of inertia is in the range $0.24-0.28$. 
\item for $S_{2,3,1}$, we have $0.65 \lesssim q_1 \lesssim 0.75$. The situation of layers $2$ and $3$ is reversed compared to $S_{2,1,3}$. Layer $2$ is more massive than layer $3$ ($\alpha_2 > \alpha_3 \approx 1$), and it rotates faster.
\item for $S_{3,1,2}$ : we have $q_1 \approx 0.75$. The core host most of the mass. It rotates faster than the other layers, it is more oblate than the surface layer. The interface $E_2$ is close to spherical. We have $\alpha_2 \approx 1$, and $\alpha_1 < \alpha_3 \approx 4$. The moment of inertia is in the range $0.27-0.33$.
\item for $S_{3,2,1}$ : the core is slightly larger than for $S_{3,1,2}$ with $q_1 \approx 0.75$ but its similar properties as for $S_{3,1,2}$. There is a global reversal between layers $2$ and $3$.
\end{itemize}

As quoted, it is reasonnably not possible to perform the same kind of analysis for all possible canonical sets. We have noticed that the $6$ configurations presented above roughly persist from one canonical set to another. However, when $Y_1$ increases (there is no more root for $Y_1 \gtrsim 0.43$; see Fig. \ref{fig:4layersy1y2y3.ps}), the equilibria involving small cores tend to disappear.

\section{Discussion}
\label{sec:discussion}


This paper investigates the conditions in which the structure at equilibrium of an inhomogeneous body made of ${\cal L}$ homogeneous, heteroeoidal layers in relative orbital motion is compatible with a given set of observational data, namely:\begin{itemize}
\item the equatorial radius $\re$,
\item the spheroidal shape of the outermost layer (through the ellipticity),
\item the total mass $M$,
\item the first gravitational moments $J_2$ to $J_{2n}$,
\item the spin rate $\omobs$ (here, attributed to the deepest ``layer'').
\end{itemize}
As shown, the problem is potentially solvable provided $n$ and ${\cal L}$ are linked by $2{\cal L}-1=n$. The full problem is self-consistent and solved exactly, within the limit of the approximation underlying the theory of nested figure (see Papers I and II). In particular, the spheroidal surface bounding the layers are {\it not necessarily homothetical} or similar, and the rotation rates, imposed by the dynamical equilibrium, are different from each other, which points represent the main originalities of the article. The results have been compared  successfully to numerical solution obtained from the SCF-method. Even though this approach does not account for complex physics appropriate to investigate planetary interiors, it can be used as a reference for numerical models.

\subsection{Summary for Jupiter}

While the assumption of homogeneous layers is hard to justify for gaseous planets, especially at the surface, we have applied the method to Jupiter, as the even moments up to $J_{12}$ are known with precision. As already quoted, we mainly intent to illustrate the impact of non-similar bounding spheroidal surfaces combined with rotationally-decoupled layers on global quantities, in a concrete case. We have considered ${\cal L} \in \{2,3,4\}$.

In the {\it two-layer problem}, we can reproduce the Jupiter's main data, including the two moments $J_2$ and $J_4$. There is no degree of freedom for ${\cal L}=2$: a single configuration matches. The core has an equatorial radius of about $0.689$. Its mass is found to be $244$ Earth masses, with a mass density of about $3$ g/cm$^3$. This value is in good agreement with what can be find in the litterature for 2-layer Jupiter models. The density jump at the interface is about $6.63$, which is far greater than what is usually found \citep{mgf16,ni20}. As a consequence, the core mass derived from this model is $5$ to $10$ times greater that was is commonly deduced. Since the core expands past half the total radius, it can be seen as the concatenation of a compact core and a dilute core. It is not surprising that only two incompressible layers fail to match the results obtained with more realistic models. The ellipticity of the core/envelope interface is about $0.310$. The core is rotating a little bit faster that the envelope, in excess of about $0.02 \%$, which makes the system in a state very close to global rotation. The uncertainties in $J_2$ and $J_4$ have no significant effect in these results. With a rotation period higher by about $3$ s, however, the synchroneous motion can be reached, as Tab. \ref{tab:jup2layers_gr} shows. The model gives very good values of the moments beyond $J_4$.

In the {\it three-layer case}, the even moments match up to $J_8$. The parameter space is wider (but tractable). The number of internal structures that match the observational data becomes basically infinite. There are two groups of equilibria. The first group is characterized by a core fractional radius $q_1$ in the range $[0.58,071]$. The mass of the core is in between $160$ to $255$ Earth masses typically. The central density is similar as for ${\cal L}=2$. The state of global rotation is reached with a few purcents at $q_1 \approx 0.59$. Below this value, the core and the surface layer rotate significantly faster than the intermediate layer, and the situation is reversed beyond this value. Again, we find that a slight increase in the rotation period of the planet permits to better approach the perfect synchronisation of all layers. 
We give in Tab. \ref{tab:drop3layers_gr} the parameters obtained for a rotational period increased by $2$ s, which seems the nominal shift. All these values can undergo some variations due to the uncertainty in $\Delta J_8$ mainly. The second groups of configurations is more singular. The core has a still a high mass, in between $200$ to $265$ Earth masses, but it rotates $10$ times faster than the intermediate layer, and it is significantly oblate. As for the 2-layer case, we only find configurations where the core is much larger and more massive than expected. Again, this can be interpreted as the unification of a compact core with a dilute core. The middle layer can be assimilated to the metallic hydrogen layer enriched with helium, and the envelope (layer 3) to the molecular hydrogen depleted in helium. One essential refinement with respect to the $2$-layer case is the distribution mass density jumps, making the $3$-layer model more satisfactory \citep{hubbard2016preliminary,miguel2016jupiter,wahl17,vazan2018jupiter,net21}. Between the core and the middle layer, the jump ranges from $1.5$ to about $4$, and it is between $2$ and $7.5$ at the middle layer-envelope interface.

The situation becomes yet more complicated in the {\it four-layer case}, which reproduces $J_{2}$ to $J_{12}$. Actually, the parameter space ($6$ dimensions) becomes difficult to probe in details. With ${\cal L}=4$, however, two main properties clearly emerge. The first result concerns the moments. There is, stricly, no solution with central values for $J_{10}$ and $J_{12}$ and the model naturally excludes a wide zone of the error box formed by the uncertainties in  $J_{10}$ and $J_{12}$. If we focus on configurations having a positive gradient of ellipticities from the center to the surface, then we find (including the error bar for $J_8$)
\begin{flalign}
  \begin{cases}
    J_{10} = +0.1921 \pm 0.0050 \times 10^{-6},\\
    J_{12} = -0.0164 \mp 0.0008 \times 10^{-6}.
    \label{eq:meanj10j12}
  \end{cases}
\end{flalign} 
From the recurrence relationship between the $J_{2n}$'s (the recursion differs with ${\cal L}$), the model even predicts values for $J_{14}$. In the same condition as above, we find
\begin{flalign}
      J_{14} = +1.47 \pm 0.11 \times 10^{-9}.
\end{flalign} 
The second important point is the opportunity of getting small cores, in size (about $10 \%$ and less) and in mass (of the order of a few Earth masses). This result, which is not permitted with $2$ and $3$ layers, is specific to the planet considered (through the $A_i$'s). Solutions involving a small and dense core surrounded by a dilute core expanding about $50-60 \%$ are in agreement to what has been recently suggested by other internal structure and formation models \citep{ni2019understanding,militzer2016understanding}. The solution presented in Tab. \ref{tab:drop4layers} and Fig. \ref{fig:drop4layers_perm12.ps} has a large density jump between the compact core and the dilute core, of the order of $10$, which value is compatible with a core mainly composed of heavy elements. The next jumps are much smaller: about $3.8$ between the second and third layer, and about $2.9$ between the third layer and the outermost one. These values are still high compared to current models, and could  be interpreted as the transition between layers hosting different chemical compounds with different molecular weights (for instance, the transition between and helium poor layer and an helium rich one).

It is worth noting that, for ${\cal L} \ge 3$, the rotational discontinuity at the interface between adjacent layers are quite large for some solutions (i.e. $\Omega_i/\Omega_{i+1} \gg 1$ or $\ll 1$). Such configurations could take place in the early phases of planet formation, depending the angular momentum of the accreted material and spin state of the body under construction. The persistence of a relative motion of high amplitude between layers is, however, hard to maintain on long term. It requires a mechanical input to fight against dissipative effects that tend to install all layers in similar rotation states (as long as any meridional circulation can be neglected).

\begin{table}
   \begin{tabular}{rr}
    &this work$^\dagger$  \\ \hline
    $q_1$ & $0.68950$  \\
    $\epsilon_1$ & $0.31018$  \\
     $b_1/a_2=q_1 \bar{\epsilon}_1$ & $0.65549$ \\
    $\alpha_1$ &  $6.63121$  \\
    $\Omega_2^2/\Omega_1^2$ & $1.00000$ \\
    $\nu_1$ & $0.76822$ \\
    $I_\Delta/MR_e^2$ & $0.26311$ \\ \hline
   \end{tabular}\\
   $^\dagger$input data : $a_2,\epsilon_2,M,J_2,J_4$ from Tab. \ref{tab:jupdata}\\
   \hspace*{+1.7cm}$\omobs=\frac{2 \upi}{\mathbf{35732.704}}$ s$^{-1}$
   \caption{Parameters of the two-layer model for Jupiter compatible with global rotation with enhancement of the rotation period; see also Tab. \ref{tab:jup2layers}.}
  \label{tab:jup2layers_gr}
\end{table}

\subsection{Perspectives}

The method can be expanded to accounted for more layers, if needed. For ${\cal L}$ layers, the solution of the $y_i$-problem consists in finding the roots of a $({\cal L}-1)$-degree polynomial, i.e.  
\begin{flalign}
P_{{\cal L}-1}(Y)=0.
  \label{eq:ply}
\end{flalign}
This make sense if more moments are available, at the expense of degeneracy. From an analytical point of view, increasing the number of homogeneous layers to mimic a continuous stratification is probably not a good option. As shown, the $4$-layer problem is already quite complicated.

 The hypothesis of homogenous layers is clearly very restrictive in the context of gaseous bodies, that is why the method presented here is not supposed to be compete with models that use sophisticated Equation-Of-State, but simply to propose a different and complementary approach to the problem and to exhibit the sensitivities. The assumption of rigidly rotating layer is yet another limitation, in particular to model the outermost layers where the dynamics is generally more complex than at great depth. In contrast, the model seems well suited to investigate rortaing rocky planets surrounded by a liquid ocean.
  
A theory of nested figures capable of accounting for non-uniform density profiles would be of major interest. While, in the details, the equations and relationships considered here would be different, the two-step method reported in the article should hold. Even, the degeneracy observed here will be reinforced if layers with different EOS can be accounted for (with polytropic indices as a new set of parameters). Another interesting point that would be worth to study is the relative motion of layers. If huge rotational jumps at interfaces seem not plausible without exciting mechanisms (at least in a stable way), the presence of rotational discontinuities clearly opens onto interesting questions. The shear between layers, if it can be maintained on longterm, is clearly a source of energy dissipation and heat release, which whould be interesting to quantify.

\begin{table}
   \begin{tabular}{rrr}
    &this work$^\dagger$ & this work$^\star$ \\ \hline
     $q_1$ & $0.59070$ & $0.59071$ \\
     $q_2$ & $0.88633$ & $0.88633$ \\
     $\epsilon_1$ & $0.29795$ & $0.29795$ \\
     $\epsilon_2$ & $0.33355$ & $0.33355$\\
     $\alpha_1$  & $3.59224$ & $3.59203$ \\
     $\alpha_2$  & $3.15317$ & $3.15316$ \\
     $\Omega_1^2/\Omega_2^2$  & $1.00009$ & $0.99998$\\
     $\Omega_2^2/\Omega_3^2$  & $1.00001$ & $1.00001$ \\
     $\nu_1$ & $0.56323$ & $0.56325$  \\
     $\nu_2$ & $0.36632$ & $0.36631$ \\    
     $I_\Delta/MR_e^2$ & $0.26351$ & $0.26351$ \\ \hline
  \end{tabular}\\
   \raggedright
   $^\dagger$input data : $a_2,\epsilon_2,M,J_2,J_4,\omobs$\\
   $^\star$input data : $a_2,\epsilon_2,M,J_2,J_4$ from Tab. \ref{tab:jupdata}\\
   \hspace*{+1.7cm}$\omobs=\frac{2 \upi}{\mathbf{35732.704}}$ s$^{-1}$
  \caption{Nominal parameters of the three-layer model for Jupiter compatible with global rotation, without (column 2) and with enhancement of the rotation period (column 3); see Tabs. \ref{tab:jupdata} and \ref{tab:jup3layers}.}
  \label{tab:drop3layers_gr}
\end{table}

\section*{Data availability} All data are incorporated into the article.

\section*{Acknowledgements}
We are grateful to A. Dutrey, S. Guilloteau, T. Guillot, W. Hubbard, E. Di Folco and C. Staelen for stimulating discussions and inputs. We thank the referee for stimulating reports. We thank the MCIA for providing computing time on the local computer.
\bibliographystyle{mn2e}


\onecolumn
\appendix

\section{The $2$-layer case for small ellipticities}
\label{app:chi2}

The case of vanishing ellipticities, which is generally associated with the slow-rotation limit, is interesting as the expressions for $\Omega_1$ and $\Omega_2$ take a simple form, which enables an explcit form for the $\chi_2$-function. At the lowest order, actually, (42) and (58) of Paper I simplify into
\begin{flalign}
  \frac{\Omega_2^2}{2 \upi G\rho_2} \approx \frac{2}{15}\epsilon_1^2 \left[2\frac{\epsilon_2^2}{\epsilon_1^2}+(\alpha_1 -1)q_1^3\left(5\frac{\epsilon_2^2}{\epsilon_1^2}-3q_1^2\right)\right]
  \label{eq:anarot_tv_o2}
\end{flalign}
for the envelope and
\begin{flalign}
  \nonumber
  & \frac{\alpha_1\Omega_1^2}{2 \upi G\rho_2} \approx \frac{\Omega_2^2}{2 \upi G\rho_2}\\
  & \qquad\qquad\qquad + \frac{2}{15}\epsilon_1^2 (\alpha_1 -1)\left[2\alpha_1 - 3\left(\frac{\epsilon_2^2}{\epsilon_1^2}-1\right)\right]
  \label{eq:anarot_tv_o1}
\end{flalign}
for the core. These expressions (type-V solutions; see Paper I) remain general in the sense that they do not assume global rotation (i.e. $\Omega_1 =\Omega_2$). Although $\rho_2$ is not knwon, we can eliminate this quantity by using (\ref{eq:j2j4mass:a}) which yields:
\begin{equation}
\frac{1}{2 \upi G\rho_2} = \frac{a_2^3 \bar{\epsilon_2}}{GM}\frac{2}{3}(1+C_1).
\label{eq:norm2}
\end{equation}
By injecting this expression in (\ref{eq:anarot_tv_o2}) and (\ref{eq:anarot_tv_o1}), we get the expression for $\chi_2(q_1)$. In the general case, we have from
\begin{flalign}
\chi_2(q_1)=\frac{GM \epsilon^2_2}{5 a_2^3 \bar{\epsilon}_2 (1+C_1) \omobs^2} \times \frac{1}{1+\frac{C_1 \compellenv}{q_1^2\sqrt{q_1^2- \ellenv^2 y_1}}}\left\{ 2 + \frac{C_1 \compellenv}{q_1^2\sqrt{q_1^2- \ellenv^2 y_1}} \left[2\frac{C_1 \compellenv}{q_1^2\sqrt{q_1^2- \ellenv^2 y_1}}-3+5\frac{y_1}{q_1^2}+q_1^3(5-3y_1) \right] \right\}- 1
\label{eq:chi2order0_typev}
\end{flalign}
For global rotation, the function simplifies into
\begin{flalign}
\chi_2(q_1)=\frac{GM \epsilon^2_2}{5 a_2^3 \bar{\epsilon}_2 (1+C_1)\omobs^2} \times  \left[ \frac{10y_1(q_1^2-y_1)}{q_1^2[2y_1+q_1^5(3y_1-5)]} -3+5\frac{y_1}{q_1^2} \right]-1.
\label{eq:chi2order0_typec}
\end{flalign}
In this case, the mass density jump is deduced from (\ref{eq:anarot_tv_o2}) and (\ref{eq:anarot_tv_o1}), namely
\begin{flalign}
  \alpha_{1C} \approx 1+ \frac{5(q_1^2-y_1)}{2y_1+q_1^5(3y_1-5)}.
\end{flalign}

\section{Formula for $s$, $d$ and $p$ for the $y_i$-problem of the $4$-layer case}
\label{app:4layer}

\begin{subnumcases}{}
  s=-\frac{(A_3-A_6)(A_1-A_7)-(A_3-A_9)(A_1-A_4)}{(A_2-A_5)(A_1-A_7)-(A_2-A_8)(A_1-A_4)},\\
  d=-\frac{(A_3-A_6)(A_2-A_8)-(A_3-A_9)(A_2-A_5)}{(A_2-A_5)(A_1-A_7)-(A_2-A_8)(A_1-A_4)},\\
  p=\frac{(A_3A_4-A_6A_1)(A_2A_7-A_1A_8)-(A_3A_7-A_1A_9)(A_2A_4-A_1A_5)}{(A_4-A_1)(A_2A_7-A_1A_8)-(A_7-A_1)(A_2A_4-A_1A_5)}
\end{subnumcases}

\end{document}